# Using Nyquist or Nyquist-Like Plot to Predict Three Typical Instabilities in DC-DC Converters

Chung-Chieh Fang

**Abstract**

By transforming an exact stability condition, a new Nyquist-like plot is proposed to predict occurrences of three typical instabilities in DC-DC converters. The three instabilities are saddle-node bifurcation (coexistence of multiple solutions), period-doubling bifurcation (subharmonic oscillation), and Neimark bifurcation (quasi-periodic oscillation). In a single plot, it accurately predicts whether an instability occurs and what type the instability is. The plot is equivalent to the Nyquist plot, and it is a useful design tool to avoid these instabilities. Nine examples are used to illustrate the accuracy of this new plot to predict instabilities in the buck or boost converter with fixed or variable switching frequency.

C.-C Fang is with Advanced Analog Technology, 2F, No. 17, Industry E. 2nd Rd., Hsinchu 300, Taiwan, Tel: +886-3-5633125 ext 3612, Email: fangcc3@yahoo.com

# Contents

## I. Introduction

Operations of nonlinear DC-DC converters in continuous conduction mode (CCM) can be described *exactly* by the switching system shown in Fig. 1 (notations explained later) [1], which leads to a sampled-data dynamics [1]. Instability occurs when there exists a sampled-data pole outside the unit circle in the complex plane. There are three ways that the sampled-data pole leaves the unit circle, thus causing three typical instabilities in DC-DC converters [1], [2], [3]. When the sampled-data pole leaves the unit circle through 1 in the complex plane, the instability is generally a saddle-node bifurcation (SNB) [2] (or pitchfork and transcritical bifurcations [4] which are less seen in DC-DC converters). The SNB generally involves coexistence of multiple solutions [5], or sudden disappearances or jumps of steady-state solutions [2]. When the pole leaves the unit circle through -1, the instability is a period-doubling bifurcation (PDB) [6], which generally involves *fast-scale* subharmonic oscillation. When the pole leaves through a point other than 1 or -1 on the unit circle, the instability is a Neimark-Sacker bifurcation (NSB), which generally involves a *slow-scale* quasi-periodic oscillation [1], [2], [3]. Many instability examples of DC-DC converters are shown in [7], [8], [9], [10], [11].

Averaged models are traditionally applied to analyze DC-DC converters. It has been known that most averaged models cannot accurately predict the occurrence of PDB (subharmonic oscillation) [6]. By considering the sampling effects and increasing the system dimension, improved averaged models can predict the occurrence of PDB [12], [13]. However, there is no *single* design-oriented tool like the Nyquist plot to predict the occurrences of *all three* instabilities. In a recent paper [14], two *different* approaches for slow-scale instability (NSB) and fast-scale instability (PDB) respectively are combined to predict instabilities in DC-DC converters.

The switching system of Fig. 1 is a well-accepted model to describe the operation of DC-DC converters. The stability of the switching system of Fig. 1 is a pure *mathematical* problem and has a closed-form solution. It is well known [15] that the condition of pole locations can be easily transformed to a characteristic equation, and the Nyquist plot determines the stability. Then, one has the next question: how to derive an exact dynamics to be used for the Nyquist plot?

In this paper, a new Nyquist-like plot, designated as “F-plot”, based on sampled-data dynamics is proposed to predict the occurrences of *all three* instabilities in a *single* plot. The F-plot is based on the *exact* switching system. Therefore, the F-plot is also exact and it predicts the exact instability boundaries of these instabilities. It is applicable to converters under either current mode control (CMC) or voltage mode control (VMC).

The remainder of the paper is organized as follows. In Sections II and III, the exact switching system and the sampled-data analysis are presented. In Sections IV and V, a new F-plot is proposed and it is equivalent to the Nyquist plot. In Sections VI-VIII, the F-plot or Nyquist plot is applied to various examples to predict PDB, SNB, and NSB, respectively. In Section IX, similar methodology is applied to constant on-time control (COTC). Conclusions are collected in Section X.

## II. Brief Review of Exact Switching System

A unique aspect about the switching system of Fig. 1 is that the ramp in VMC or CMC is represented by the same signal $h(t)$ [1], [16], [17], [18]. The operation of VMC and CMC is briefly reviewed here to make this paper self-contained. In VMC, the reference signal $v_r$ controls the output voltage $v_o$. In CMC, $v_r$ is generally replaced by $i_c$ to denote a control signal to control the (peak) inductor current $i_L$. Denote the source voltage as $v_s$. In the model, $A_1, A_2 \in \mathbf{R}^{N\times N}$, $B_1, B_2 \in \mathbf{R}^{N\times 2}$, $C, E_1, E_2 \in \mathbf{R}^{1\times N}$, and $D \in \mathbf{R}^{1\times 2}$ are constant matrices, where $N$ is the system dimension. Within a clock period $T$, the dynamics is switched between two stages, $S_1$ and $S_2$. Switching occurs when the ramp signal $h(t) := V_l + (V_h - V_l)(\frac{t}{T} \bmod T)$ intersects with the compensator output $y := Cx + Du \in \mathbf{R}$, where $h(t)$ varies from a low value $V_l$ to a high value $V_h$. Denote the ramp amplitude as $V_m = V_h - V_l$, and let the ramp slope be $\dot{h}(d) = m_a$. Denote the switching frequency as $f_s := 1/T$ and let $\omega_s := 2\pi f_s$.

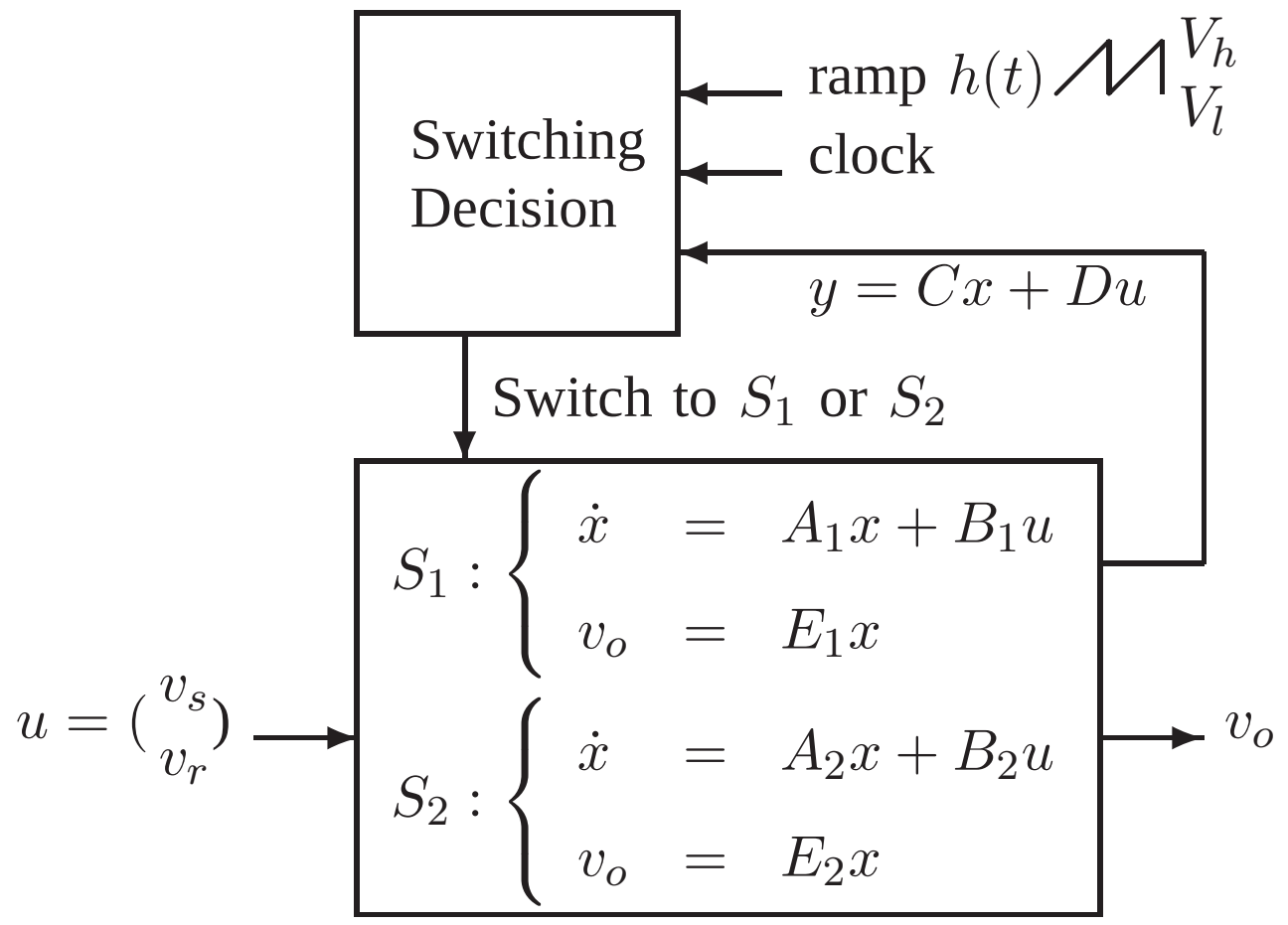

Figure 1. Exact switching system for a DC-DC converter in CCM.

## III. Small-Signal Analysis

Let the steady-state duty cycle be $D$ and let $d := DT$. Let the $T$-periodic orbit be $x^0(t)$. Let $\dot{x}^0(d^-) = A_1x^0(d) + B_1u$ and $\dot{x}^0(d^+) = A_2x^0(d) + B_2u$ denote the time derivative of $x^0(t)$ at $t = d^-$ and $d^+$, respectively. Let $y^0(t) = Cx^0(t) + Du$. One has $\dot{y}^0(t) = C\dot{x}^0(t)$. Confusion of notations for capacitance $C$ and duty cycle $D$ with the matrices $C$ and $D$ can be avoided from the context.

In steady state,

$$x^0(d) = e^{A_1 d}x^0(0) + \int_0^d e^{A_1\sigma} d\sigma B_1 u \tag{1}$$

$$x^0(0) = e^{A_2(T-d)}x^0(d) + \int_0^{T-d} e^{A_2\sigma} d\sigma B_2 u \tag{2}$$

$$y^0(d) = Cx^0(d) + Du = h(d) \tag{3}$$

Solving (1)-(3) gives $D$ and $x^0(d)$.

Generally the controller includes an integrator (with a pole at zero), making $A_1$ (or $A_2$) and $I - e^{A_1T}$ non-invertible. In that case, the pole at zero can be replaced by a very small number, then $A_1$ and $I - e^{A_1T}$ are invertible. Therefore, the invertibility of $A_1$ or $I - e^{A_1T}$ is not critical and can be resolved. This statement about invertibility of a matrix is not repeated later.

Using a hat $\hat{}$ to denote small perturbations (e.g., $\hat{x}_n = x_n - x^0(0)$), where $x_n$ is the sampled state at $t = nT$. From [1], [16], [19], [20], the linearized closed-loop dynamics is

$$\hat{x}_{n+1} = \Phi\hat{x}_n := e^{A_2(T-d)}\Big(I - \frac{(\dot{x}^0(d^-) - \dot{x}^0(d^+))C}{\dot{y}^0(d^-) - \dot{h}(d)}\Big)e^{A_1 d}\hat{x}_n \tag{4}$$

Instability occurs when there exists an eigenvalue $z$ (also the sampled-data pole) of $\Phi$ outside the unit circle.

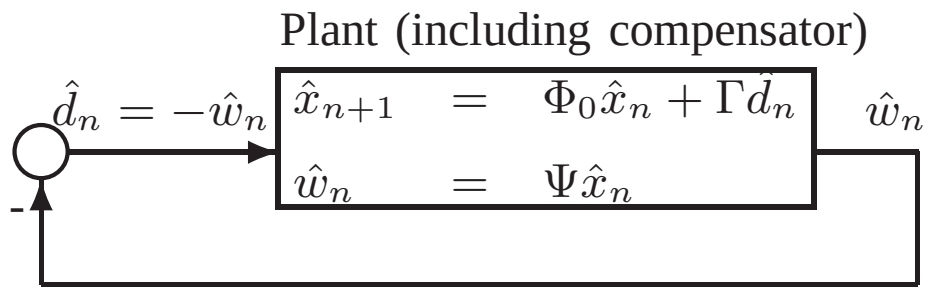

Figure 2. A plant with unity negative feedback.

## IV. Discrete-Time Nyquist Plot

Let $\Phi = \Phi_0 - \Gamma\Psi$, where

$$\Phi_0 = e^{A_2(T-d)} e^{A_1 d} \tag{5}$$

$$\Gamma = e^{A_2(T-d)} (\dot{x}^0(d^-) - \dot{x}^0(d^+)) \tag{6}$$

$$\Psi = \frac{Ce^{A_1 d}}{C\dot{x}^0(d^-) - m_a} \tag{7}$$

The dynamics (4) can be transformed into a plant (having an input $\hat{d}_n$ and an output $\hat{w}_n$, shown in Fig. 2) with a *unity negative* feedback ($\hat{d}_n = -\hat{w}_n$)

$$\begin{aligned} \hat{x}_{n+1} &= \Phi_0 \hat{x}_n + \Gamma \hat{d}_n \\ \hat{d}_n &= -\Psi \hat{x}_n \end{aligned} \tag{8}$$

Although it looks like that (8) is arbitrarily transformed from (4), the dynamics (8) does have a meaning. It is exactly the power stage (but including the compensator) dynamics [20] with a PWM modulator gain $1/(C\dot{x}^0(d^-) - m_a)$, or $1/(C\dot{x}^0(d^-) - m_a)T$ if the duty cycle is used as the input. The modulator gain agrees with [12] which can predict PDB. It differs from $1/m_a T$ based on traditional averaging [21], [22] which cannot predict PDB.

The past research generally focuses on averaged dynamics. The Nyquist plot of the averaged dynamics without considering the sampling effect cannot predict PDB. Also, the past research focuses on (4), without transforming it into (8) which could lead to the Nyquist plot.

Let the loop gain transfer function of (8) be $\mathcal{N}(z)$. The characteristic equation is

$$\mathcal{N}(z) = \Psi(zI - \Phi_0)^{-1}\Gamma = -1 \tag{9}$$

The (discrete-time) Nyquist plot is $\mathcal{N}(e^{j\omega T})$ as a function of $\omega$ swept from 0 to $\omega_s/2$ for a half plot (or from $-\omega_s/2$ to $\omega_s/2$ for a full plot).

## V. Exact Stability Condition and the "F-Plot"

Although using the Nyquist plot is enough to predict instability, a new plot is proposed to give a different perspective. Based on [1, p. 46] and [23], [18], suppose $z$ is not an eigenvalue of $\Phi_0$, then $z$ is an eigenvalue of $\Phi$ if and only if the following critical condition holds:

$$\mathcal{M}(z) := C\dot{x}^0(d^-) + Ce^{A_1 d}(zI - \Phi_0)^{-1}\Gamma = m_a \tag{10}$$

The proof is as follows. Suppose $z$ is not an eigenvalue of $\Phi_0$. Then

$$\begin{aligned}\det[zI-\Phi] &= \det[zI-\Phi_0]\det[I+(zI-\Phi_0)^{-1}\Gamma\Psi] \\ &= \det[zI-\Phi_0](1+\Psi(zI-\Phi_0)^{-1}\Gamma)\end{aligned}$$

Then $\det[zI-\Phi]=0$ requires that $\mathcal{N}(z)=\Psi(zI-\Phi_0)^{-1}\Gamma=-1$, which leads to (10) by using (5)-(7).

Note that (10) is equivalent to the characteristic equation (9) through a *scaled translation*:

$$\mathcal{M}(z)=(C\dot{x}^0(d^-)-m_a)\mathcal{N}(z)+C\dot{x}^0(d^-) \tag{11}$$

Although the stability condition (10) looks unfamiliar, a special case of (10) for PDB in CMC would be familiar. When the switching frequency is high, $e^{A_1 d}\approx e^{A_2(T-d)}\approx I$. Then the critical condition (10) with $z=-1$ (for PDB) becomes

$$\frac{1}{2}(C\dot{x}^0(d^-)+C\dot{x}^0(d^+))=\frac{1}{2}(\dot{y}^0(d^-)+\dot{y}^0(d^+))=m_a \tag{12}$$

In CMC, the (inductor current) slopes $\dot{y}^0(d^-)$ and $\dot{y}^0(d^+)$ are generally expressed in most textbooks [21, p. 448] as $-m_1$ and $m_2$, respectively. Then, (12) corresponds exactly to the well known minimum ramp slope $m_a=(m_2-m_1)/2$, or equivalently $(m_a-m_2)/(m_a+m_1)>-1$, required to stabilize the converter [21], [24]. From (4), (10), or (12), one sees that the stability is closely related to the signal *slopes* [1], [20].

Instability occurs when the pole $z$ leaves the unit circle, or equivalently, $z$ satisfies the critical condition (10). One can define a plot based on the critical condition (10) with $z$ being swept around the unit circle. Let $z=e^{i\theta}$ (on the unit circle), and define an "F-plot" in the complex plane as

$$F(\theta)=\mathcal{M}(e^{i\theta})=(C\dot{x}^0(d^-)-m_a)\mathcal{N}(e^{i\theta})+C\dot{x}^0(d^-) \tag{13}$$

which is a scaled translation of Nyquist plot. From (13), if $\mathcal{N}(e^{i\theta})=-1$, one has $F(\theta)=m_a$. The "-1 point" in the Nyquist plot is translated to the "$m_a$ point," $(m_a,0)$, in the F-plot.

The idea behind the F-plot is exactly the *same* as Nyquist plot. As $z$ is swept along the unit circle (and $\theta:=\omega T$ is swept from $-\pi$ to $\pi$), one can determine the number of poles outside the unit circle (and hence the stability) based on the F-plot. Like the Nyquist plot, $\mathbf{Im}[F(\theta)]=-\mathbf{Im}[F(-\theta)]$ and the F-plot is symmetric with respect to the real axis. One can make the F-plot for $\theta\in[0,\pi]$ instead of $[-\pi,\pi]$.

Let $\mathbb{P}$ be the number of open-loop poles of the plant in Fig. 2 outside the unit circle. Generally, all eigenvalues of $\Phi_0$ are inside (or on) the unit circle [20], then $\mathbb{P}=0$. Let $\mathbb{Z}$ be the number of roots of the characteristic equation $\mathcal{N}(z)$. Let $\mathbb{N}$ be the number of (clockwise) encirclements of Nyquist plot $\mathcal{N}(e^{j\omega T})$ around the -1 point. From [15], $\mathbb{N}=\mathbb{Z}-\mathbb{P}=\mathbb{Z}$, and the following numbers are also equal to $\mathbb{N}$:

1) The number of poles for the dynamics (4) outside the unit circle.
2) The number of poles for the dynamics (8) outside the unit circle.
3) The number of eigenvalues of $\Phi=\Phi_0-\Gamma\Psi$ outside the unit circle.
4) The number of encirclements of the F-plot around the $m_a$ point.

In terms of the pole $z$ for (4) or (8), the pole $z$ satisfies the following *equivalent* equations

1) $\det[zI-\Phi]=\det[zI-\Phi_0+\Gamma\Psi]=0$,
2) the characteristic equation, $\mathcal{N}(z)=\Psi(zI-\Phi_0)^{-1}\Gamma=-1$, and
3) the critical condition, $\mathcal{M}(z)=C\dot{x}^0(d^-)+Ce^{A_1 d}(zI-\Phi_0)^{-1}\Gamma=m_a$.

The critical conditions of PDB, SNB, and NSB are summarized in Table I. For example, PDB occurs when $\lambda=-1$, $F(\pi)=m_a$, or $\mathcal{N}(-1)=-1$. The procedure of stability analysis is as follows. First, solve

Table I

CRITICAL CONDITIONS OF PDB, SNB, AND NSB.

| Bifurcation | Sampled-data pole | F-plot | Nyquist plot |
|---|---|---|---|
| PDB | $z = -1$ | $F(\pi) = m_a$ | $\mathcal{N}(-1) = -1$ |
| SNB | $z = 1$ | $F(0) = m_a$ | $\mathcal{N}(1) = -1$ |
| NSB | $z = e^{i\theta}$, $\theta \neq 0$ or $\pi$ | $F(\theta) = m_a$ | $\mathcal{N}(e^{i\theta}) = -1$ |

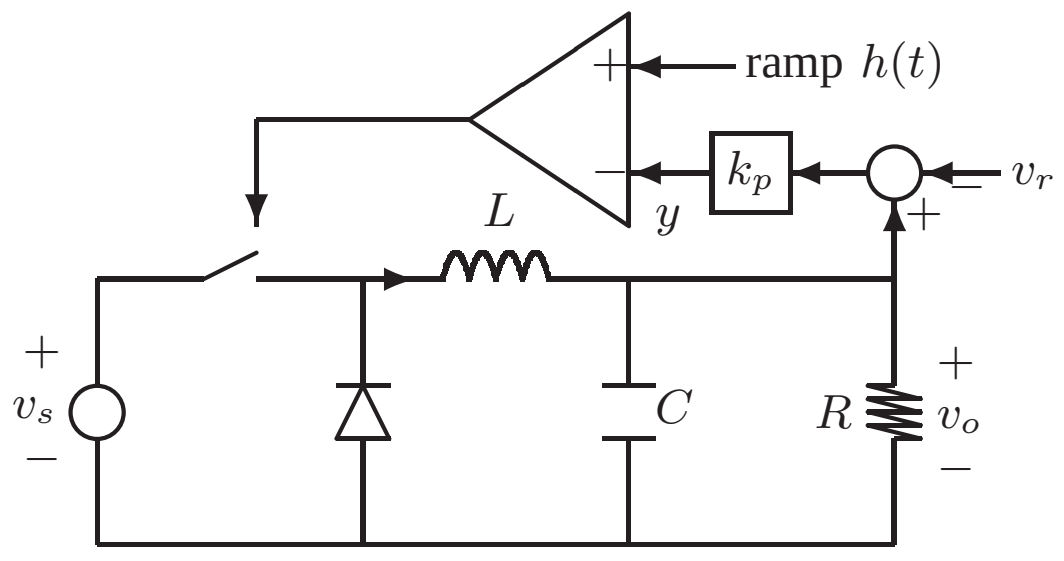

Figure 3. Buck converter under PVMC.

(1)-(3) and obtain $D$ and $x^0(d)$, which can be also obtained approximately from the average analysis. Then, make the F-plot based on (10) and (13). Using (5)-(9), one can also make Nyquist or Bode plot.

In the following nine examples, various control schemes (such as average current mode control and type-III compensator) for buck or boost converters are used. The F-plot is compared with Nyquist and Bode plots. All of these plots accurately predict the three typical instabilities.

The F-plot starts from $F(0)$ (marked as *) and ends at $F(\pi)$ (marked as ○). The other half of symmetric part for $\theta \in [-\pi, 0]$ is not plotted for briefness. The $m_a$ point, located at $(m_a, 0)$ in the complex plane, is marked as $+$. In each example, two F-plots (as solid and dotted lines, respectively) are made *before* and *after* the bifurcation to show the onset of different instabilities predicted by the F-plot.

## VI. PREDICTION OF PERIOD-DOUBLING BIFURCATION (PDB)

**Example 1.** (*The F-plot accurately predicts PDB and shows the required ramp slope to stabilize the converter*.) Consider a widely studied buck converter with proportional voltage mode control (PVMC) from [6] shown in Fig. 3. The proportional feedback gain is $k_p = 8.4$. The converter parameters are $T = 400$ $\mu$s, $L = 20$ mH, $C = 47$ $\mu$F, $R = 22$ $\Omega$, $V_l = 3.8$, $V_h = 8.2$, and $v_r = 11.3$ V. The PDB is known to occur when $v_s = 24.5$.

**Time-domain simulation.** First, let $v_s = 24$. The converter is stable with a $T$-periodic orbit shown in Fig. 4(a). Next, let $v_s = 25$. The $T$-periodic orbit is unstable. Subharmonic oscillation occurs with a $2T$-periodic orbit shown in Fig. 4(b).

**F-plot.** Here, the ramp slope is $m_a = V_m/T = 11000$. First, let $v_s = 24$, the converter is stable. Solving (1)-(3) gives $D = 0.5$. The F-plot (Fig. 5) does not encircle the $m_a$ point (if it is plotted for $\theta \in [-\pi, \pi]$ which includes the other half of symmetric part of the F-plot). At $v_s = 24.5$, $F(\pi)$ touches the $m_a$ point and PDB occurs.

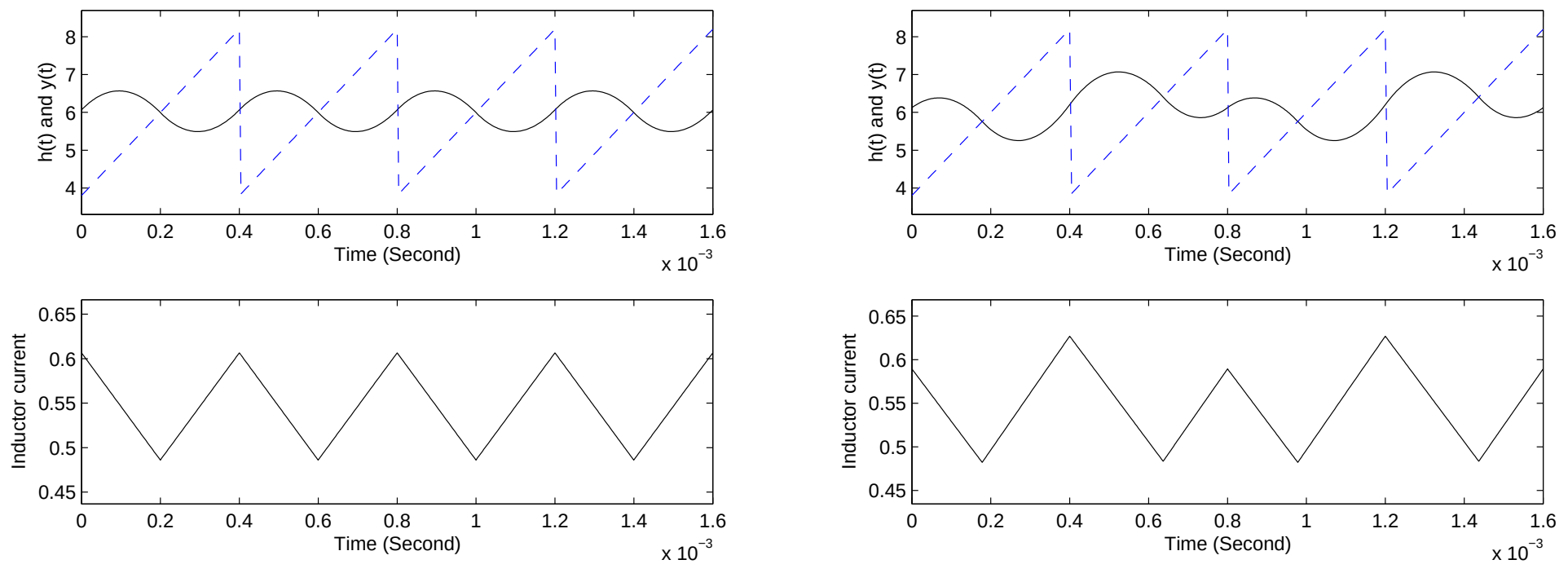

(a) (b)

Figure 4. Time-domain simulations. (a) Stable $T$-periodic orbit, $v_s = 24$; (b) Subharmonic oscillation occurs with a $2T$-periodic orbit, $v_s = 25$.

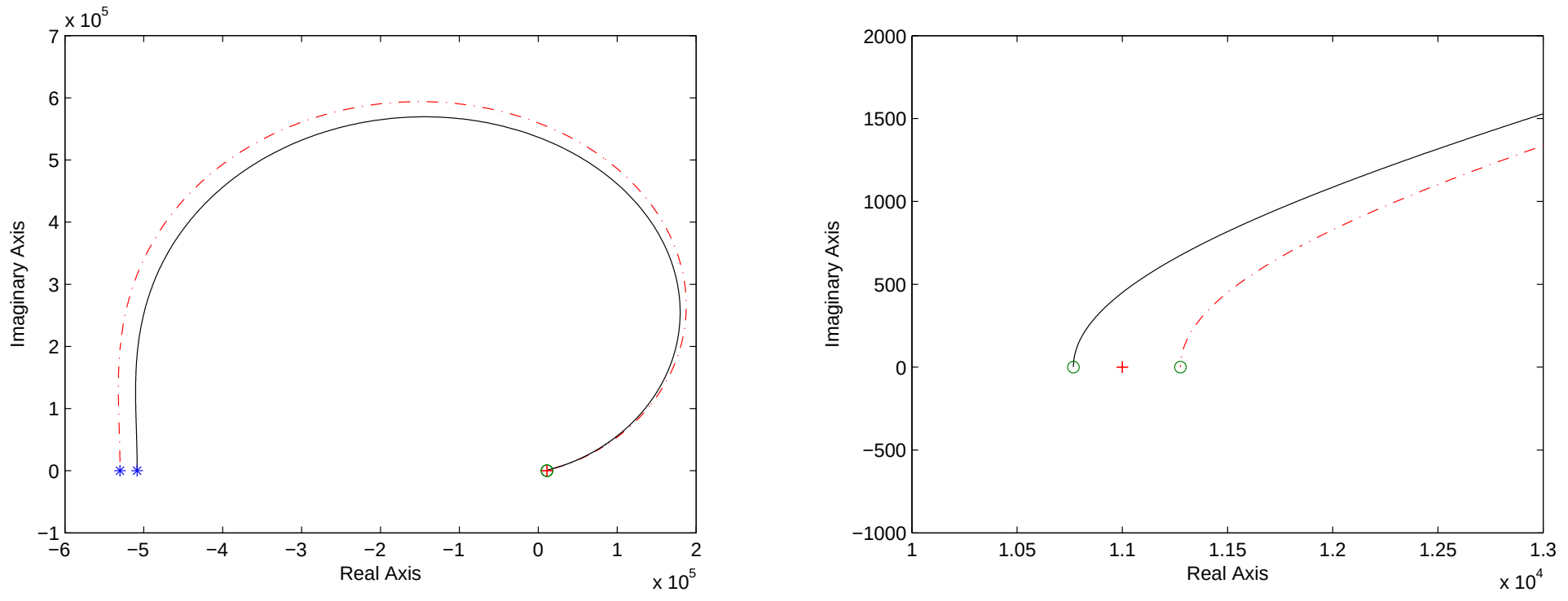

(a) (b)

Figure 5. (a) F-plot for $\theta \in [0, \pi]$, (b) blown-up view of (a); solid line for $v_s = 24$ (stable) and dotted line for $v_s = 25$ (unstable).

Next, let $v_s = 25$, the converter is unstable. Solving (1)-(3) gives $D = 0.48$. The F-plot (Fig. 5) encircles the $m_a$ point (if it is plotted for $\theta \in [-\pi, \pi]$). The F-plot accurately predicts the occurrence of PDB. The F-plot also shows how to adjust the ramp slope to stabilize the converter with $v_s = 25$. The F-plot in Fig. 5 shows that if $m_a > F(\pi) = 11276$ the converter is stabilized. To keep the original operating condition at $v_s = 25$ and $D = 0.48$, let $V_l = 3.6856$ and $V_h = 8.3056$. Now, $m_a = 11550$. The sampled-data poles are $-0.8202 \pm 0.0803j$ (inside the unit circle), confirming the stability.

**Nyquist/Bode plot.** First, let $v_s = 24$. The Nyquist plot (Fig. 6) does not encircle the -1 point. The Bode plot (Fig. 7) shows a very small gain margin of 0.0933 dB.

Next, let $v_s = 25$. The Nyquist plot (Fig. 6) encircles the -1 point. The Nyquist plot starting from $\mathcal{N}(1)$ (for $\omega = 0$) to $\mathcal{N}(-1)$ (for $\omega = \omega_s/2$) is the lower half in Fig. 6. Here, $\mathcal{N}(-1)$ is closer to the -1 point than $\mathcal{N}(1)$, indicating that it is a PDB instead of an SNB. The Nyquist plot accurately predicts the occurrence of PDB. The Bode plot is similar to Fig. 7 but with a gain margin of -0.108 dB at the subharmonic frequency $\omega_s/2 = 7854$ rad/s. □

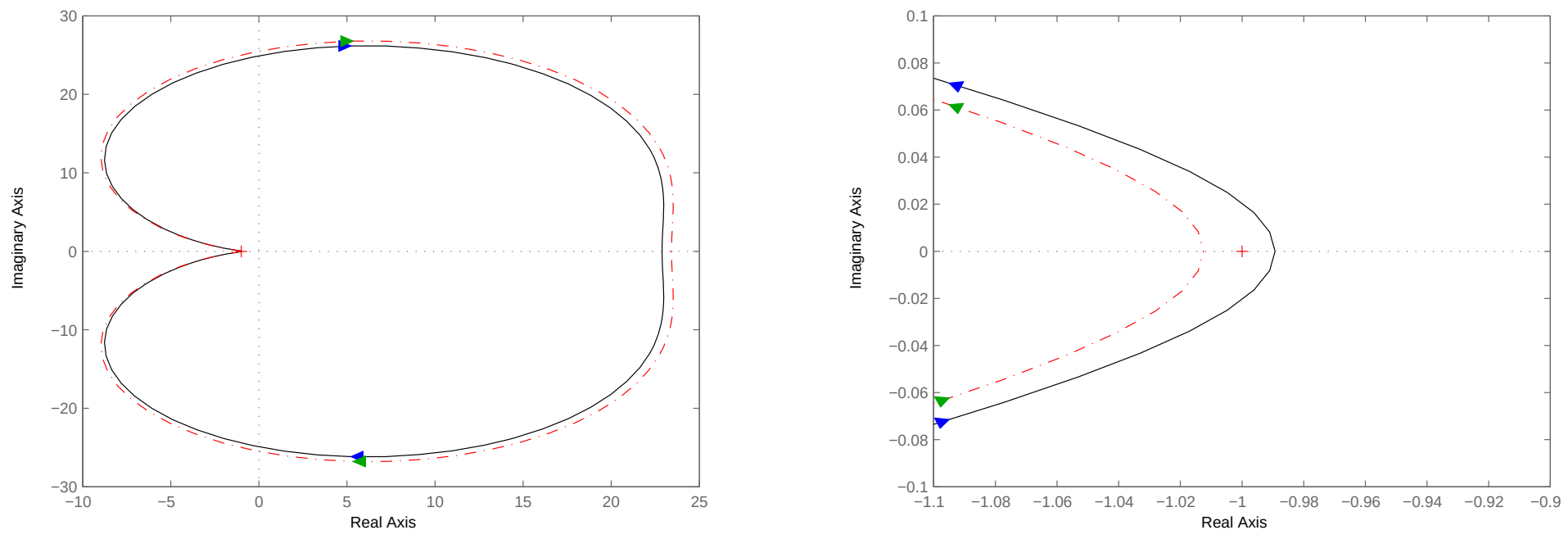

(a) (b)

Figure 6. (a) Nyquist plot, (b) blown-up view of (a); solid line for $v_s = 24$ (stable) and dotted line for $v_s = 25$ (unstable).

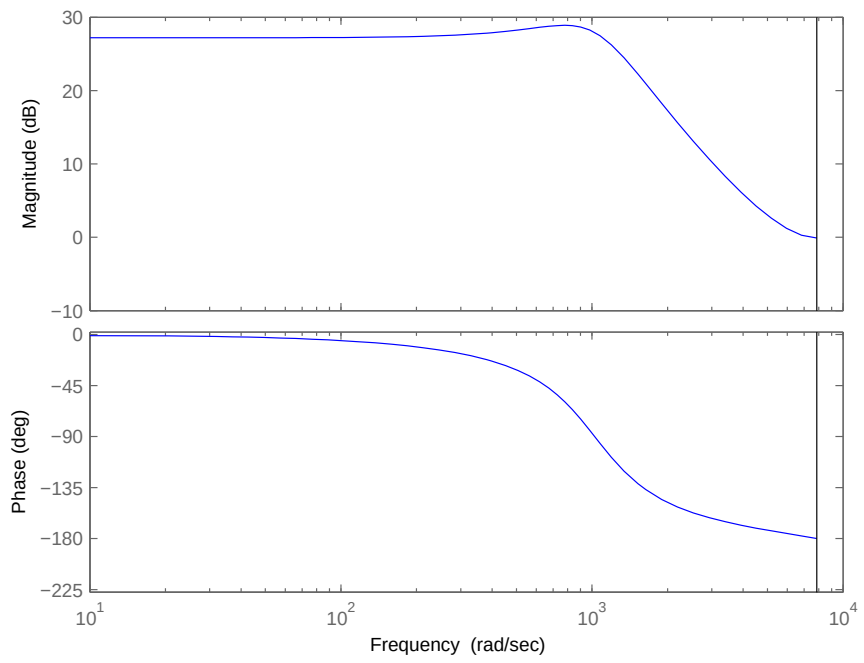

Figure 7. Bode plot for $v_s = 24$.

The next example shows that PDB occurs not only in peak current mode control but also in average current mode control (ACMC). The PDB instability in ACMC was reported ten years ago [25] but it was unnoticed. The instability phenomenon is very interesting because the PDB instability occurs only in a window value of a pole. Since the value of the pole does not affect the duty cycle but affects the ripple size, this example also shows that the PDB instability is unrelated to the ripple size. For details see [7], [8], [11]. The instability is verified by three independent approaches: time-domain simulation, the sampled-data pole trajectories, and F/Nyquist/Bode plots.

**Example 2.** (*The F-plot accurately predicts the subharmonic oscillation window in a buck converter under ACMC.*) Consider an ACMC buck converter [26, p. 114] shown in Fig. 8. The ACMC type-II compensator is

$$G_c(s) = \frac{K_c(1 + \frac{s}{\omega_z})}{(s + \delta)(1 + \frac{s}{\omega_p})} \tag{14}$$

The converter parameters are $v_s = 14$ V, $v_o = 5$ V, $v_r = 0.5$, $f_s = 50$ kHz, $L = 46.1$ $\mu$H, $C = 380$ $\mu$F with ESR $R_c = 0.02$ $\Omega$, $R = 1$ $\Omega$, $R_s = 0.1$ $\Omega$, $V_l = 0$, $V_h$=1, $m_a = 50000$, $K_c = 75506$, and $\omega_z = 5652.9$ rad/s. Let $\delta = 1$ be the integrator pole which is small and it does not affect much the dynamics.

**Time-domain simulation.** The compensator pole $\omega_p$ is varied from $0.14\omega_s$ to $0.81\omega_s$. An unstable *window* of $\omega_p$ between $0.18\omega_s$ and $0.49\omega_s$ was found and reported in [25]. When $\omega_p$ is inside the window, the subharmonic oscillation occurs.

For example, let $\omega_p = 0.15\omega_s$. The converter is stable (Fig. 9). Let $\omega_p = 0.49\omega_s$. The converter is unstable with subharmonic oscillation (Fig. 10). Next, let $\omega_p = 0.81\omega_s$. The converter is stable again (Fig. 11).

**Independent sampled-data analysis.** The sampled-data pole trajectories for $\omega_p/\omega_s \in (0.1, 0.8)$ are shown in Fig. 12. There are four poles. Two poles are fixed around 0.88, and 0.95 ($\approx e^{\frac{-T}{RC}}$). A pole leaves the unit circle through -1 when $\omega_p = 0.18\omega_s$, and enters the unit circle when $\omega_p = 0.49\omega_s$. This explains exactly the instability window of $\omega_p$.

**F-plot.** Solving (1)-(3) gives $D = 0.3571$. First, let $\omega_p = 0.15\omega_s$, the converter is stable. The F-plot is shown in Fig. 13. The non-smoothness of the curve is due to the small integrator pole $\delta$. For $\theta \in (0, \pi]$, the curve is actually smooth. When $\theta$ is close to 0, the F-plot trajectory suddenly jumps from $F(0) = -5 \times 10^9$ to a large value and forms an abrupt bent. The blown-up views of the F-plot are shown in Fig. 14. The F-plot does not encircle the $m_a$ point. The sampled-data poles are 0.8868, 0.9521, and $-0.6406 \pm 0.1557j$ (inside the unit circle), also confirming the stability.

Next, let $\omega_p = 0.49\omega_s$, the converter is unstable. The F-plot is shown in Fig. 15 and it encircles the $m_a$ point. The sampled-data poles are -1.0080, -0.0511, 0.8826, 0.9533, also confirming the instability.

To show the instability window, let $\omega_p = 0.5\omega_s$, the converter is stable again. The F-plot is shown in Fig. 16 and it does not encircle the $m_a$ point. The sampled-data poles are 0.0487, 0.9936, 0.8825, 0.9533, also confirming the stability. The F-plot accurately predicts the instability window.

**Nyquist/Bode plot.** First, let $\omega_p = 0.15\omega_s$. The Nyquist plot (Fig. 17) does not encircle the -1 point. The Bode plot (Fig. 18) shows a positive gain margin. Second, let $\omega_p = 0.49\omega_s$. The Nyquist plot (Fig. 19) encircles the -1 point. The Bode plot (Fig. 20) shows a negative gain margin. Third, let $\omega_p = 0.5\omega_s$. The Nyquist plot (Fig. 21) does not encircle the -1 point. The Bode plot (Fig. 22) shows a positive gain margin. □

**Example 3.** (*With phase margin of* $38.9°$ *based on the averaged model, subharmonic oscillation still occurs.*) A converter with a type-III compensator is shown in Fig. 23. A type-III compensator [27, p.

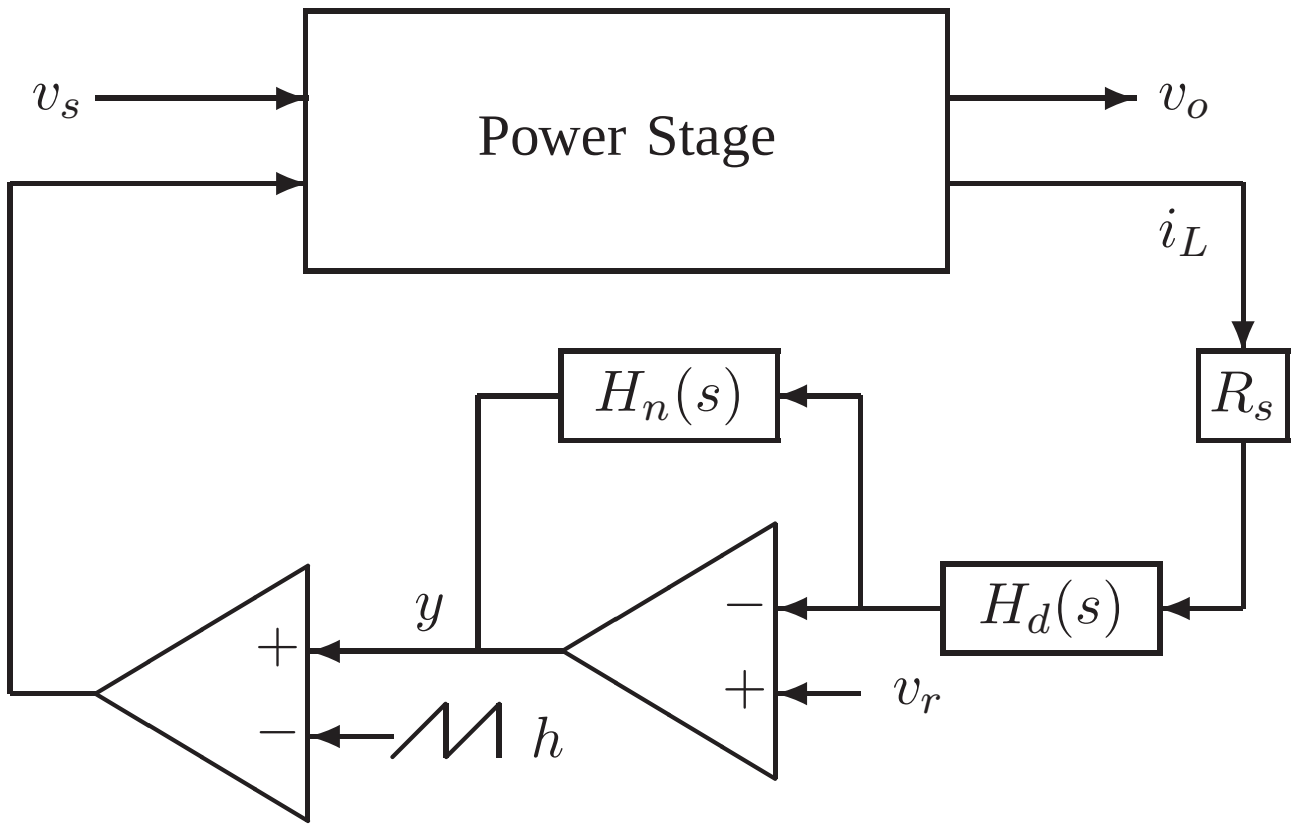

Figure 8. System diagram of an ACMC DC-DC converter.

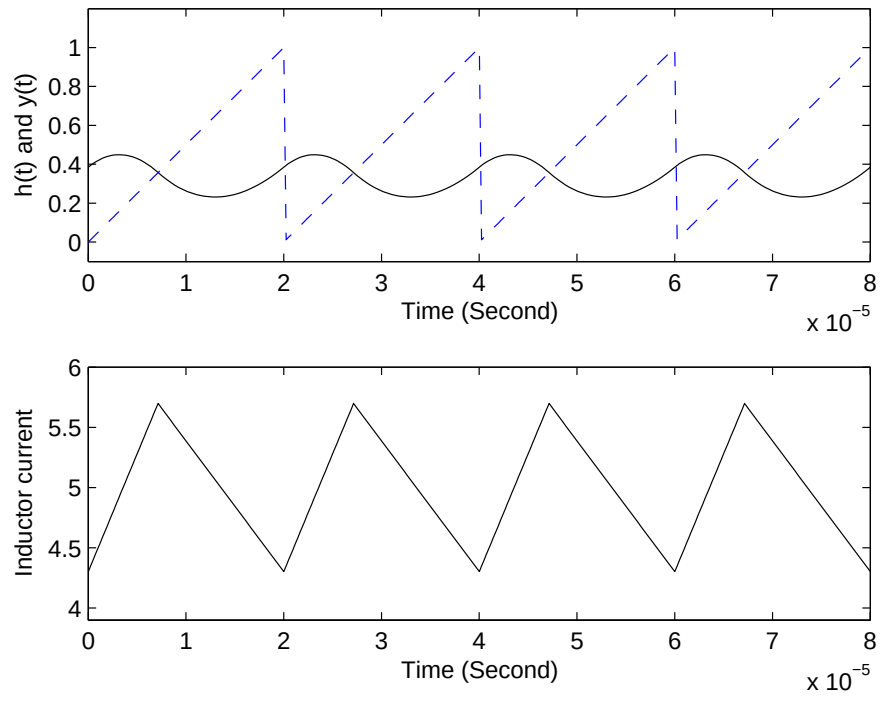

Figure 9. Stable $T$-periodic solution, $\omega_p = 0.15\omega_s$.

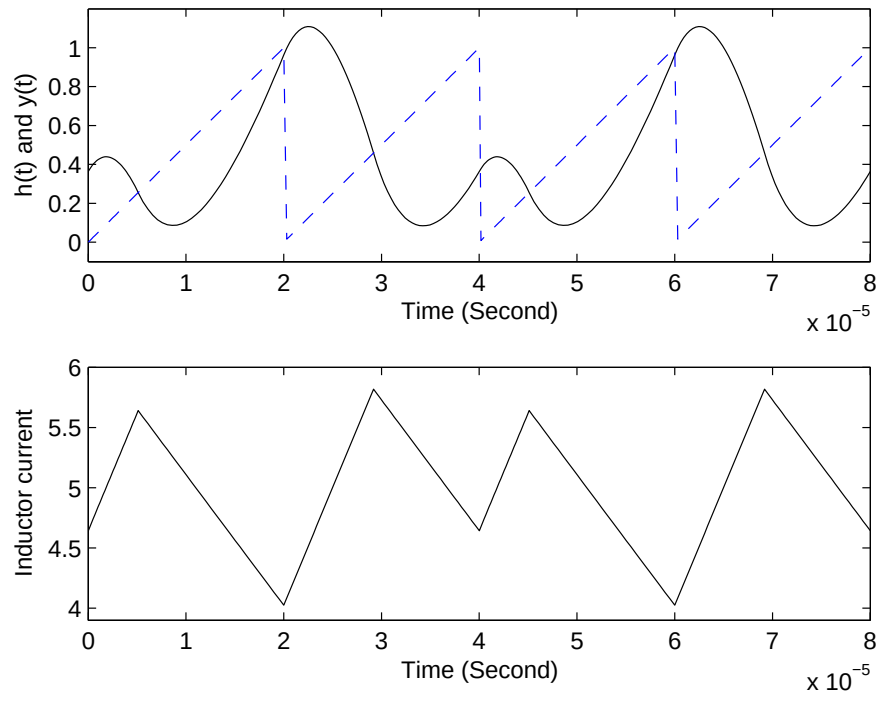

Figure 10. Subharmonic oscillation, $\omega_p = 0.49\omega_s$.

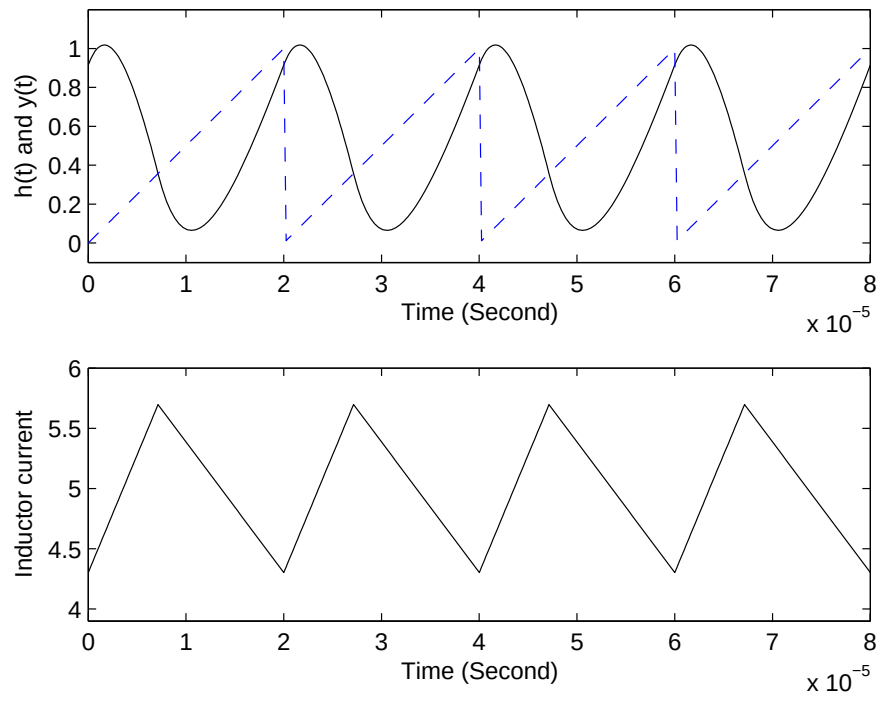

Figure 11. Stable $T$-periodic solution, $\omega_p = 0.81\omega_s$.

261] has three poles, two zeros, and a gain $K_c$, with a transfer function

$$G_c(s) = \frac{H_n(s)}{H_d(s)} = \frac{K_c(1 + \frac{s}{z_1})(1 + \frac{s}{z_2})}{(s + \delta)(1 + \frac{s}{\omega_p})(1 + \frac{s}{p_2})} \tag{15}$$

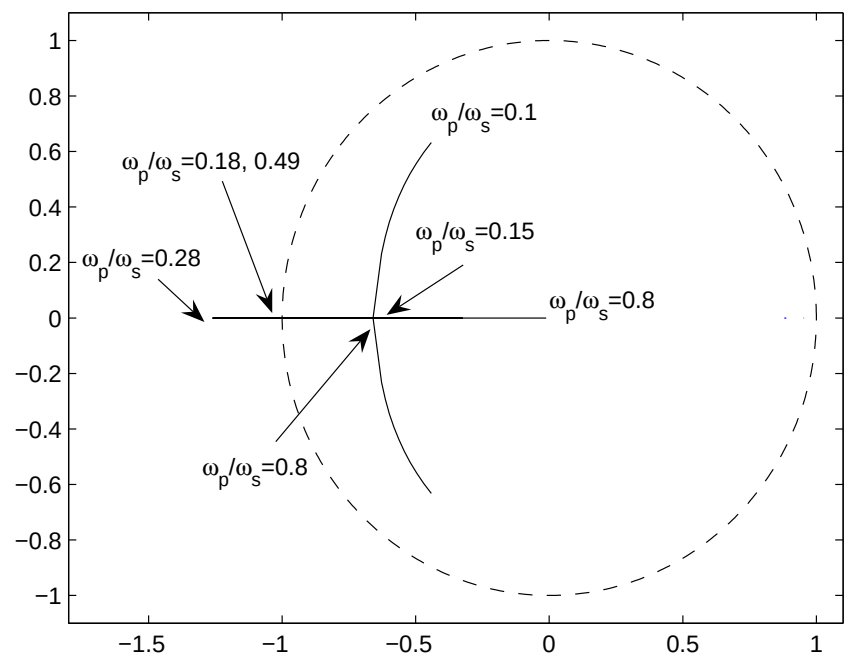

Figure 12. Sampled-data pole trajectories for $\omega_p/\omega_s \in (0.1, 0.8)$.

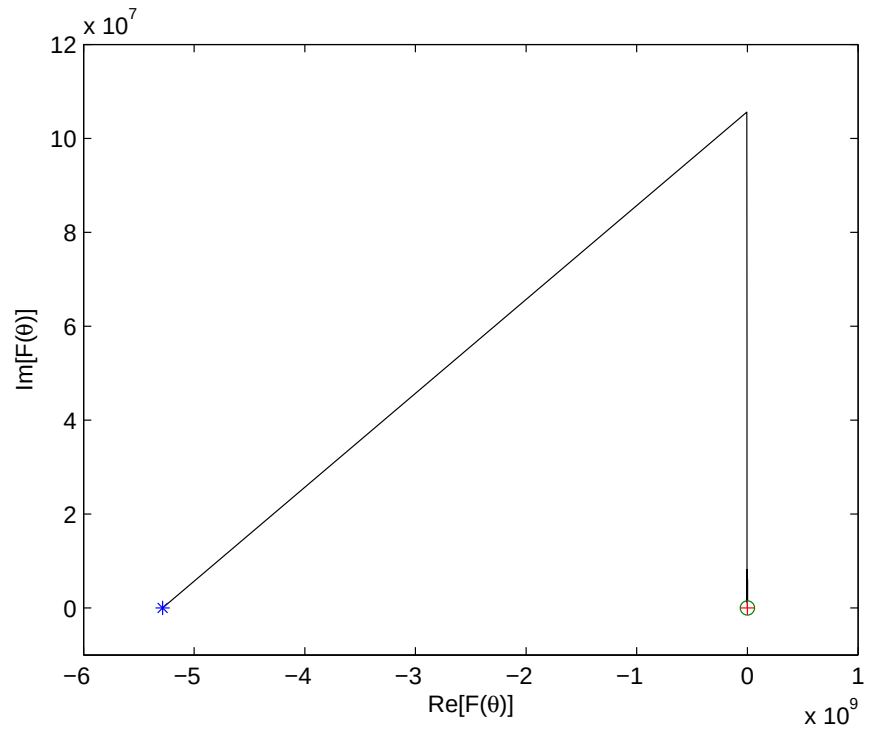

Figure 13. The F-plot for $\theta \in [0, \pi]$, $\omega_p = 0.15\omega_s$.

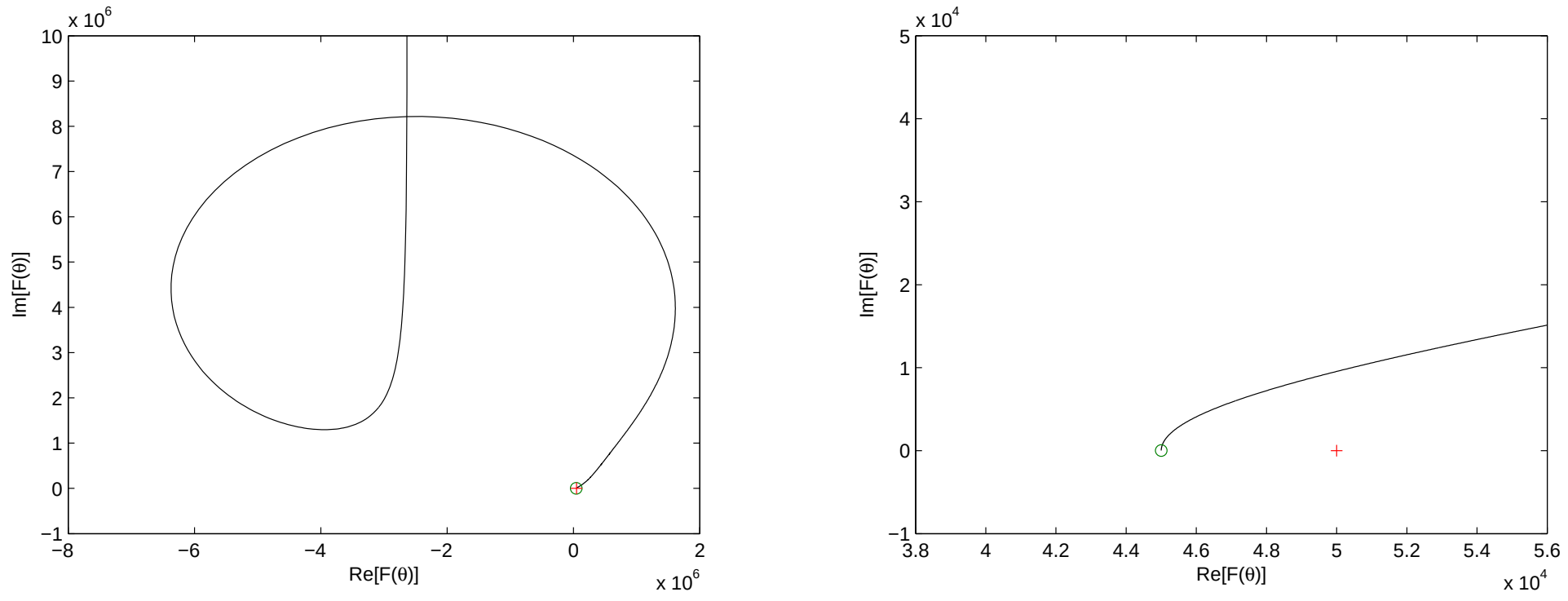

(a) (b)

Figure 14. The blown-up views of F-plot shows that the F-plot does not encircle the $m_a$ point, $\omega_p = 0.15\omega_s$.

A typical guideline [27, p. 412] popular in industry to set the parameters of the compensator is as follows. Set one pole at $\delta \approx 0$ (as an integrator), and set $p_2 = 1/R_cC$. Set the gain $K_c$ to adjust the

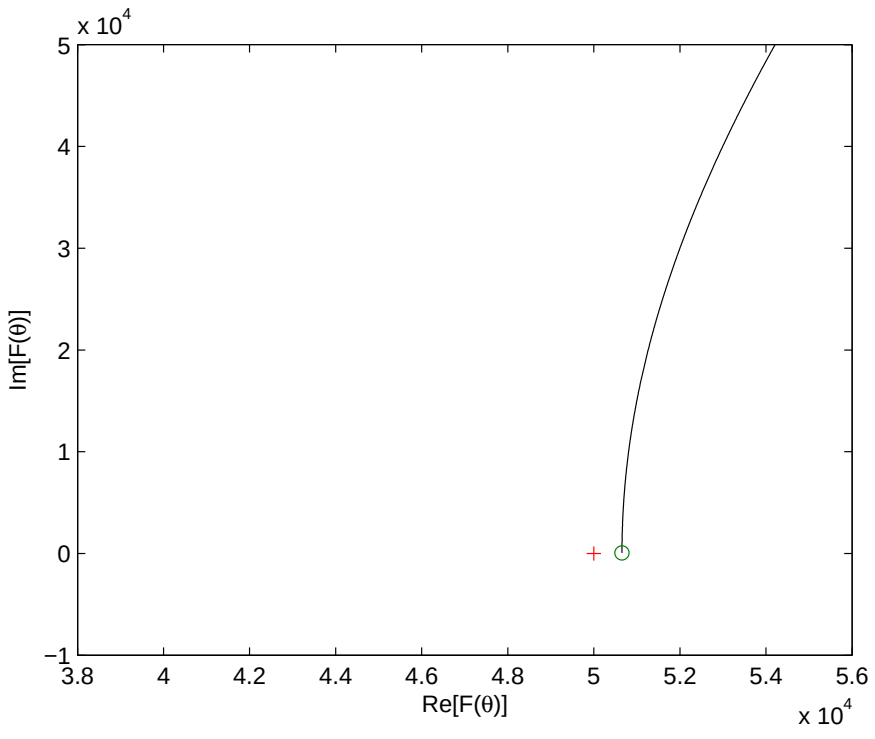

Figure 15. The blown-up view of F-plot shows that the $m_a$ point is encircled by the F-plot, $\omega_p = 0.49\omega_s$.

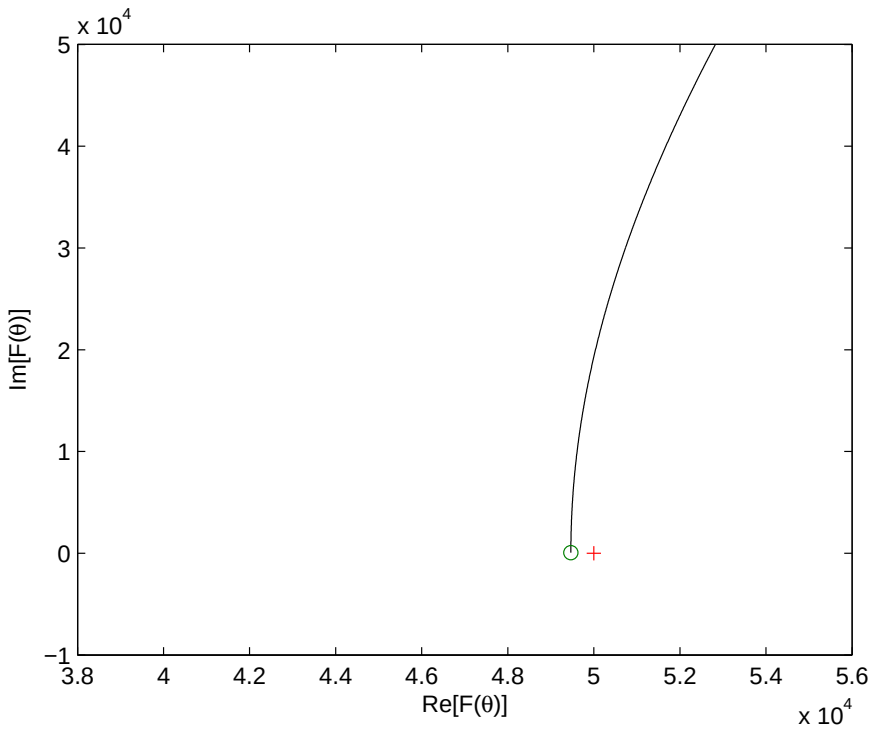

Figure 16. The blown-up view of F-plot shows that the $m_a$ point is not encircled by the F-plot, $\omega_p = 0.5\omega_s$.

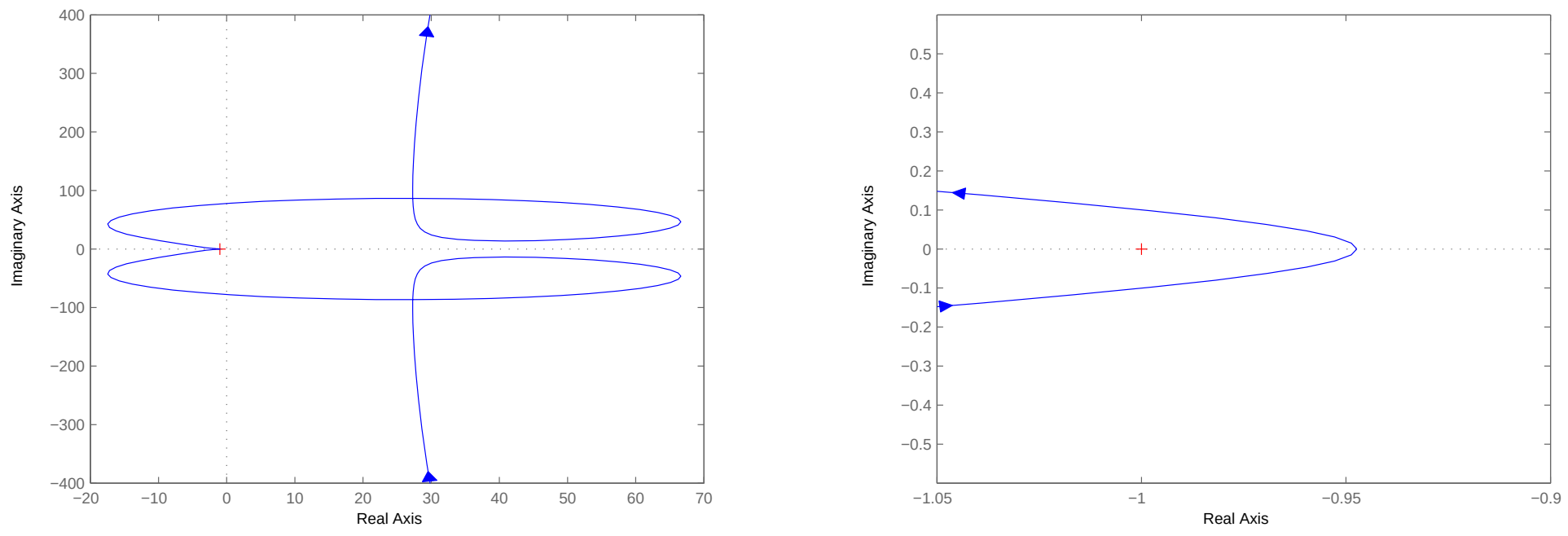

(a) (b)

Figure 17. (a) The Nyquist plot for $\omega_p = 0.15\omega_s$, (b) blown-up view of (a) shows that the Nyquist plot does not encircle the -1 point.

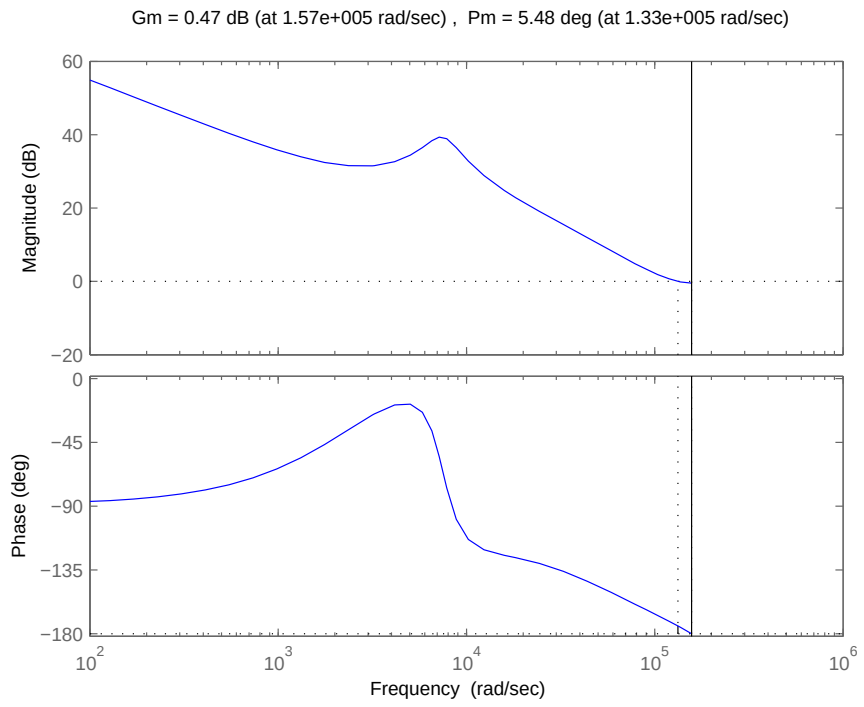

Figure 18. The Bode plot shows a positive gain phase margin, $\omega_p = 0.15\omega_s$.

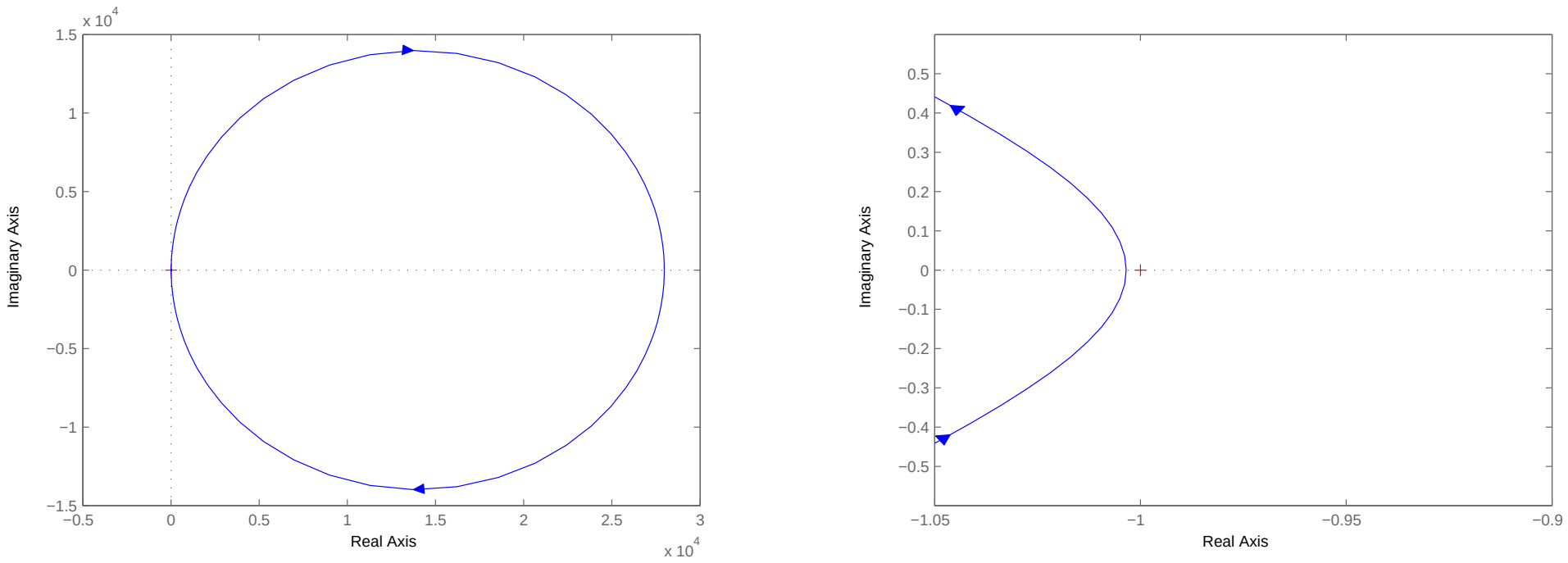

(a) (b)

Figure 19. (a) The Nyquist plot for $\omega_p = 0.15\omega_s$, (b) blown-up view of (a) shows that the Nyquist plot encircles the -1 point.

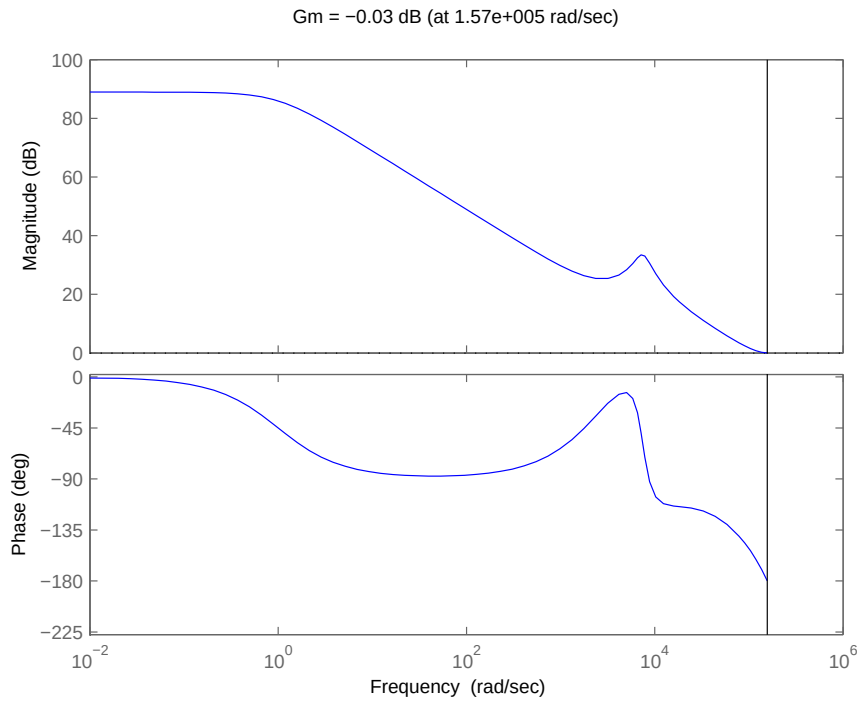

Figure 20. The Bode plot shows a negative gain phase margin, $\omega_p = 0.49\omega_s$.

phase margin and the crossover frequency. Let $z_1 = \kappa_z/\sqrt{LC}$ and $z_2 = 1/\sqrt{LC}$, where $\kappa_z$ is a zero scale factor to have additional flexibility to adjust the phase margin and the crossover frequency. The zero scale factor $\kappa_z$ used in industry typically varies between 0.1 and 1.2. As will be shown later, a smaller value of $\kappa_z$ may lead to subharmonic oscillation. Taking into account the above guidelines, the compensator has a transfer function

$$G_c(s) = \frac{K_c(1 + \frac{\sqrt{LC}s}{\kappa_z})(1 + \sqrt{LC}s)}{(s + \delta)(1 + \frac{s}{p_1})(1 + R_cCs)} \tag{16}$$

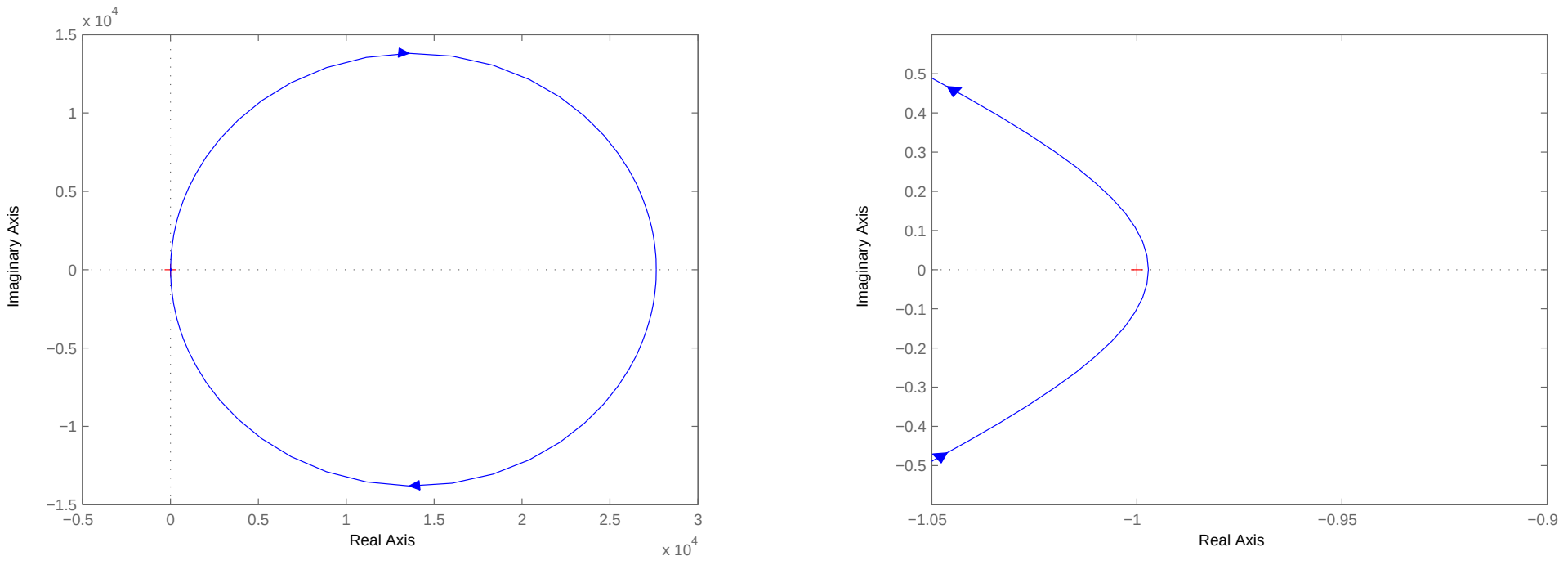

(a) (b)

Figure 21. (a) The Nyquist plot for $\omega_p = 0.15\omega_s$, (b) blown-up view of (a) shows that the Nyquist plot does not encircle the -1 point.

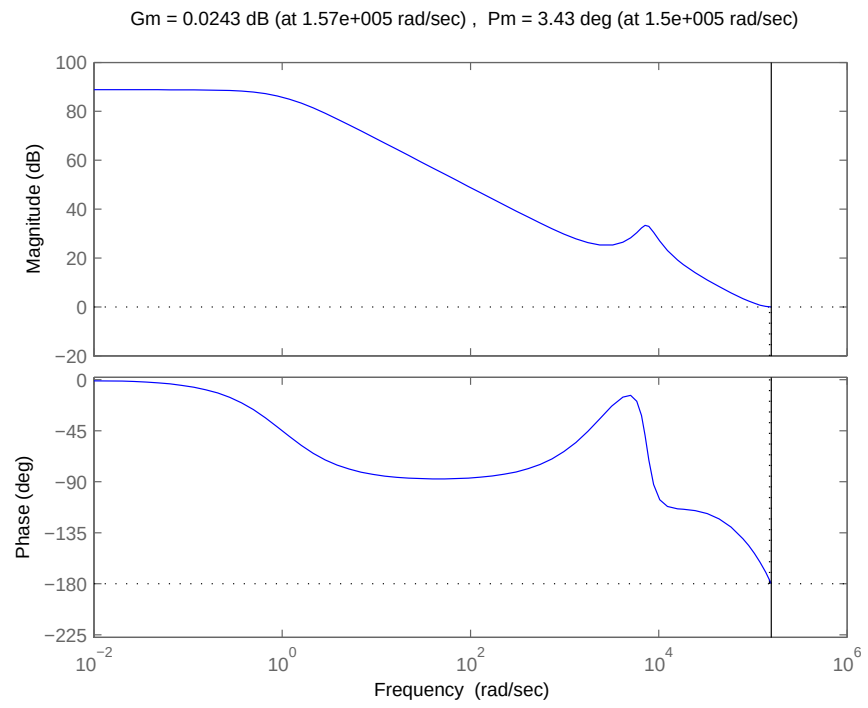

Figure 22. The Bode plot shows a positive gain phase margin, $\omega_p = 0.5\omega_s$.

Consider a buck converter with the type-III compensator (16). Exactly the same parameters as in a technical note [28] are used: $f_s = 1/T = 300$ kHz, $L = 900$ nH, $C = 990$ $\mu$F, $R = 0.4$ $\Omega$, $R_c = 5$ m$\Omega$, $v_r = 3.3$ V, $V_m = 1.5$ V, $K_c = 7.78 \times 10^4$, $z_1 = 1/2\sqrt{LC} = 1.675 \times 10^4$, $z_2 = 1/\sqrt{LC} = 3.35 \times 10^4$, $\omega_p = \omega_s/2 = 9.425 \times 10^5$, and $p_2 = 1/R_cC = 2.02 \times 10^5$. Let $\delta = 1$ be the integrator pole which is small and it does not affect much the dynamics.

Simulation (Fig. 24) shows that subharmonic oscillation occurs when $v_s = 16$ V ($D \approx 0.206$). This is also confirmed by the exact sampled-data analysis with a sampled-data pole at -1 when the subharmonic oscillation occurs. Based on the averaged model for $v_s = 16$, the loop gain frequency response (Fig. 25) shows a phase margin of $38.9°$. The frequency response also shows an infinite gain margin because the phase never reaches -180°, which means no matter how much $v_s$ increases (to increase the loop gain), the converter is expected to be stable based on the averaged model. However, subharmonic oscillation still occurs when $v_s = 16$. □

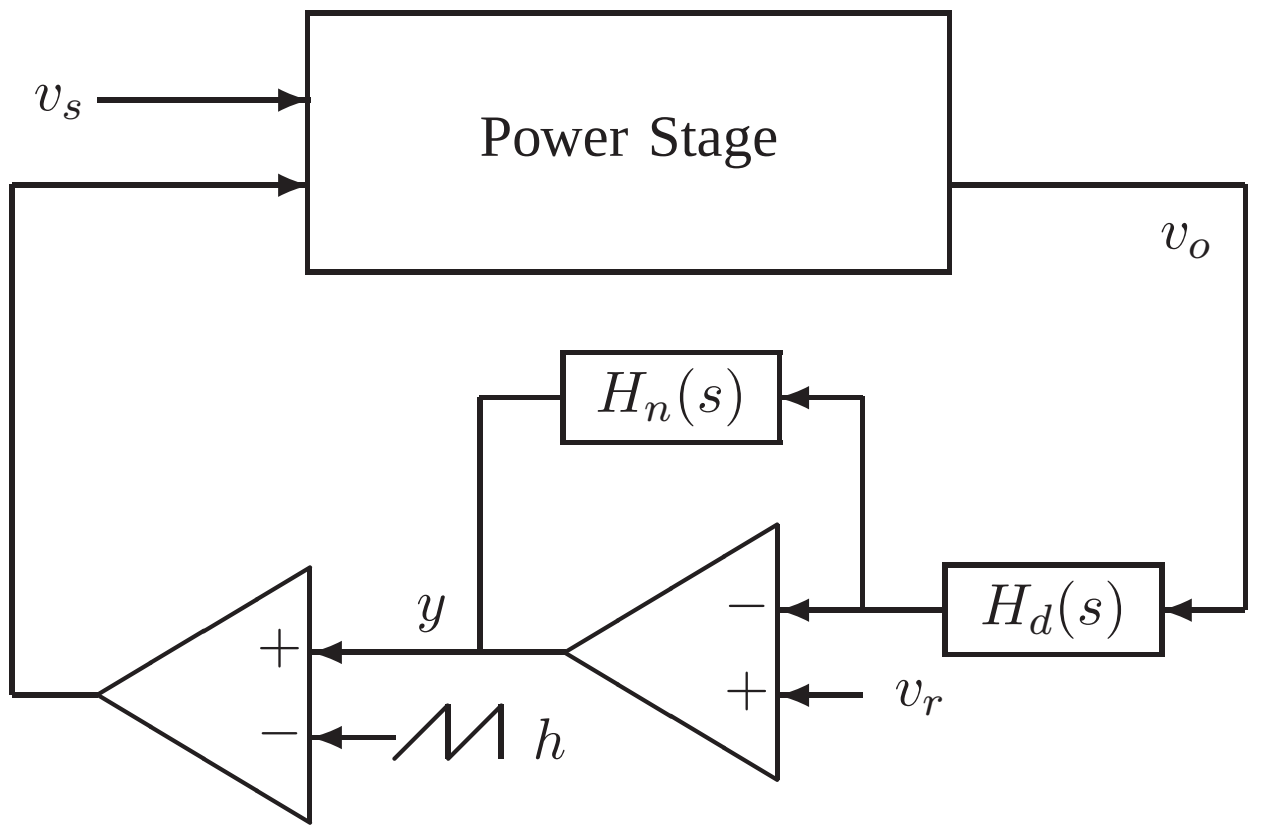

Figure 23. A DC-DC converter with a type-III compensator $G_c(s) = H_n(s)/H_d(s)$.

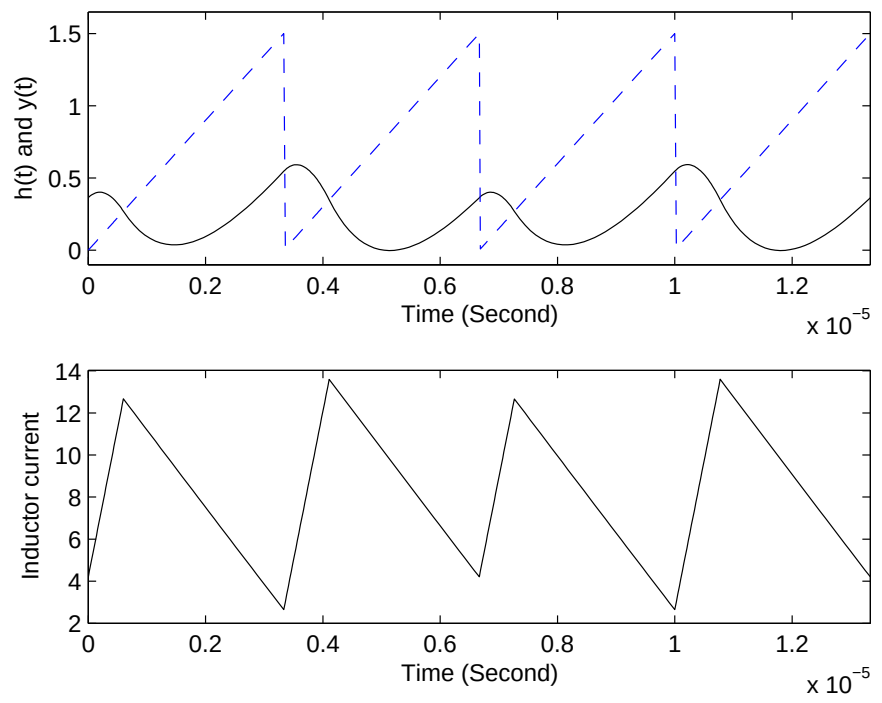

Figure 24. Signal waveforms showing the subharmonic oscillation.

**Example 4.** (*Unstable window of* $p_1$, *unrelated to the ripple size of* $y^0(t)$.) Consider again Example 3 where $p_1 = \omega_s/2$. Now, vary $p_1$ from $0.1\omega_s$ to $0.6\omega_s$.

**Unstable window.** An unstable window of $p_1 \in (0.23, 0.5)\omega_s$ is found. The value of $p_1$ adjusts the ripple size of $y^0(t)$. A larger $p_1$ leads to a larger ripple of $y^0(t)$. In [14], it is hypothesized that the ripple size of $y^0(t)$ is related to subharmonic oscillation. The following simulation shows that the ripple size of $y^0(t)$ is *unrelated* to subharmonic oscillation.

**Time-domain simulation.** The unstable window of $p_1$ is confirmed by time-domain simulation. For $p_1 = 0.2\omega_s$, the ripple size of $y$ is small, and the converter is stable (Fig. 26). For $p_1 = 0.24\omega_s$, the ripple size of $y$ is larger, and the converter is unstable (Fig. 27). For $p_1 = 0.6\omega_s$, the ripple size of $y$ is even larger, but the converter is stable (Fig. 28). Comparing Figs. 26-28, the ripple size of $y^0(t)$ is unrelated to subharmonic oscillation. This shows a counter-example for the hypothesis proposed in [14] that the ripple size of $y^0(t)$ is related to the subharmonic oscillation.

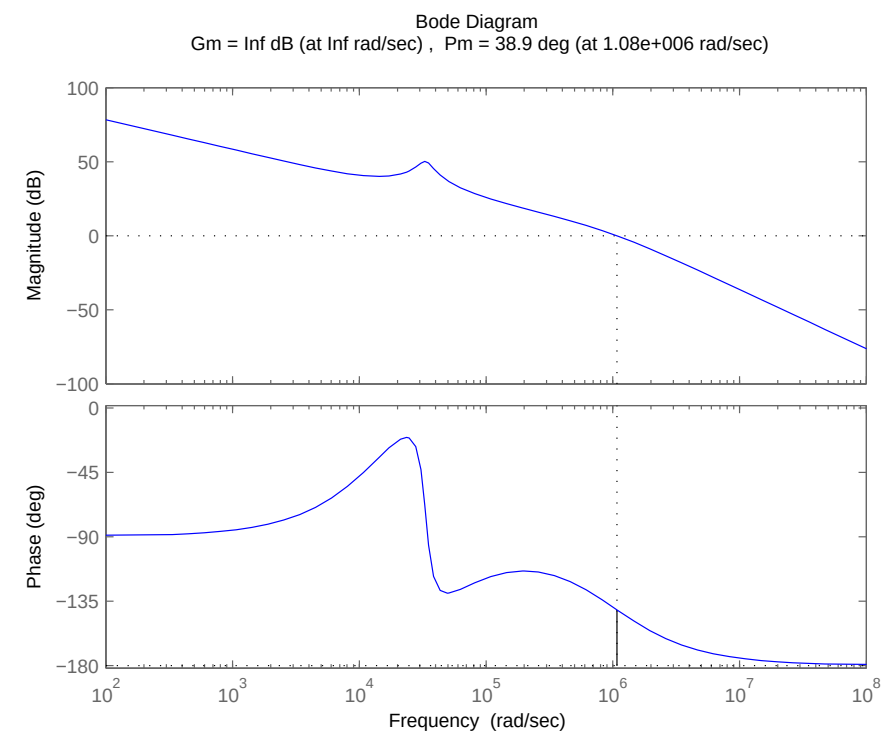

Figure 25. Loop gain frequency response shows a phase margin of $38.9^\circ$ and an infinite gain margin based on the *averaged* model, but the subharmonic oscillation still occurs.

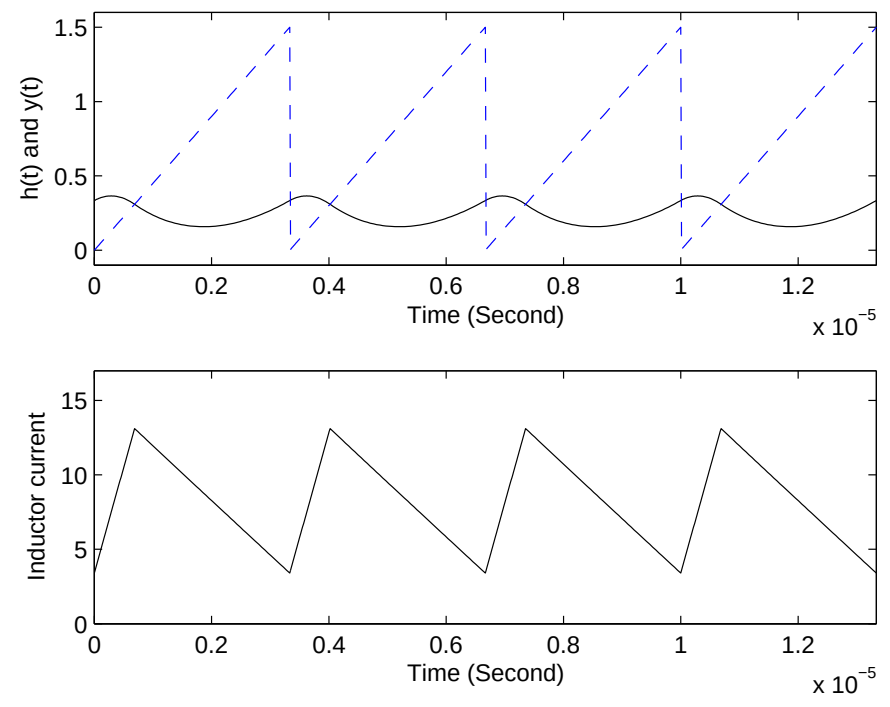

Figure 26. The converter is stable, $p_1 = 0.2\omega_s$.

**Confirmation by sampled-data pole trajectories.** The unstable window of $p_1$ is also confirmed by the sampled-data pole trajectories. The sampled-data pole trajectories for $0.1\omega_s < p_1 < 0.6\omega_s$ are shown in Fig. 29. Three poles are fixed around 0.9485, 0.8853, and 0.51. A pole leaves the unit circle through -1 when $p_1 = 0.23\omega_s$, and enters the unit circle when $p_1 = 0.5\omega_s$. This explains exactly the unstable window of $p_1$.

**Confirmation by F/Nyquist/Bode plots.** Here, $m_a = V_m/T = 45000$. First, let $\omega_p = 0.2\omega_s$, and the converter is stable. The F-plot (Fig. 30) does not encircle the $m_a$ point. The Nyquist plot (Fig. 31) does not encircle the $-1$ point. The Nyquist path goes to infinity and *connects* to form a loop (not shown). *This statement is not repeated later when the Nyquist path goes to infinity.* The Bode plot (Fig. 32) shows positive gain/phase margins.

Second, let $\omega_p = 0.5\omega_s$, and the converter is unstable due to PDB. The F-plot (Fig. 33) encircles the $m_a$ point. The Nyquist plot (Fig. 34) encircles the $-1$ point. The Bode plot (Fig. 35) shows a negative gain margin.

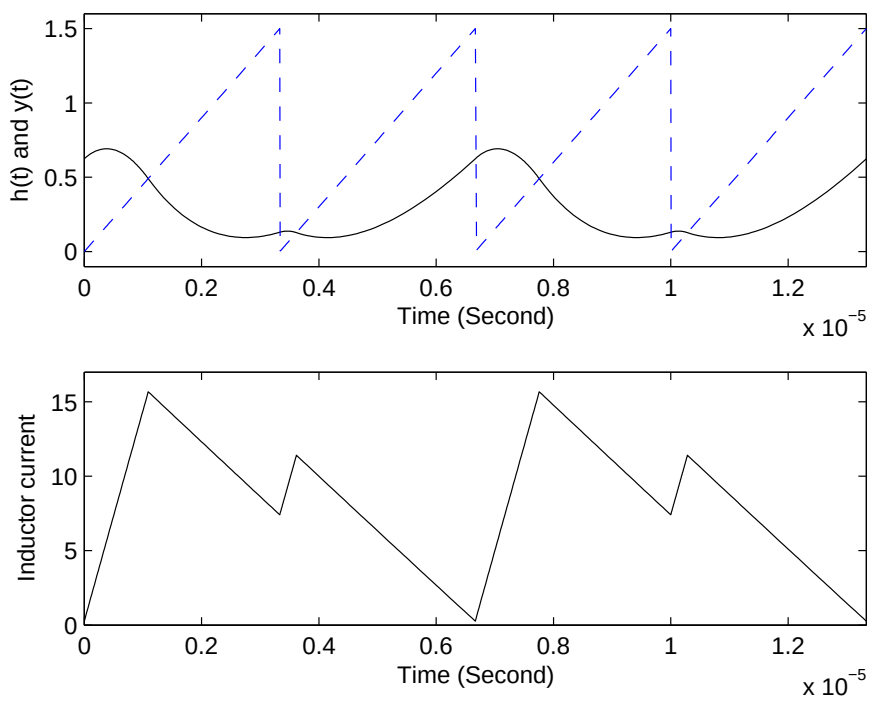

Figure 27. Subharmonic oscillation, $p_1 = 0.24\omega_s$.

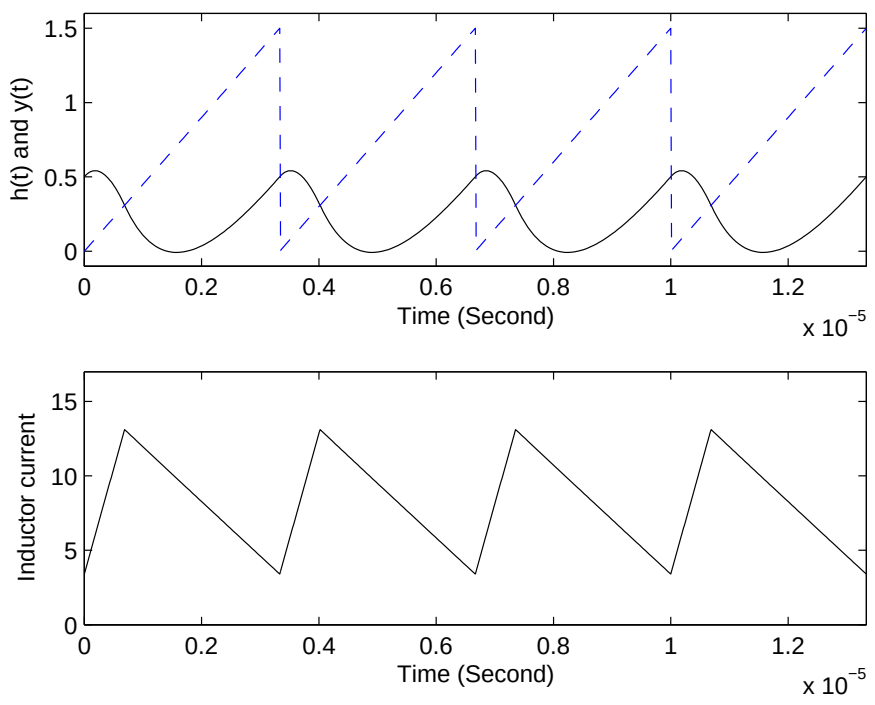

Figure 28. The converter is *stable* with a *larger* ripple of $y^0(t)$, $p_1 = 0.6\omega_s$.

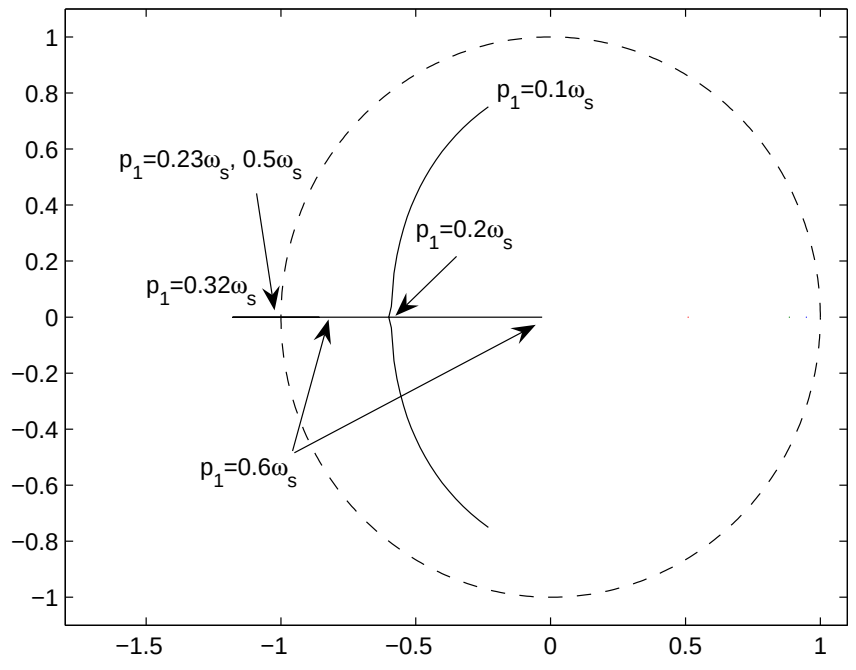

Figure 29. Sampled-data pole trajectories for $0.1\omega_s < p_1 < 0.6\omega_s$.

Third, let $\omega_p = 0.6\omega_s$, and the converter is stable. The F-plot (Fig. 36) does not encircle the $m_a$

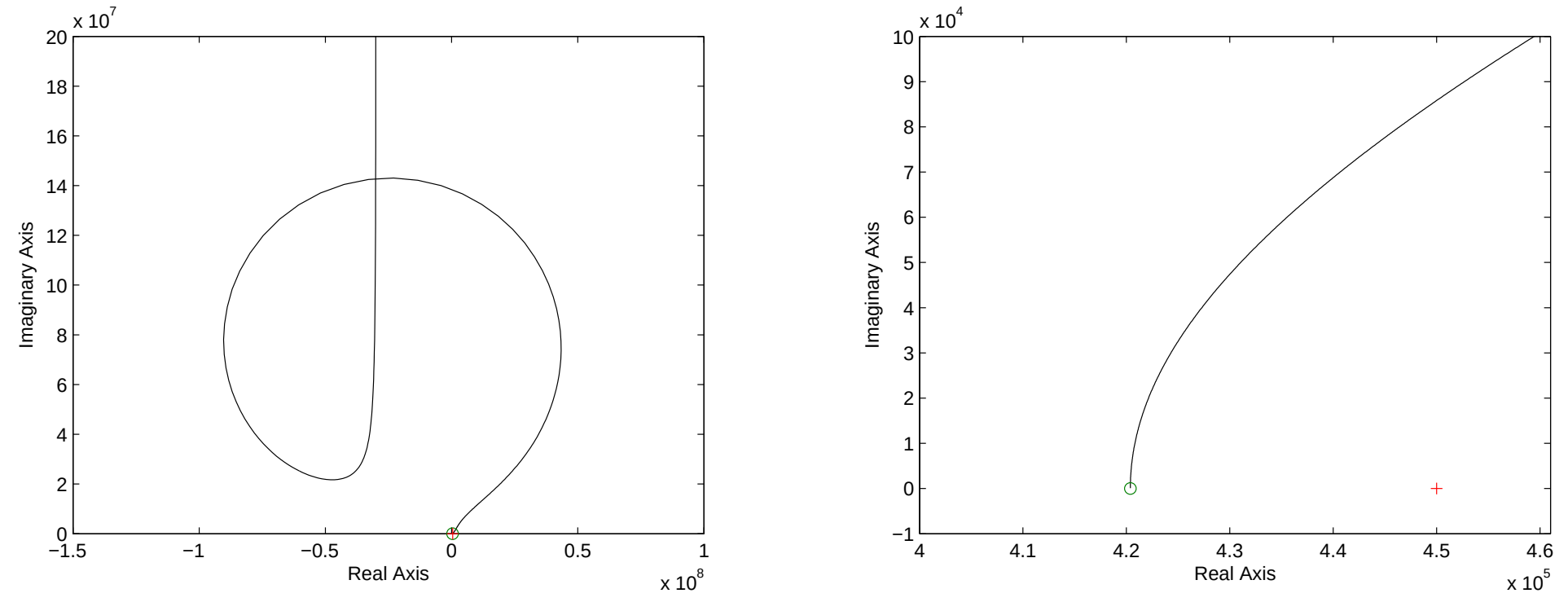

(a) (b)

Figure 30. (a) F-plot for $\theta \in [0, \pi]$, $\omega_p = 0.2\omega_s$; (b) blown-up view of (a) shows that the F-plot does not encircle the $m_a$ point.

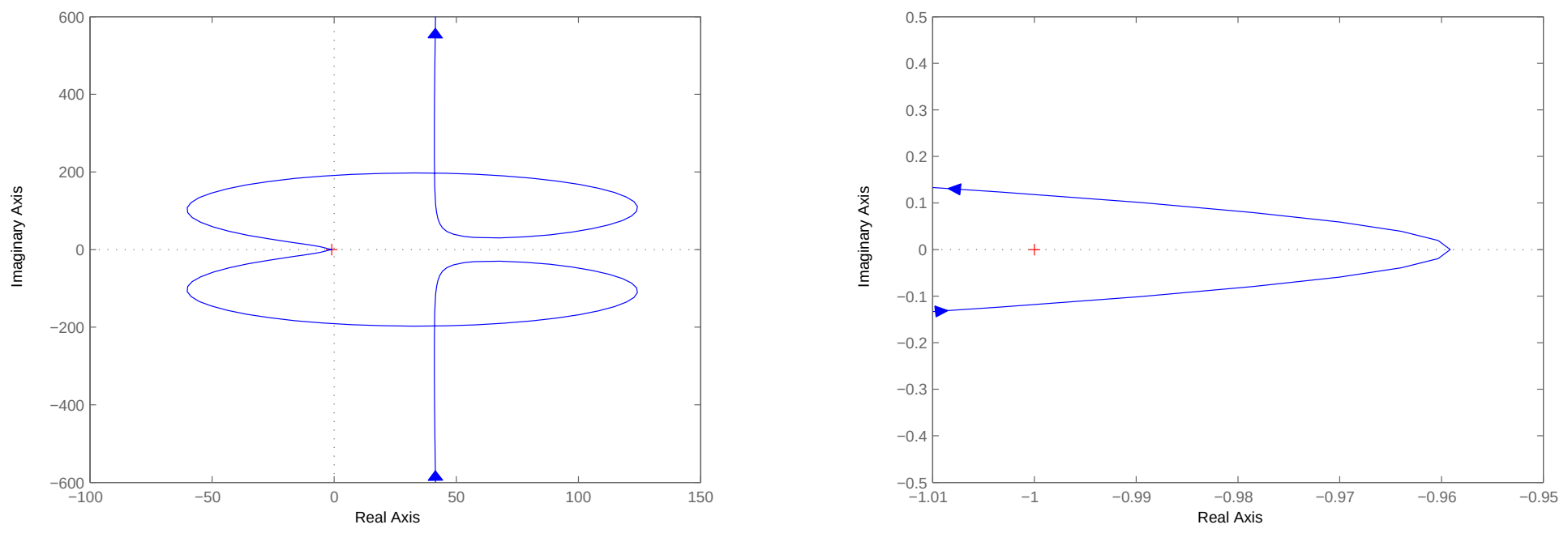

(a) (b)

Figure 31. (a) Nyquist plot, $\omega_p = 0.2\omega_s$; (b) blown-up view of (a) shows that the Nyquist plot does not encircle the -1 point.

point. The Nyquist plot (Fig. 37) does not encircle the $-1$ point. The Bode plot (Fig. 38) shows positive gain/phase margins.

In this example, the unstable window of $p_1$ is verified by three different approaches: time-domain simulation, the sampled-data pole trajectories, and F/Nyquist/Bode plots. □

## VII. Prediction of Saddle-Node Bifurcation (SNB)

**Example 5.** (*The F-plot accurately predicts SNB.*) Consider a boost converter from [29], [30] shown in Fig. 39. The converter parameters are $T = 2$ $\mu$s, $v_s = 4$ V, $L = 5.24$ $\mu$H, $C = 0.2$ $\mu$F, $R = 16$ $\Omega$, $v_r = 0.48$, $V_l = 0$, and $V_h = 1$. Let the feedback gains for the current and voltage loops respectively be $k_i = -0.1$ and $k_v = 0.01$. The switch is on when $h(t) < y := v_r - k_i i_L - k_v v_o$ and it is off when otherwise.

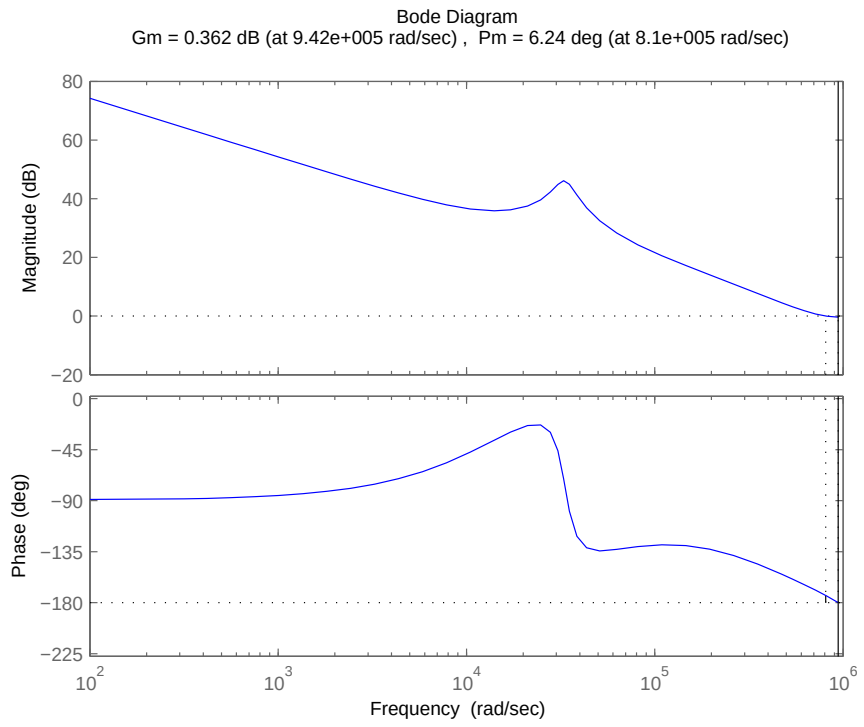

Figure 32. The Bode plot shows positive gain/phase margins, $\omega_p = 0.2\omega_s$.

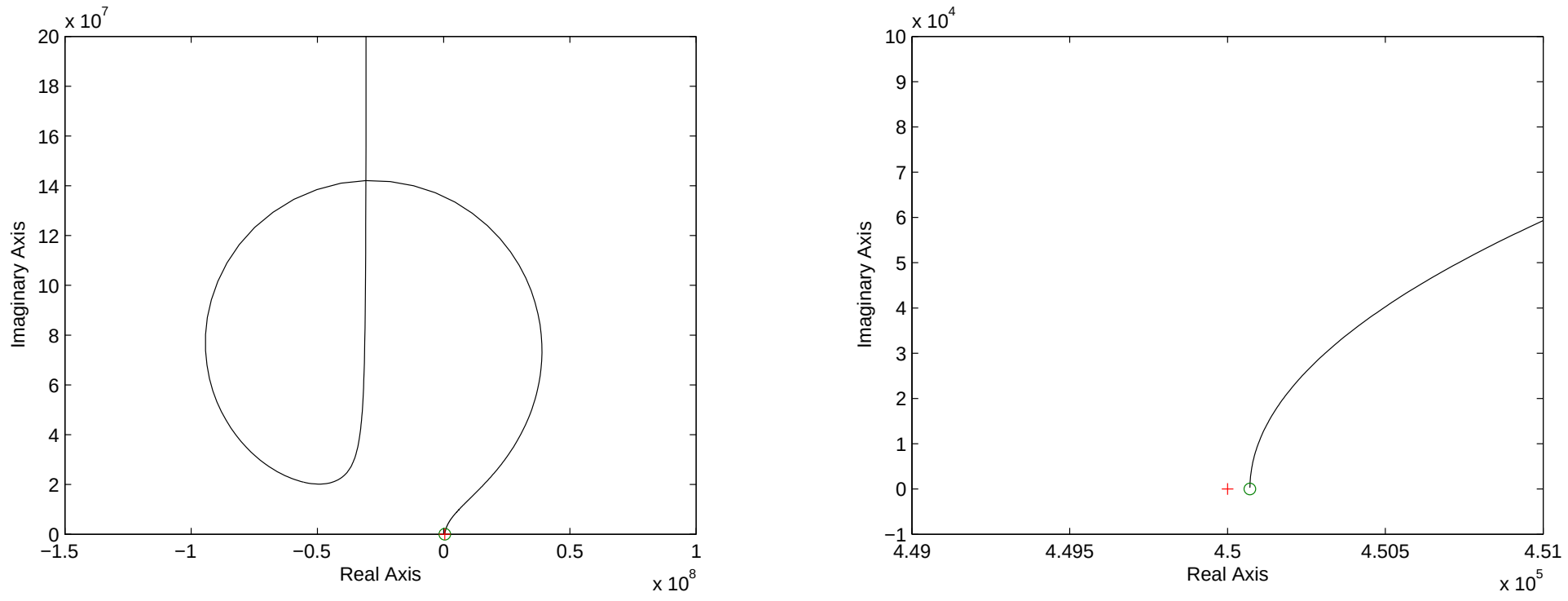

(a) (b)

Figure 33. (a) F-plot for $\theta \in [0, \pi]$, $\omega_p = 0.5\omega_s$; (b) blown-up view of (a) shows that the F-plot encircles the $m_a$ point.

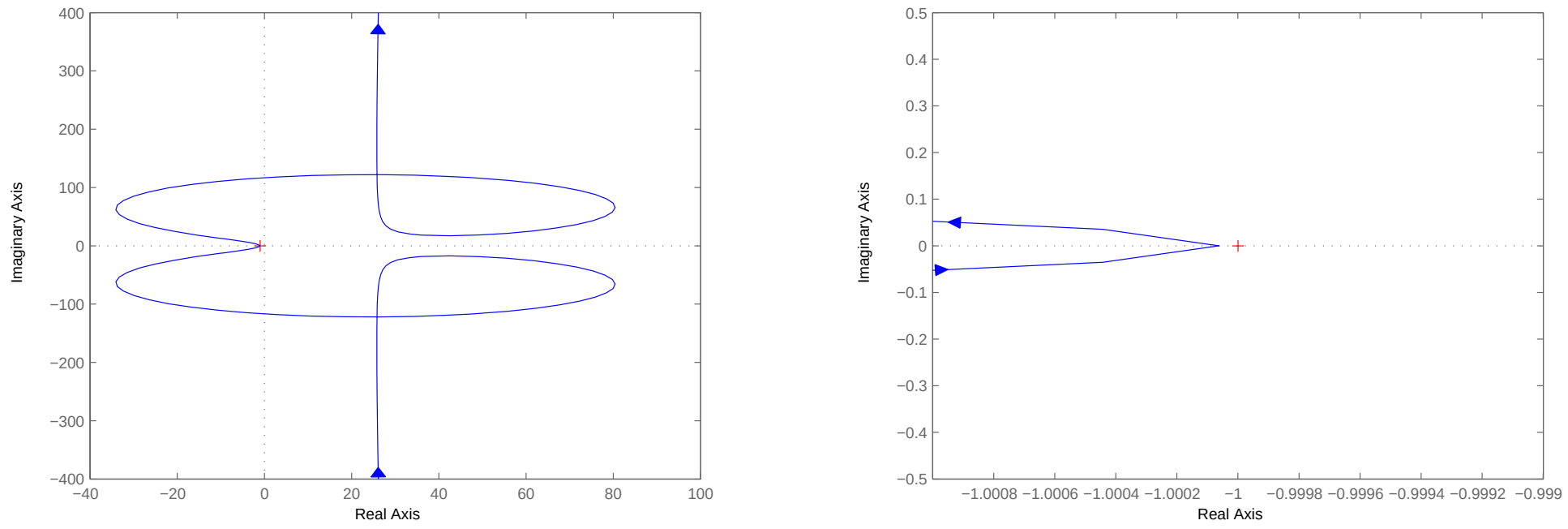

(a) (b)

Figure 34. (a) Nyquist plot, $\omega_p = 0.5\omega_s$; (b) blown-up view of (a) shows that the Nyquist plot encircles the -1 point.

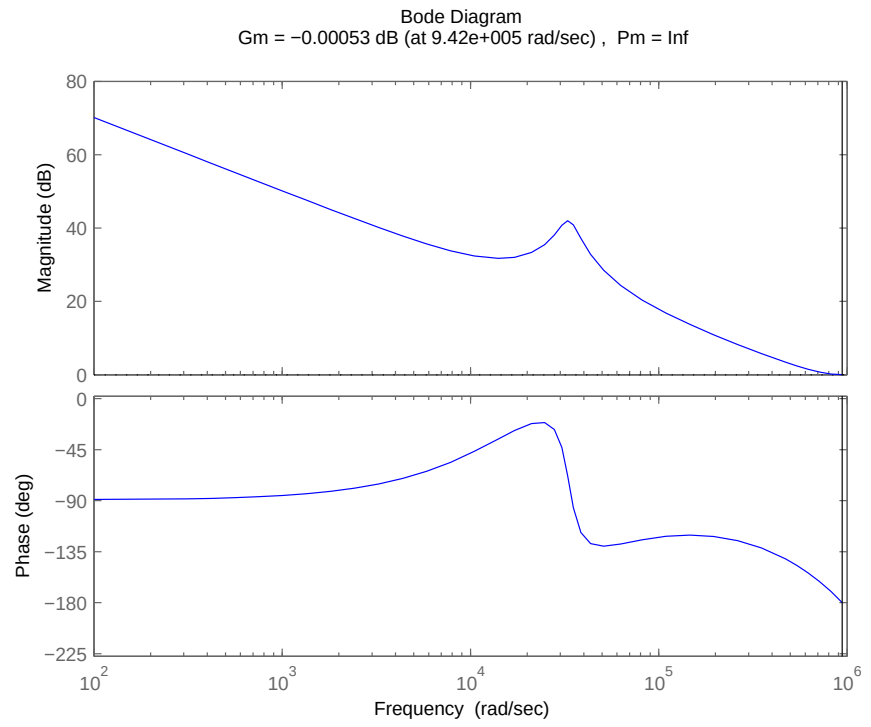

Figure 35. The Bode plot shows a negative gain margin, $\omega_p = 0.5\omega_s$.

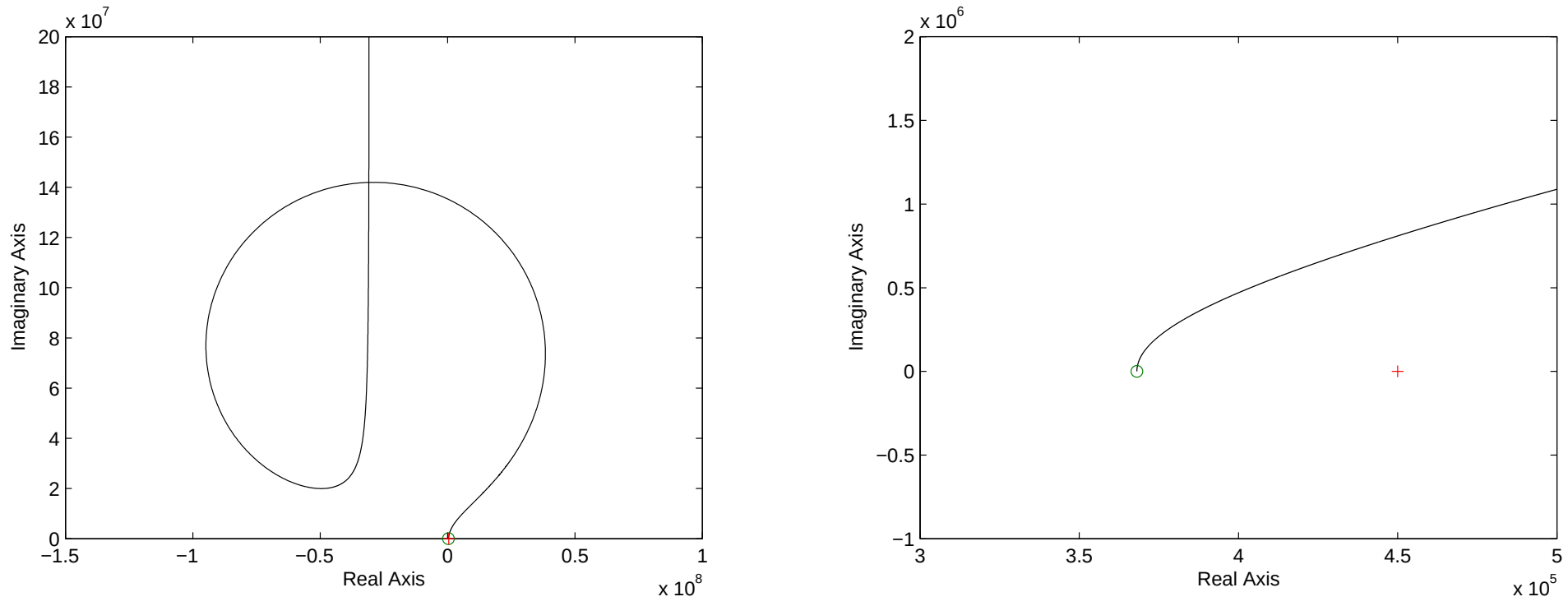

(a) (b)

Figure 36. (a) F-plot for $\theta \in [0, \pi]$, $\omega_p = 0.6\omega_s$; (b) blown-up view of (a) shows that the F-plot does not encircle the $m_a$ point.

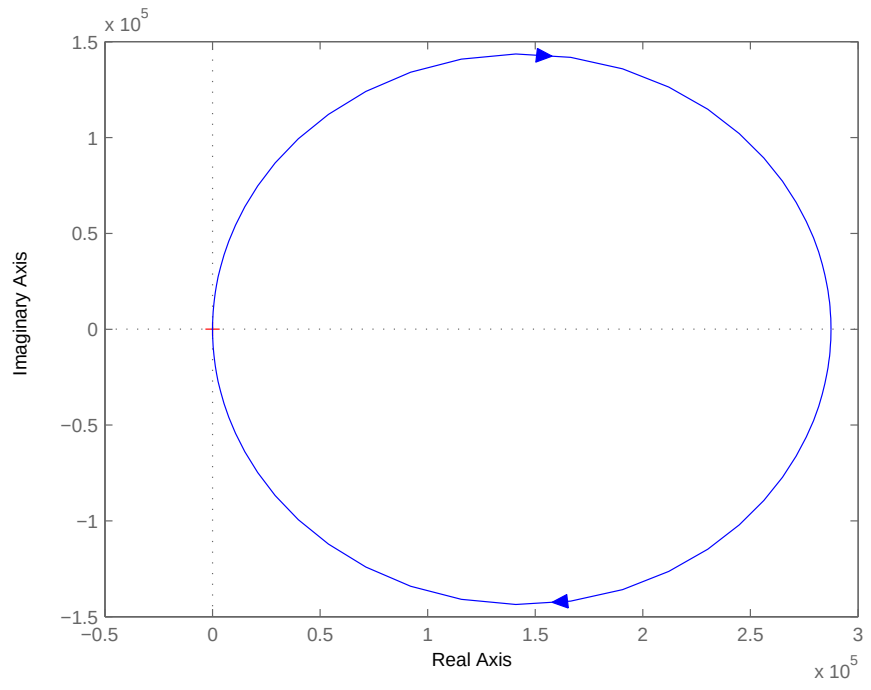

Figure 37. Nyquist plot does not encircle the -1 point, $\omega_p = 0.6\omega_s$.

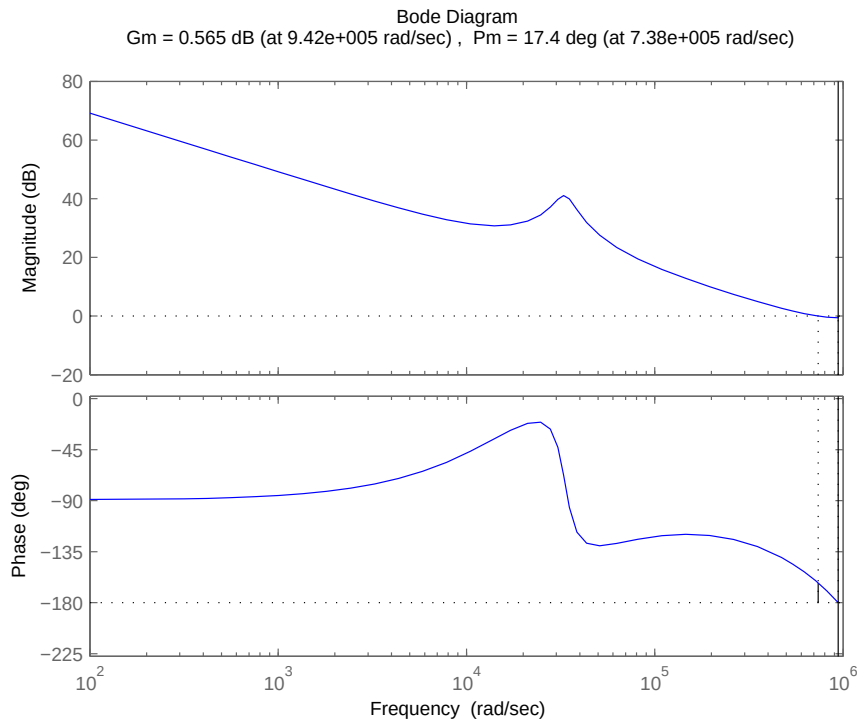

Figure 38. The Bode plot shows positive gain/phase margins, $\omega_p = 0.6\omega_s$.

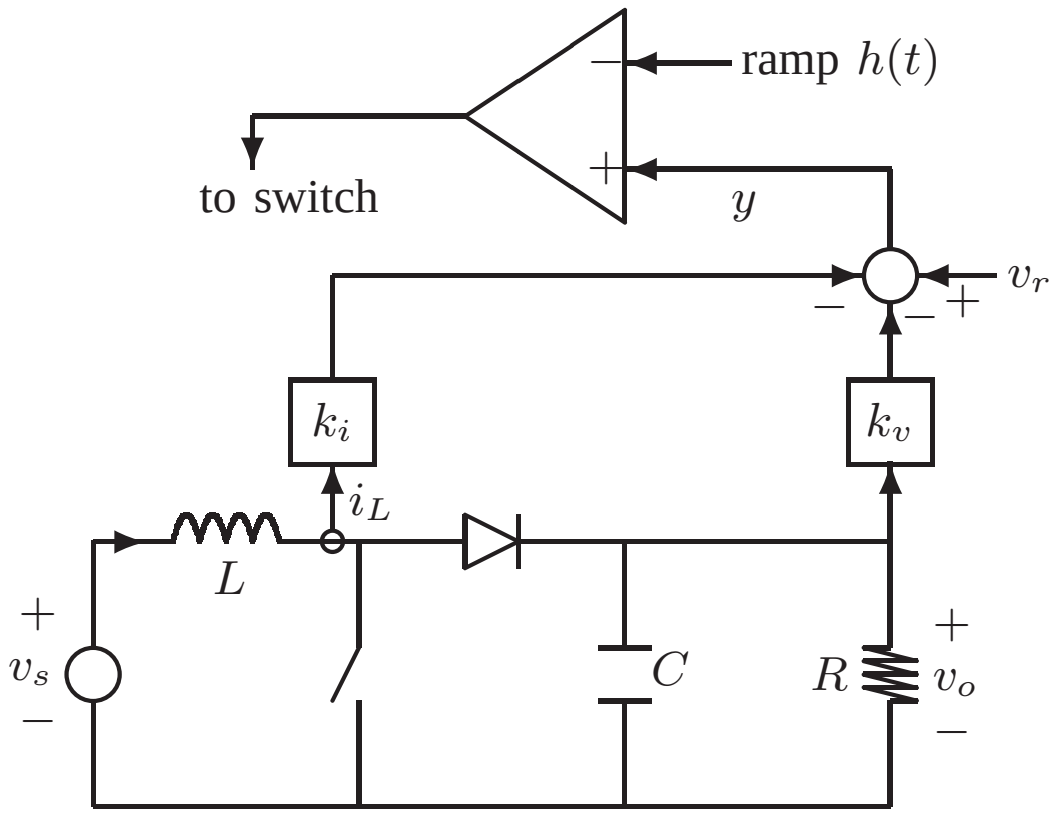

Figure 39. Boost converter under multi-loop state feedback.

**Time-domain simulation.** Three solutions coexist: a *stable* $T$-periodic orbit with $D = 0.586$ (solved from (1)-(3)), an unstable $T$-periodic orbit with $D = 0.71$, and a *stable* DC solution with $(i_L, v_o) = (v_s/r_L, 0)$, where $r_L$ is inductor resistance. The converter is thus *bistable*. The two $T$-periodic orbits are shown in Fig. 40. The final state is sensitive to the initial condition. First, let the initial condition be $(i_L(0), v_o(0)) = (2.2, 16.2)$, close to the unstable $T$-periodic orbit shown in Fig. 40. In about $50T$, the state trajectory moves toward the stable $T$-periodic orbit as shown in Fig. 41. The stable orbit *attracts* the state trajectory. Next, let $(i_L(0), v_o(0)) = (2.3, 16.2)$, *still* close to the unstable orbit. The state trajectory moves away from the unstable orbit as shown in Fig. 41, agreed with [30]. The unstable orbit *expels* the state trajectory toward the stable DC solution. After $t > 8T$, the switch is always on (since $y(t) > h(t)$), and the inductor current increases toward $v_s/r_L$.

The bifurcation diagram using $v_r$ as the bifurcation parameter is shown in Fig. 42 with the DC solution omitted. One sees that SNB occurs at $v_r = 0.496$ and $D = 0.65$ when the two $T$-periodic orbits coalesce. For $v_r > 0.496$, no $T$-periodic orbit exists. For $v_r = 0.48$, the bifurcation diagram shows exactly the two $T$-periodic orbits with $D = 0.586$ and $D = 0.71$.

**F-plot.** Here, the ramp slope is $m_a = V_m/T = 500000$. First, let $D = 0.586$. The F-plot (Fig. 43) does not encircle the $m_a$ point. The sampled-data poles are $0.8045 \pm 0.4510j$ (inside the unit circle), also confirming the stability.

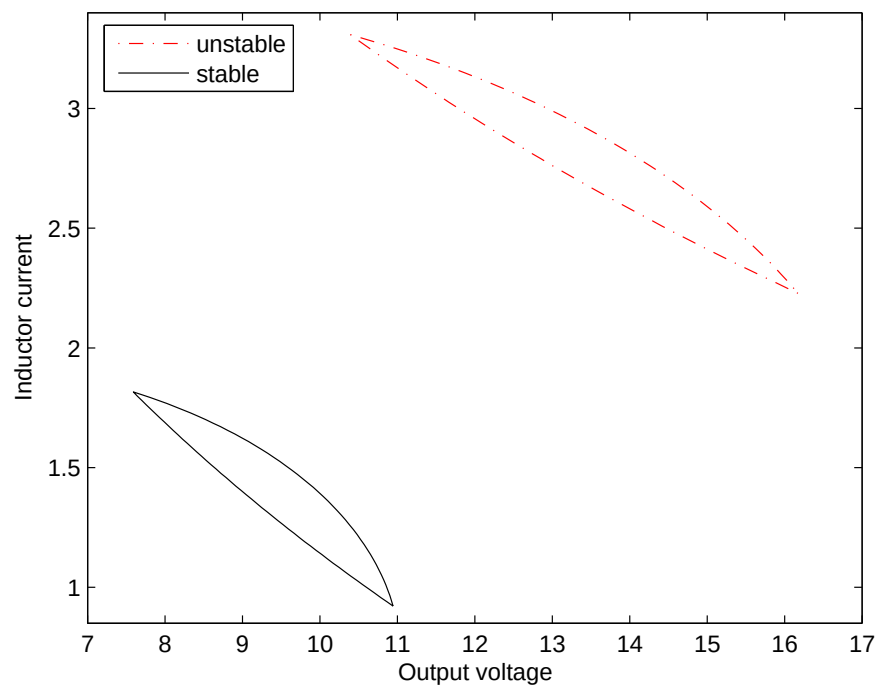

Figure 40. Two coexisting $T$-periodic orbits in state space.

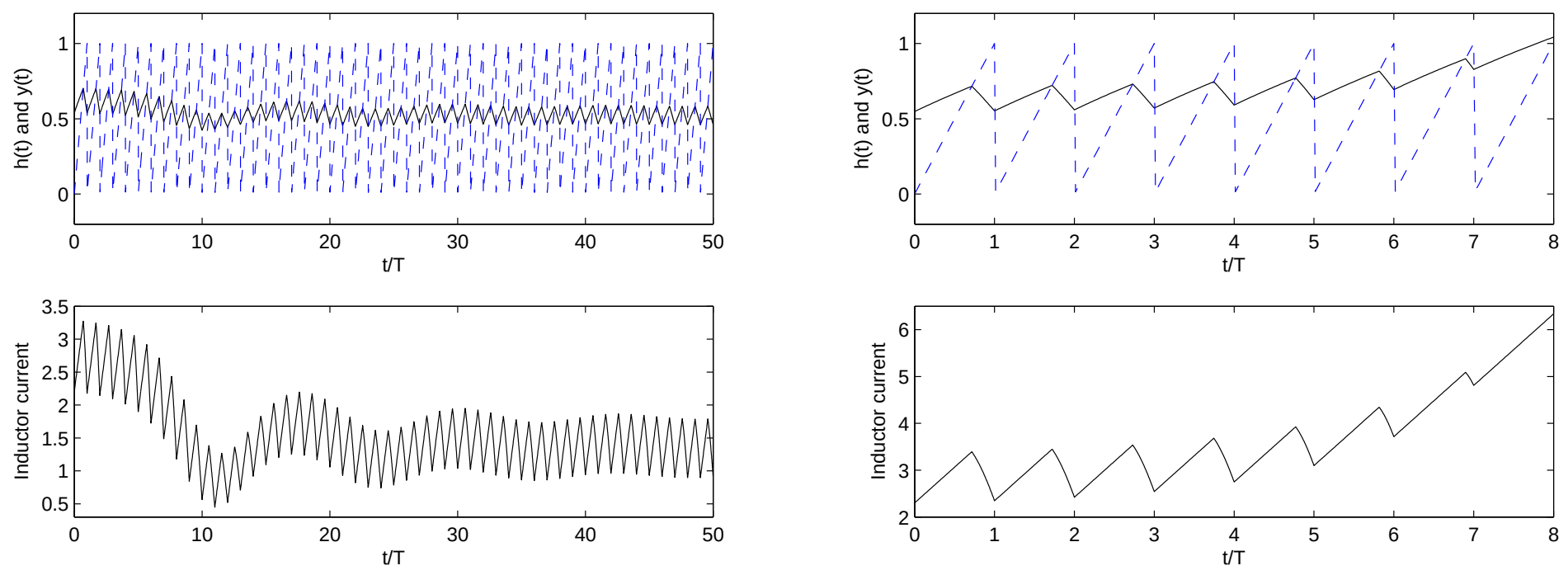

(a) (b)

Figure 41. Stability is sensitive to the initial condition. (a) The state trajectory starting from $(i_L(0), v_o(0)) = (2.2, 16.2)$ moves toward the *stable* orbit; (b) The state trajectory starting from $(i_L(0), v_o(0)) = (2.3, 16.2)$ moves away from the *unstable* orbit.

Note that the F-plot (such as Fig. 43) associated with SNB is a little different from the F-plot (such as Fig. 5) associated with PDB. In Fig. 43, $F(0)$, which is associated with SNB, is closer to the $m_a$ point than $F(\pi)$, and the F-plot for $\theta \in [0, \pi]$ is almost below the real axis. In Fig. 5, $F(\pi)$, which is associated with PDB, is closer to the $m_a$ point than $F(0)$, and the F-plot for $\theta \in [0, \pi]$ is above the real axis. However, in the both plots, they are plotted clockwise as $\theta$ increases.

Next, let $D = 0.71$. The F-plot (Fig. 43) encircles the $m_a$ point. The F-plot accurately predicts the occurrence of SNB. The sampled-data poles are 1.5891 and 0.6501, also confirming the instability.

**Nyquist/Bode plot.** First, let $D = 0.586$. The Nyquist plot (Fig. 44(a)) does not encircle the -1 point. The Bode plot (Fig. 44(b)) starts from -6 dB, and it increases to 2 dB with the phase decreased from $180°$ (away from the -1 point, thus implying stability).

Next, let $D = 0.71$. The Nyquist plot (Fig. 44(a)) encircles the -1 point. Different from Example 1, the Nyquist plot starting from $\mathcal{N}(1)$ (for $\omega = 0$) to $\mathcal{N}(-1)$ (for $\omega = \omega_s/2$) is the upper half in Fig. 44(a). Here, $\mathcal{N}(1)$ is closer to the -1 point than $\mathcal{N}(-1)$, indicating that it is an SNB instead of a PDB. The Nyquist plot accurately predicts the occurrence of SNB. The Bode plot ((Fig. 44(b))) starts from 5 dB with the phase decreased from $180°$ (encircling the -1 point, thus implying instability). $\square$

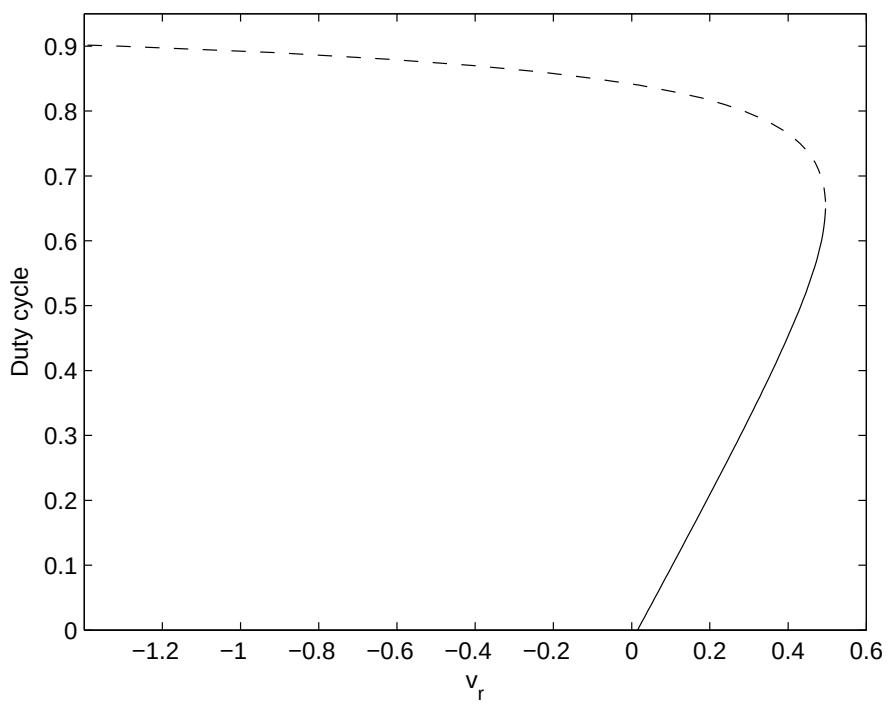

Figure 42. Bifurcation diagram showing stable (solid) and unstable (dashed) solutions. SNB occurs at $v_r = 0.496$ and $D = 0.65$.

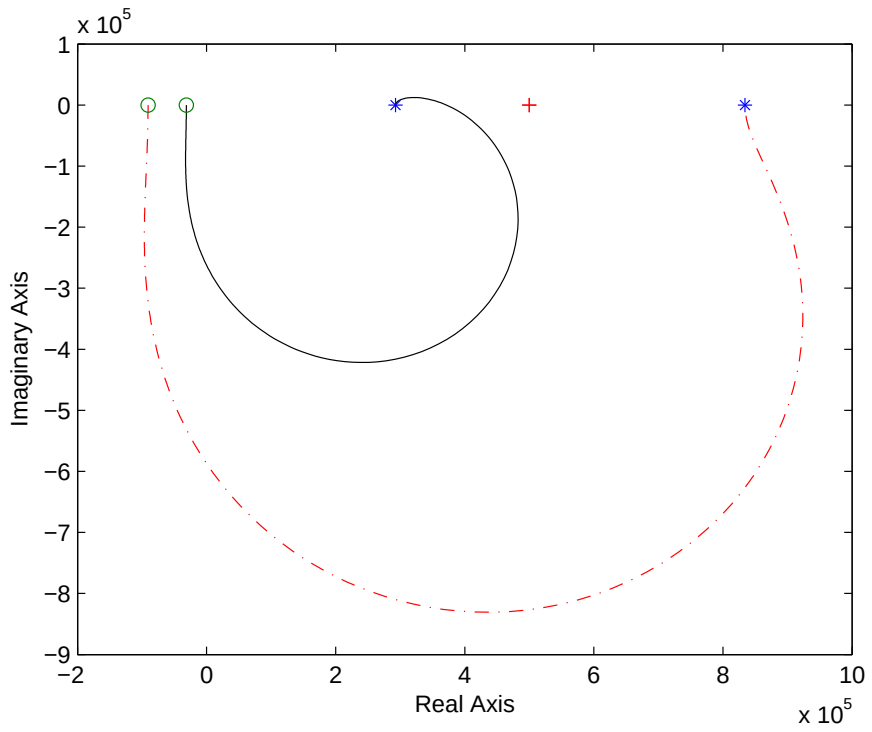

Figure 43. F-plot, solid line for $D = 0.586$ (stable) and dotted line for $D = 0.71$ (unstable).

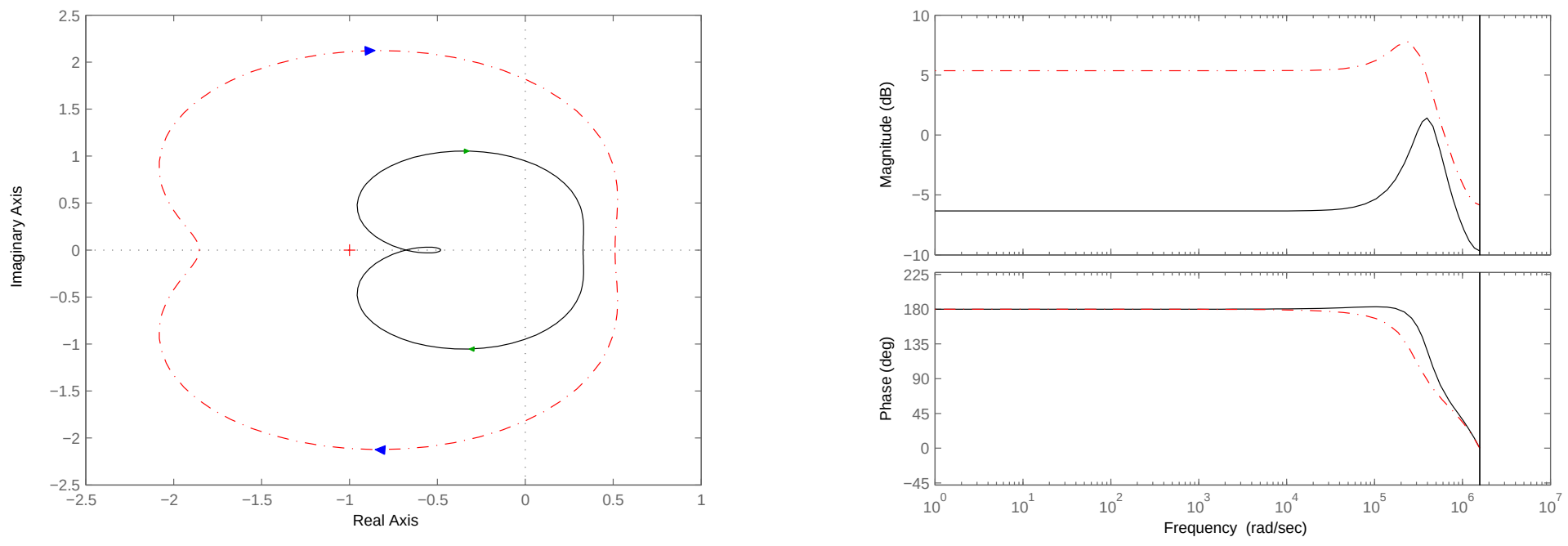

(a) (b)

Figure 44. (a) Nyquist plot, (b) Bode plot, solid line for $D = 0.586$ (stable) and dotted line for $D = 0.71$ (unstable).

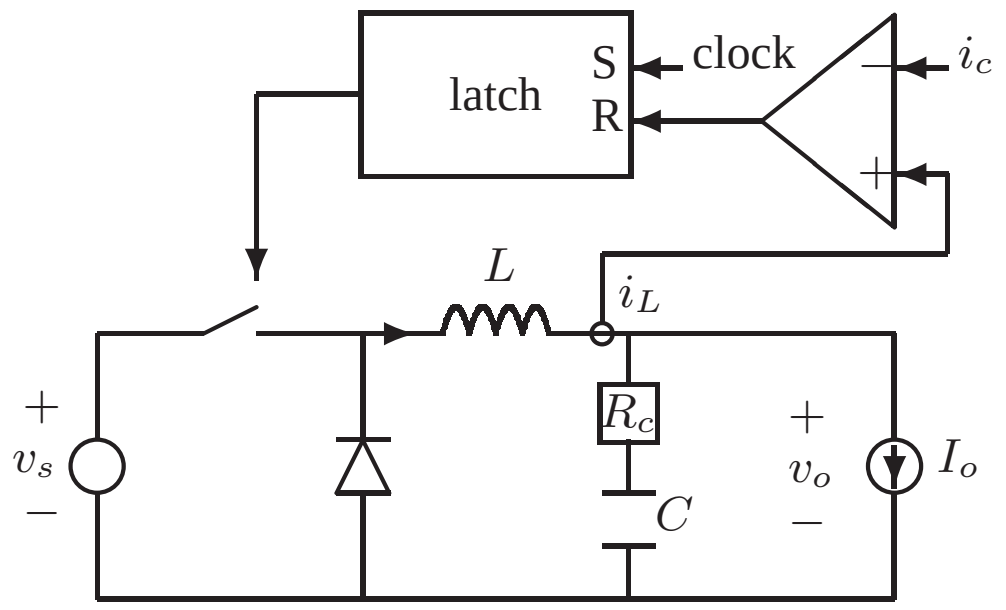

Figure 45. A buck converter under current-mode control with CCL

The next example from [18] shows that using a constant current load (CCL, such as an LED) in a CMC buck converter may cause an SNB instability. The instability is verified by four independent approaches: bifurcation diagram, coexistence of multiple solutions, the sampled-data pole trajectories, and F/Nyquist/Bode plots.

**Example 6.** (*Current mode control (CMC) buck converter with CCL.*) Consider a CMC buck converter (Fig. 45) with the following parameters [18]: $I_o = 1$ A, $v_s = 10$ V, $f_s = 1$ MHz, $R_c = 50$ m$\Omega$, $L = 10$ $\mu$H, and $C = 20$ $\mu$F. No ramp is used here and $m_a = 0$.

**Bifurcation diagram.** Let $i_c$ (peak inductor current control signal) be the bifurcation parameter. The bifurcation diagram is shown in Fig. 46. SNB occurs when $i_c = 1.125$, $D = 0.4998$, and $v_o = 5$. Note that the focus of this paper is on the $T$-periodic solution. Those non-$T$-periodic attractors are not shown in the bifurcation diagram to prevent detraction of the focus on SNB. The PDB occurs when $D = 0.5006$. As reported in [18], both PDB and SNB occur simultaneously around $D = 0.5$ if $R_c = 0$.

**Coexisting multiple solutions.** Generally in SNB, there is a hysteretic loop as shown in Fig. 46. In the figure, the upper solid line is for the operation when the switch is always on (hence, $D = 1$), and the dashed line and the lower solid line are for unstable and stable $T$-periodic solutions respectively with duty cycle less than 1. For $i_c < 1$, the switch is always off ($D = 0$). For $1 < i_c < 1.125$, there are three solutions: one stable $T$-periodic solution (with $D < 0.4998$), one unstable $T$-periodic solution (with $D > 0.4998$), and the third (stable) DC solution being that the switch is always on ($D = 1$). Take $i_c = 1.12$, for example. The two $T$-periodic solutions, with duty cycles 0.4 and 0.6, respectively, are shown in Fig. 47. The two solutions have the same peak inductor current 1.12, the same average inductor current and the same current ripple amplitude.

**Sampled-data pole trajectories.** Each solution has two poles shown in Figs. 48 and 49, respectively. The stable solution (with $D < 0.4998$) has two sampled-data poles inside the unit circle. The unstable solution (with $D > 0.4998$) has two poles outside the unit circle for $D > 0.5006$ (where PDB occurs), and it has one pole outside the unit circle and one pole inside the unit circle for $0.4998 < D < 0.5006$. When the converter operates with a periodic solution and $i_c$ is increased a little above 1.125, the output voltage will *jump up* from 5 V to 10 V. Similarly, when the converter operates with $D = 1$ and $i_c$ is decreased a little below 1, the output voltage will *jump down* from 10 V to 0 V. The jumping up and down forms a hysteretic loop shown in Fig. 46.

If SNB is not noticed, one may believe the converter is stable for $i_c < 1.125$, which is wrong. For $1 < i_c < 1.125$, the converter is actually bistable, which is a type of instability. There are two stable modes, one mode with $D = 1$ and the other with $D < 0.5$. The stable mode with $D < 0.5$ is $T$-periodic and is *locally* stable. One mode may jump to the other mode. This example illustrates the importance to uncover the existence of SNB.

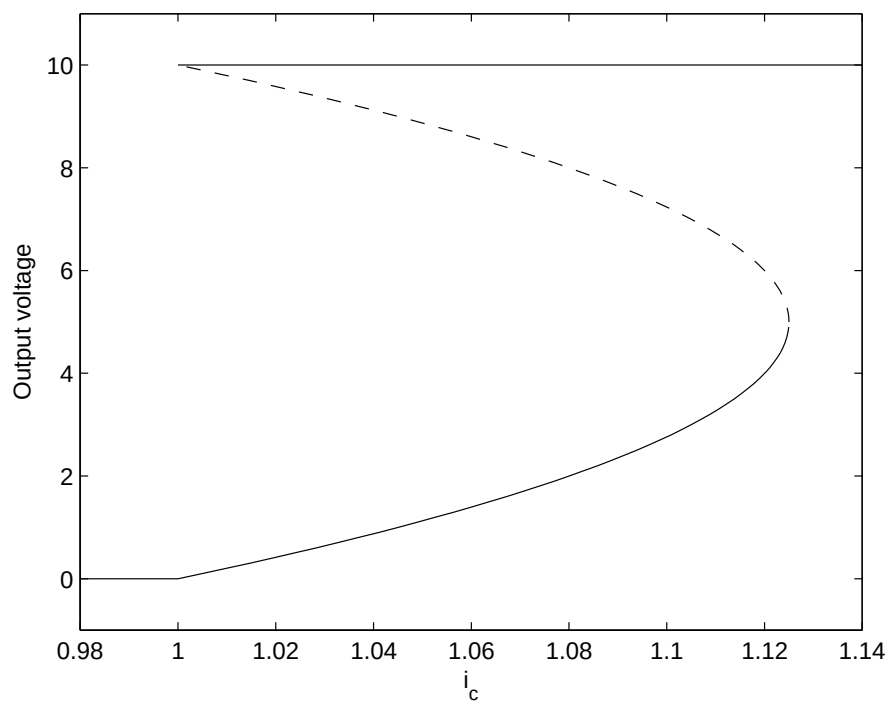

Figure 46. Bifurcation diagram shows the occurrence of SNB at $i_c = 1.125$, solid line for stable solution, dashed line for unstable solution

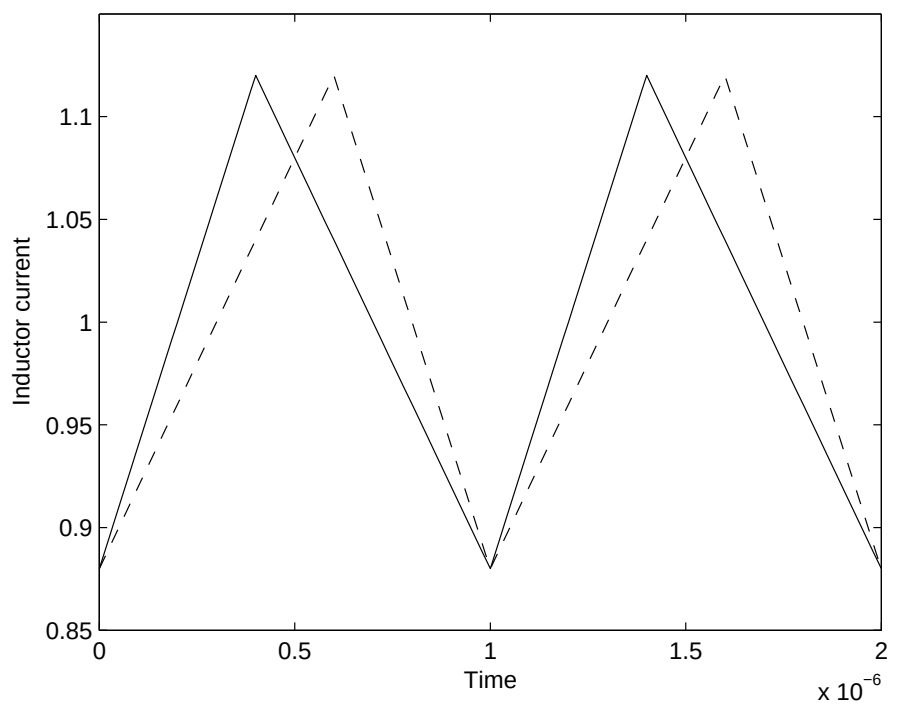

Figure 47. Coexisting two steady-state inductor currents, solid line for stable solution, dashed line for unstable solution

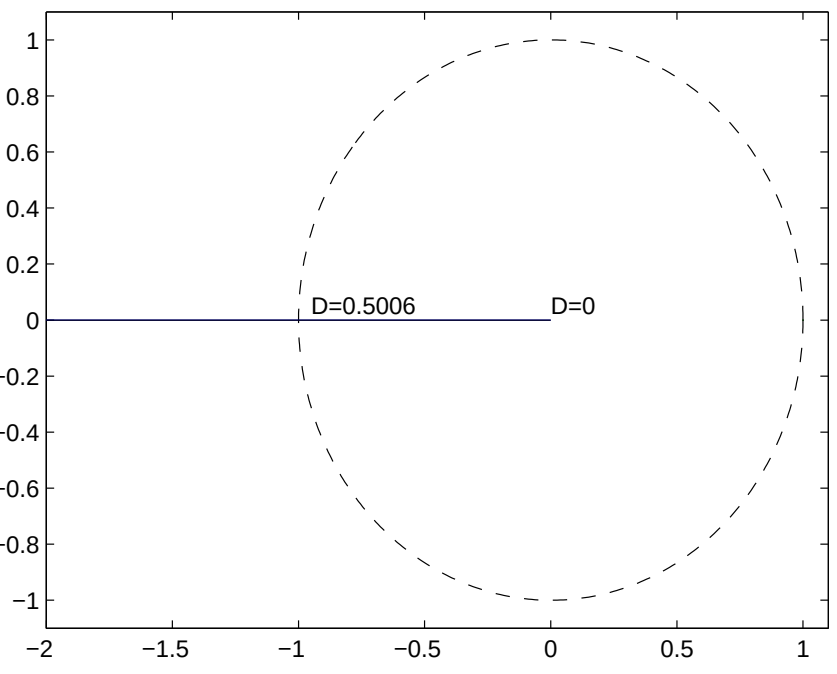

Figure 48. The locus of the first pole as $D$ varies

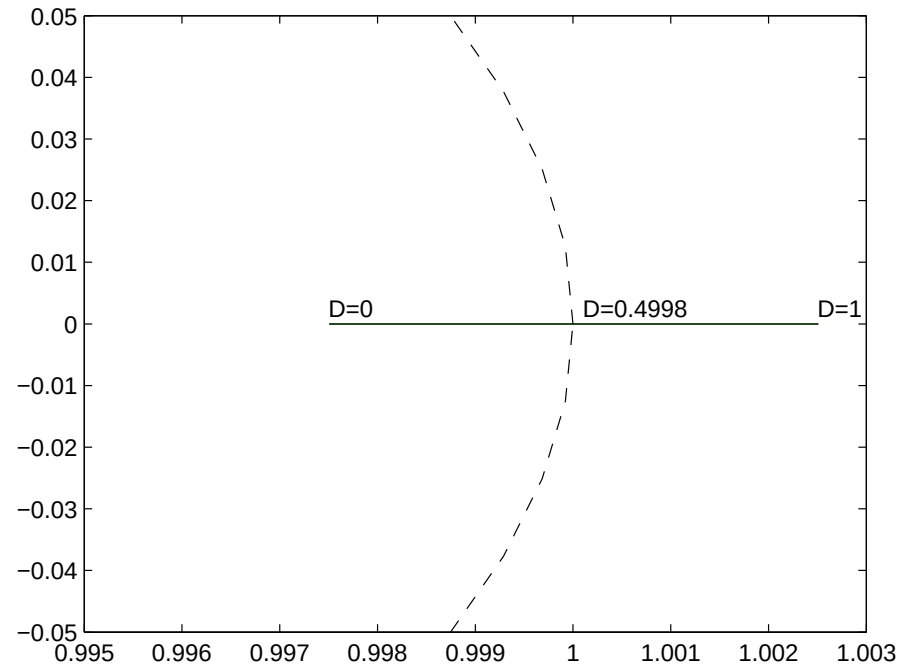

Figure 49. The locus of the second pole as $D$ varies

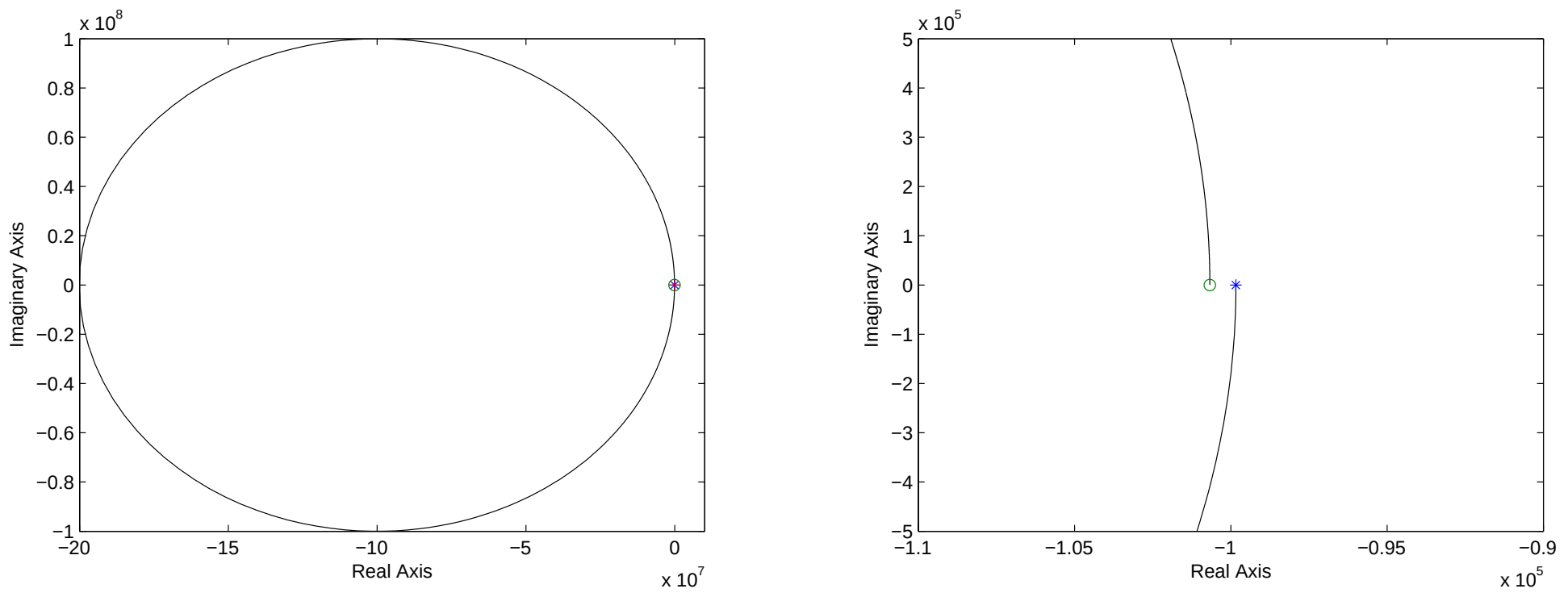

(a) (b)

Figure 50. (a) F-plot for $\theta \in [0, \pi]$, $D = 0.4$; (b) blown-up view of (a) shows that the F-plot does not encircle $(0, 0)$.

**F-plot.** First, let $D = 0.4$, the converter is stable. Since no ramp is used, $m_a = 0$. The F-plot (Fig. 50) does not encircle the $m_a$ point.

Next, let $D = 0.6$, the converter is unstable. The F-plot (Fig. 51) encircles the $m_a$ point *twice* (if a full plot for $\theta \in [-\pi, \pi]$ is made). The two encirclements are due to the fact that one pole is greater than 1 (for SNB) and the other less than -1 (for PDB). Note that here $F(0) \approx F(\pi)$, and the *double-loop* F-plot encircles the $m_a$ point simultaneously, implying the simultaneous occurrence of SNB and PDB around $D = 0.5$. If $F(0)$ and $F(\pi)$ are apart, then two encirclements of the F-plot around the $m_a$ point generally imply the occurrence of NSB with a pair of complex poles crossing the unit circle.

**Nyquist/Bode plot.** First, let $D = 0.4$, the Nyquist plot (Fig. 52) does not encircle the $-1$ point. The Bode plot (Fig. 53) shows positive gain/phase margins. Next, let $D = 0.6$, the Nyquist plot (Fig. 52, actually having two loops close to each other) encircles the $-1$ point *twice*. The Bode plot (Fig. 53) shows a negative gain margin of -1.93 dB. □

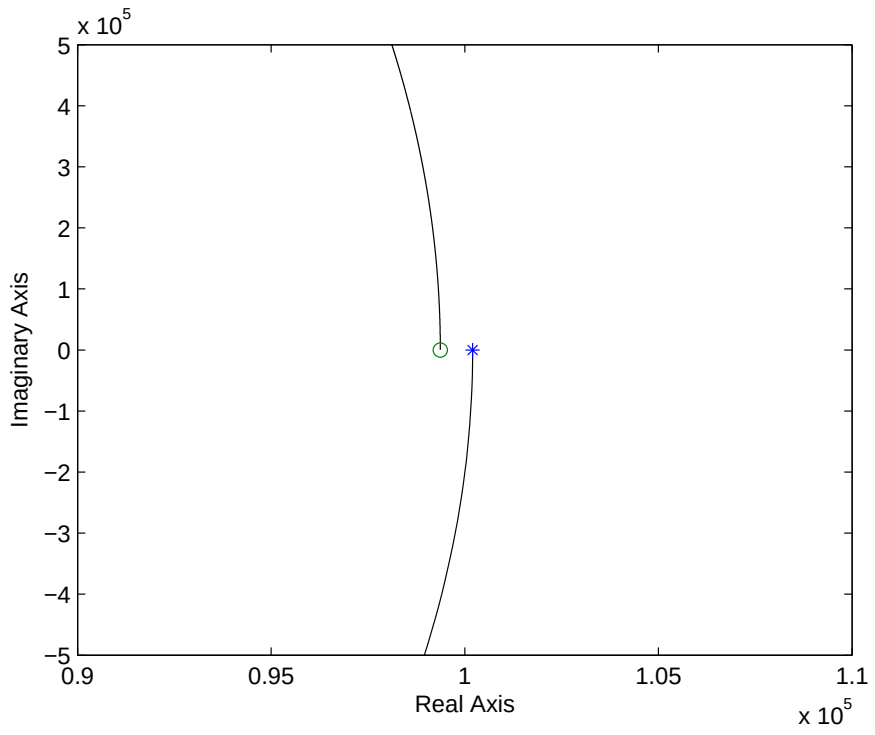

Figure 51. The F-plot encircles $(0, 0)$, $D = 0.6$.

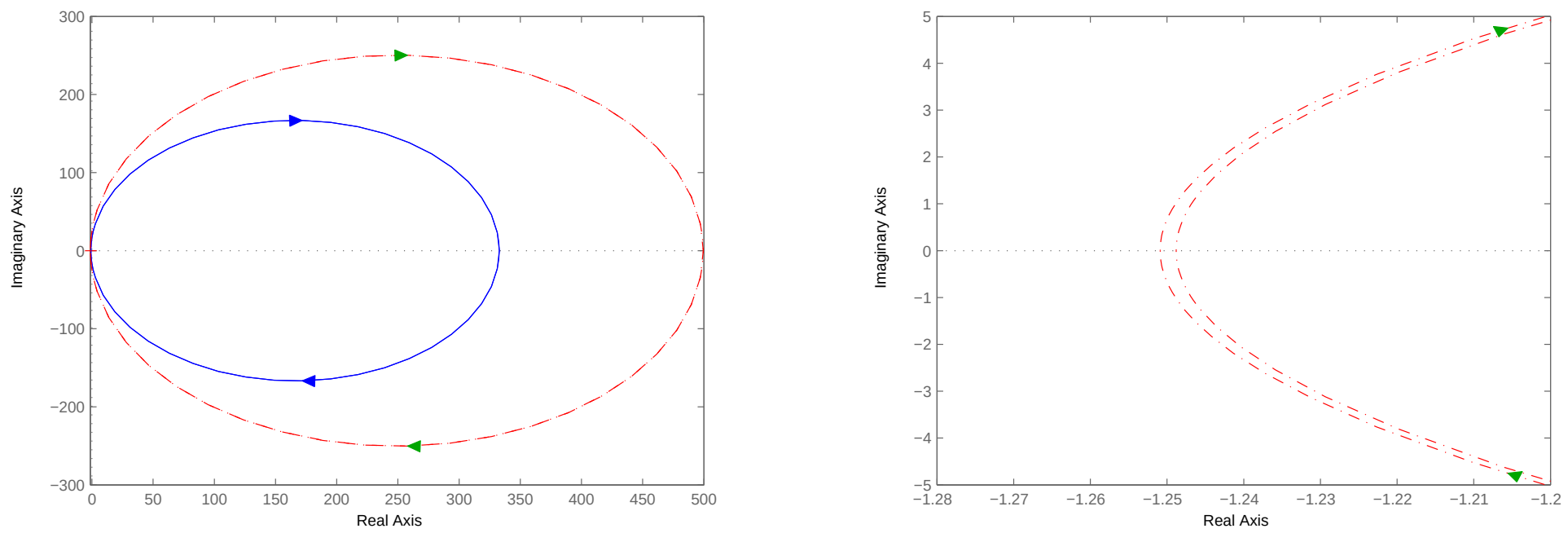

(a) (b)

Figure 52. (a) Nyquist plot, solid line for $D = 0.4$ (stable) and dotted line for $D = 0.6$ (unstable); (b) blown-up view of (a) shows that the Nyquist plot encircles the -1 point twice.

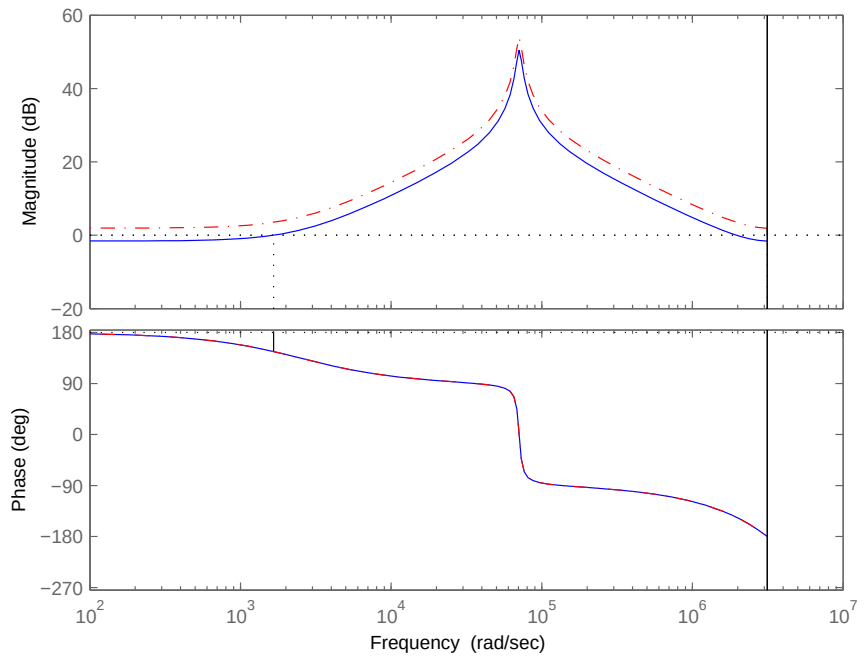

Figure 53. The Bode plot, solid line for $D = 0.4$ (stable) and dotted line for $D = 0.6$ (unstable).

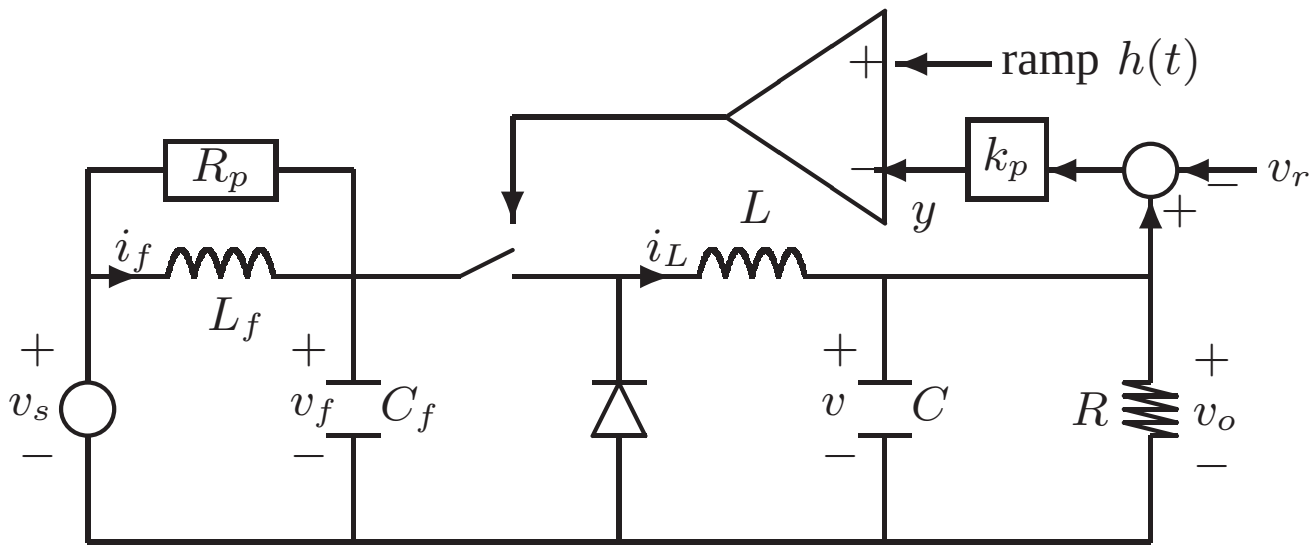

Figure 54. Buck converter with input filter.

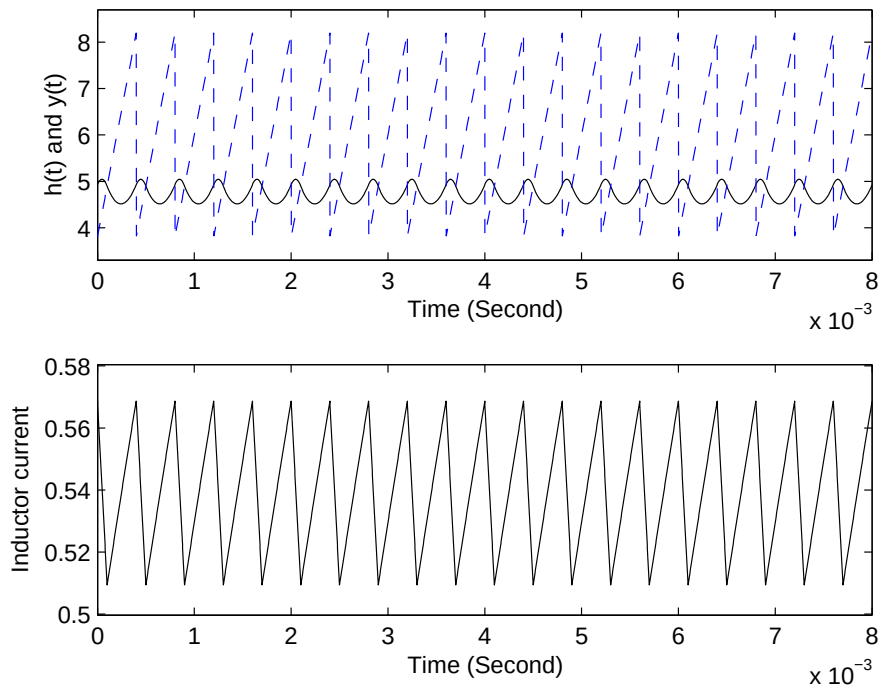

Figure 55. Stable $T$-periodic orbit, $R_p = 34$.

## VIII. Prediction of Neimark-Sacker Bifurcation (NSB)

**Example 7.** (*The F-plot accurately predicts NSB and the associated oscillation frequency.*) Consider a buck converter with input filter from [2] shown in Fig. 54 with $v_s = 15.8$ V. The power stage parameters are the same as in Example 1. The input filter parameters are $L_f$=2.5 mH and $C_f$=160 $\mu$F, and $R_p$ is varied from 1 Ω to 100 Ω. The NSB occurs when $R_p = 38.85$ [2].

**Time-domain simulation.** First, let $R_p = 34$. The converter is stable with a stable $T$-periodic orbit shown in Fig. 55. Next, let $R_p = 39$. The $T$-periodic is unstable. The state trajectory moves toward a quasi-periodic orbit shown in Fig. 56. The "period" is around $10T$. The NSB may lead to a route to border-collision bifurcation or chaos [31]. The eigenvalue trajectory of $\Phi$ as $R_p$ varies from 0 to 100 is shown in [2, Fig. 5.43]. One pair of eigenvalues is almost fixed at $-0.5963 \pm 0.5301j$, while the other pair moves as $R_p$ varies. The NSB occurs around $R_p = 38.85$, when a pair of eigenvalues $0.8087 \pm 0.5883j$ crosses the unit circle.

**F-plot.** First, let $R_p = 34$. Solving (1)-(3) gives $D = 0.79$. The F-plot (Fig. 57) does not encircle the $m_a$ point (because if a horizontal line is drawn from the $m_a$ point to the right, it does not intersect with the F-plot). As $R_p$ increases close to 38.85 when NSB occurs, $F(\theta) = m_a$ (and the F-plot touches the $m_a$ point), where $\theta \neq 0$ or $\pi$.

Next, let $R_p = 39$. The F-plot (Fig. 57) encircles the $m_a$ point twice (because if a horizontal line is drawn from the $m_a$ point to the right, it intersects twice with the F-plot for $\theta \in [-\pi, \pi]$). The two encirclements agree with the fact that there are two unstable complex poles. The F-plot accurately predicts

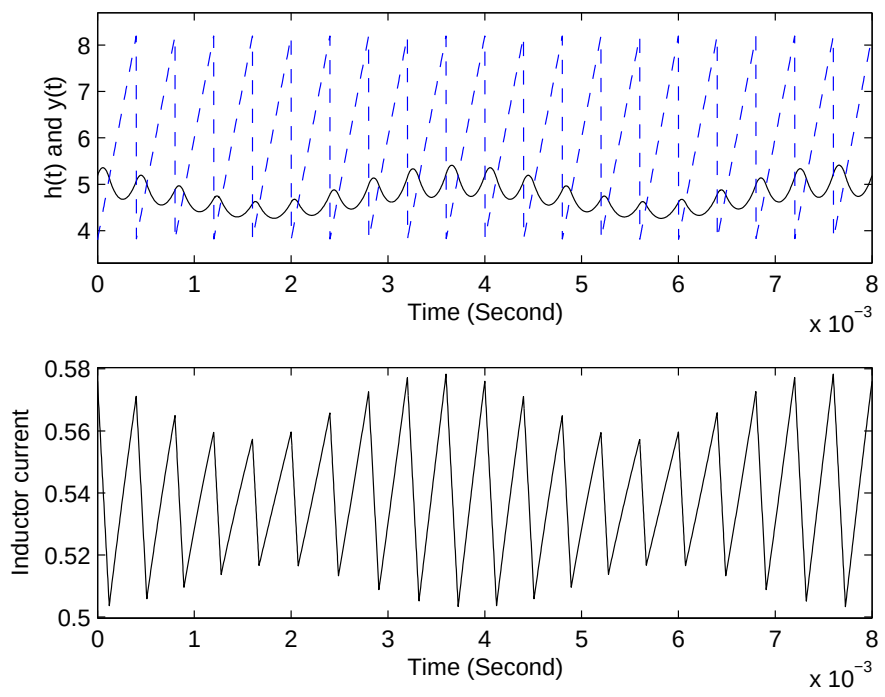

Figure 56. Quasi-periodic solution with a "period" around $10T$, $R_p = 39$.

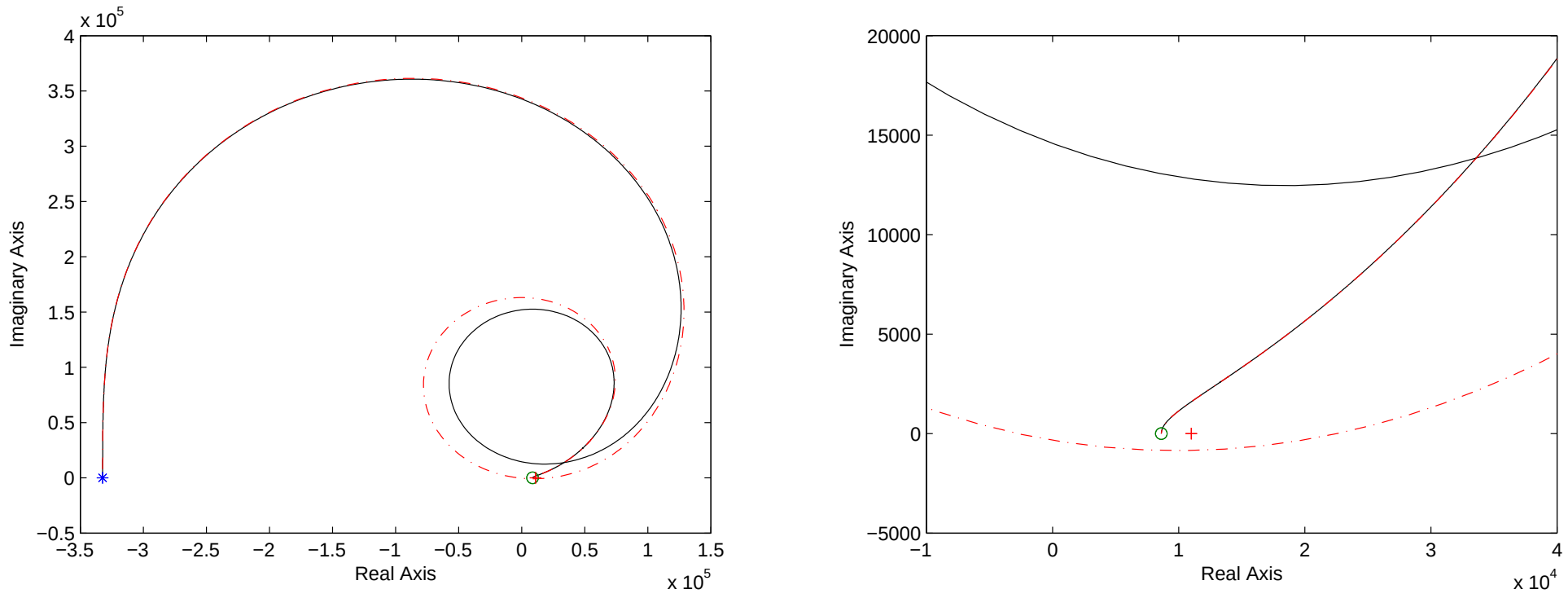

(a) (b)

Figure 57. (a) F-plot for $\theta \in [0, \pi]$, (b) blown-up view of (a); solid line for $R_p = 34$ (stable) and dotted line for $R_p = 39$ (unstable).

the occurrence of NSB.

The F-plot also predicts the oscillation frequency when the instability occurs. On the F-plot (Fig. 57), $F(0.629) = 11010 - 464j$ is close to the $m_a$ point. The expected oscillation frequency, agreed with Fig. 56, is $0.629/T = 1573 \approx \omega_s/10$ which is close to the resonance frequency of the input filter $1/\sqrt{L_f C_f} = 1581$ rad/s.

**Nyquist/Bode plot.** First, let $R_p = 34$. The Nyquist plot (Fig. 58) does not encircle the -1 point. The Bode plot (Fig. 59) shows positive gain and phase margins. Next, let $R_p = 39$. The Nyquist plot (Fig. 58) encircles the -1 point twice. The Nyquist plot accurately predicts the occurrence of NSB. The Bode plot (Fig. 59) shows a negative phase margin exactly at 1573 rad/s. The Bode plot intersects with the $-180°$ line twice, also showing the two encirclements. □

**Example 8.** (*Accurate prediction of NSB in a buck converter with a type-III compensator.*) Consider again Example 3 where $R_c = 5$ mΩ. Now, change it to $R_c = 0.427$ mΩ. Based on the sampled-data analysis, the poles are $-0.276 \pm 0.9618j$, 0.9477, 0.8884, and 0.0259. The complex poles outside the unit circle indicates the occurrence of NSB. The time-domain simulation (Fig. 60) shows a quasi-periodic solution with a "period" around $3.4T$ (corresponding to a frequency of 554400 rad/s), confirming the occurrence of NSB. The NSB may lead to a route to border-collision bifurcation or chaos [31].

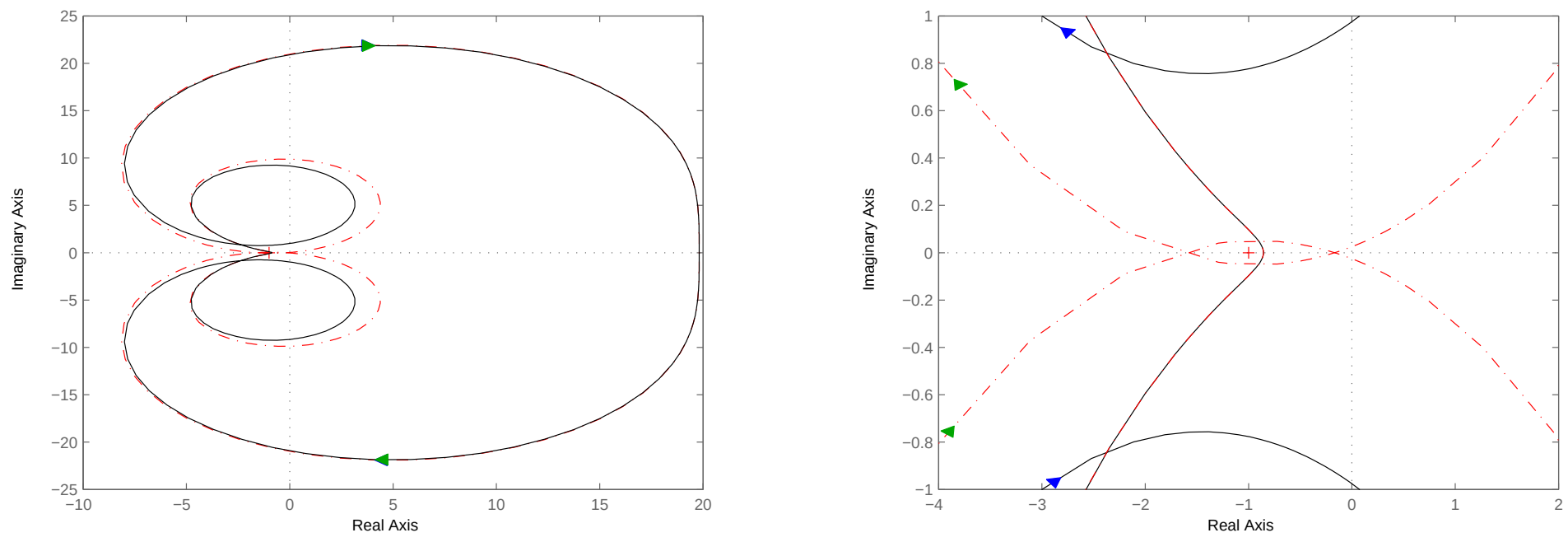

(a) (b)

Figure 58. (a) Nyquist plot, (b) blown-up view of (a); solid line for $R_p = 34$ (stable) and dotted line for $R_p = 39$ (unstable).

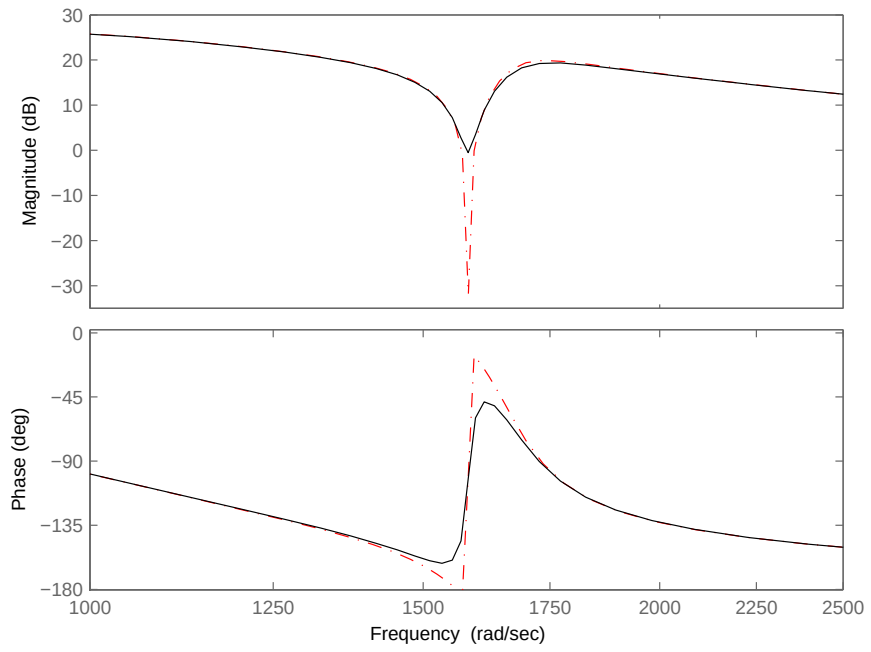

Figure 59. Bode plot (blown-up view), solid line for $R_p = 34$ (stable) and dotted line for $R_p = 39$ (unstable).

**Confirmation by F/Nyquist/Bode plots.** The F-plot (Fig. 61) encircles the $m_a$ point *twice* (if a full plot for $\theta \in [-\pi, \pi]$ is made). The Nyquist plot (Fig. 62) encircles the $-1$ point twice. The Bode plot (Fig. 63) shows negative gain/phase margins. It also shows an oscillation frequency around 554400 rad/s, agreed with the time-domain simulation. The Bode plot intersects with the $-180°$ line twice, also showing the two encirclements. □

## IX. Similar Methodology Applied to Constant On-Time Control (COTC)

For COTC, from [10], the linearized sampled-data dynamics is

$$\hat{x}_{n+1} = \Phi \hat{x}_n \tag{17}$$

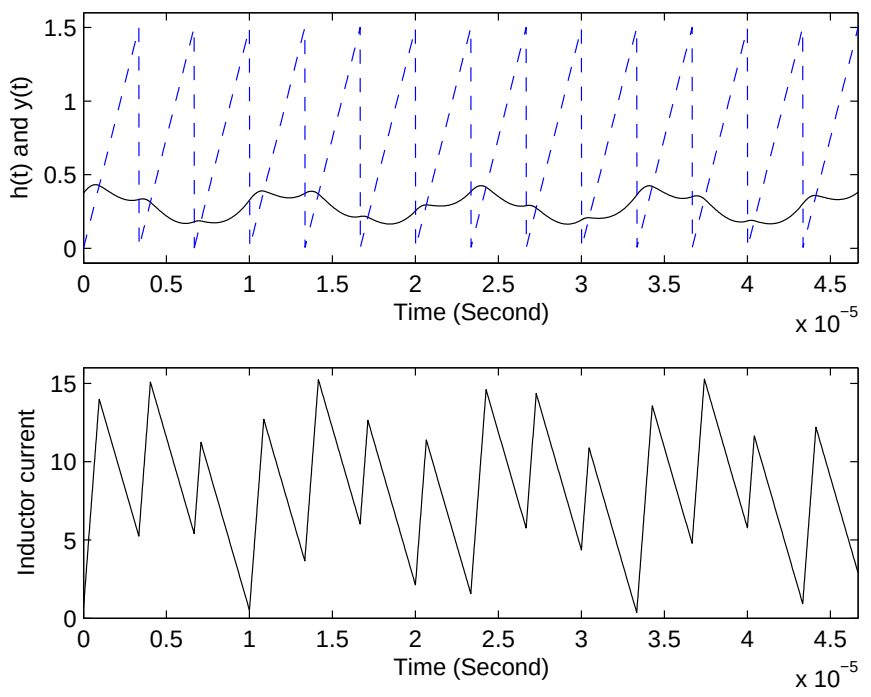

Figure 60. Quasi-periodic solution with a "period" around $3.4T$, $R_c = 0.427$ mΩ.

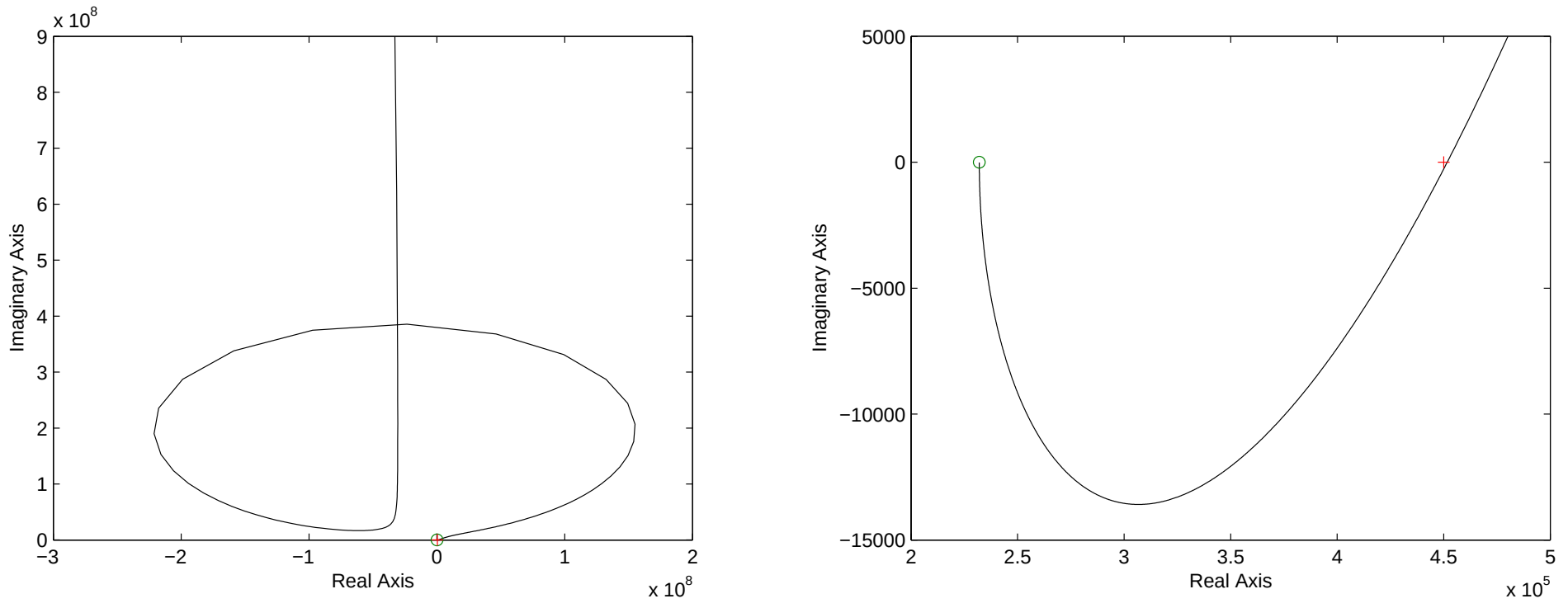

(a) (b)

Figure 61. (a) F-plot for $\theta \in [0, \pi]$, $R_c = 0.427$ mΩ; (b) blown-up view of (a) shows that the F-plot encircles the $m_a$ point.

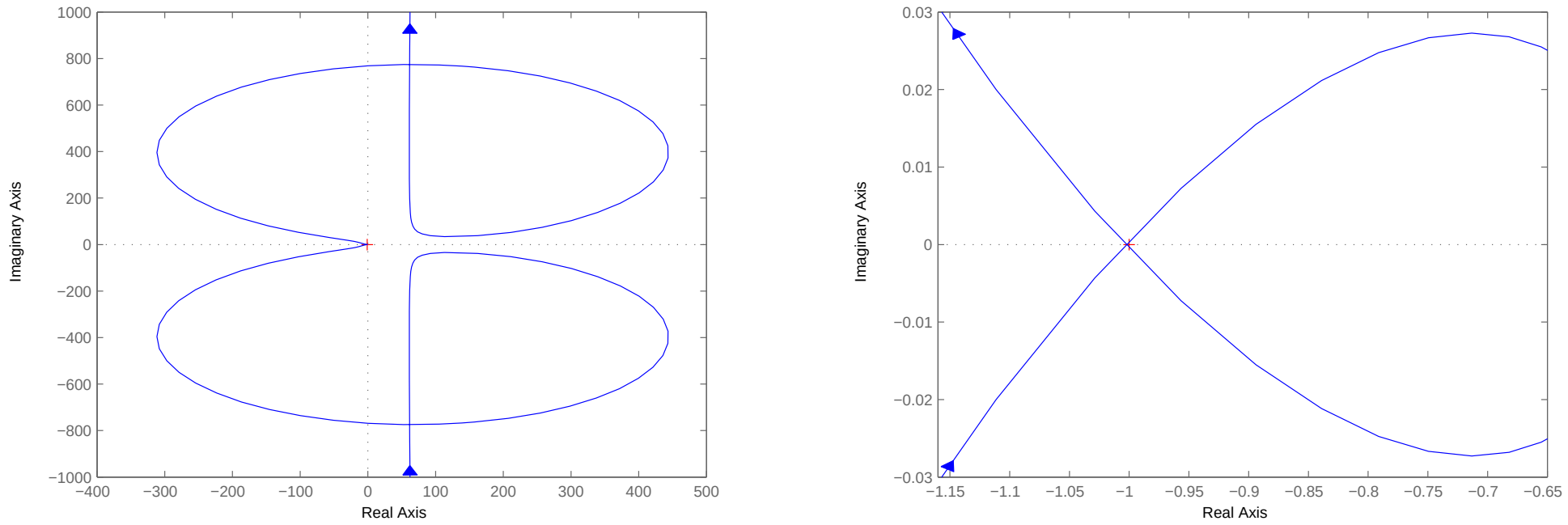

(a) (b)

Figure 62. (a) Nyquist plot, $R_c = 0.427$ mΩ; (b) blown-up view of (a) shows that the Nyquist plot encircles the -1 point.

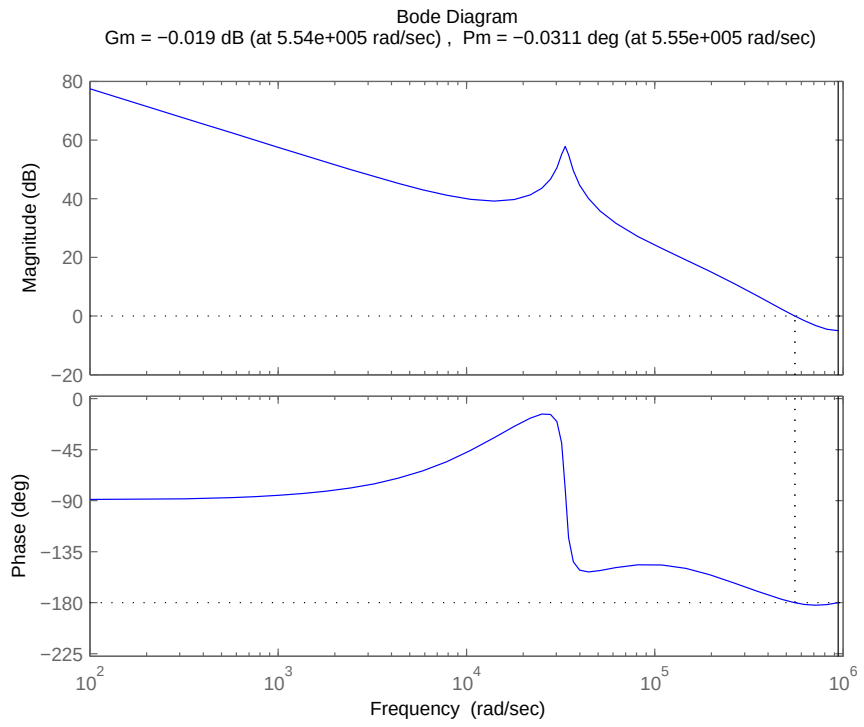

Figure 63. The Bode plot shows negative gain/phase margins, $R_c = 0.427$ m$\Omega$.

where

$$\begin{aligned}\Phi &= \frac{\partial f}{\partial x_n} - \frac{\partial f}{\partial T_n}(\frac{\partial g}{\partial T_n})^{-1}\frac{\partial g}{\partial x_n}\Big|_{(x_n,u_n,T_n)=(x^0(0),u,T)} \\ &= (I - \frac{\dot{x}^0(0^-)C}{C\dot{x}^0(0^-) - m_a})e^{A_2(T-d)}e^{A_1 d} \end{aligned} \tag{18}$$

From [10], one has the following result. Suppose that $z$ is not an eigenvalue of $e^{A_2(T-d)}e^{A_1 d}$. Then $z$ is an eigenvalue of $\Phi$ if and only if

$$C(I - z^{-1}e^{A_1 d}e^{A_2(T-d)})^{-1}\dot{x}^0(0^-) = m_a \tag{19}$$

The proof is as follows. Suppose $z$ is not an eigenvalue of $e^{A_2(T-d)}e^{A_1 d}$, then

$$\begin{aligned}\det[zI - \Phi] &= \det[zI - e^{A_2(T-d)}e^{A_1 d}] \cdot \\ &\quad \det[I + (zI - e^{A_2(T-d)}e^{A_1 d})^{-1}(\frac{\dot{x}^0(0^-)C}{C\dot{x}^0(0^-) - m_a})e^{A_2(T-d)}e^{A_1 d} \\ &= \det[zI - e^{A_2(T-d)}e^{A_1 d}] \cdot \\ &\quad (1 + Ce^{A_2(T-d)}e^{A_1 d}(zI - e^{A_2(T-d)}e^{A_1 d})^{-1}(\frac{\dot{x}^0(0^-)}{C\dot{x}^0(0^-) - m_a}))\end{aligned}$$

Therefore, $z$ is an eigenvalue of $\Phi$ if and only if

$$1 + Ce^{A_2(T-d)}e^{A_1 d}(zI - e^{A_2(T-d)}e^{A_1 d})^{-1}(\frac{\dot{x}^0(0^-)}{C\dot{x}^0(0^-) - m_a}) = 0$$

which can be rearranged as

$$C\dot{x}^0(0^-) + C(ze^{-A_2(T-d)}e^{-A_1 d} - I)^{-1}\dot{x}^0(0^-) = m_a \tag{20}$$

leading to (19) based on the matrix equality $I + (ze^{-A_2(T-d)}e^{-A_1 d} - I)^{-1} = (I - z^{-1}e^{A_1 d}e^{A_2(T-d)})^{-1}$.

As discussed above, instability occurs when there exists an eigenvalue (also the sampled-data pole) $z$ of $\Phi$ outside the unit circle. Let $z = e^{i\theta}$ (on the unit circle), and denote the left side of (19) as a function of $\theta$, designated here as an “F-Plot” in the complex plane,

$$F(\theta) = C(I - e^{-i\theta}e^{A_1 d}e^{A_2(T-d)})^{-1}\dot{x}^0(0^-) \tag{21}$$

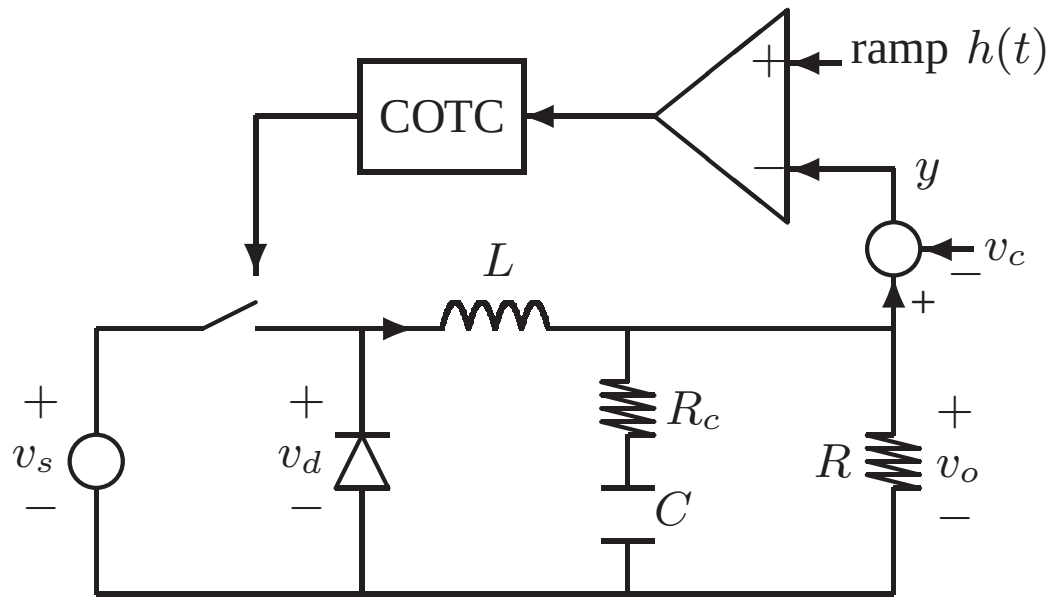

Figure 64. A voltage mode COTC buck converter.

Like the Nyquist plot, $\mathbf{Im}[F(\theta)] = -\mathbf{Im}[F(2\pi - \theta)]$ and the F-plot is symmetric with respect to the real axis in the complex plane. One can make the F-plot for $\theta \in [0, \pi]$ instead of $[0, 2\pi]$. Also like the Nyquist plot, instability occurs when the F-plot encircles the point $(m_a, 0)$ ("$m_a$ point") in the complex plane.

For stability analysis, the dynamics (17) with $\hat{u}_n = 0$ is equivalent to

$$\begin{aligned} \hat{x}_{n+1} &= \Phi_0 \hat{x}_n + \Gamma \hat{d}_n &:=& \; e^{A_2(T-d)} e^{A_1 d} \hat{x}_n + \dot{x}^0(0^-) \hat{d}_n \\ \hat{d}_n &= -\Psi \hat{x}_n &:=& \; -\tfrac{C e^{A_2(T-d)} e^{A_1 d}}{C \dot{x}^0(0^-) - m_a} \hat{x}_n \end{aligned} \tag{22}$$

or equivalent to an open-loop plant (with an input $\hat{d}_n$ and an output $\Psi \hat{x}_n$) as shown in (22) with a *unity negative* feedback. Let the loop gain transfer function of (22) be $\mathcal{N}(z)$. One has $\mathcal{N}(z) = \Psi(zI - \Phi_0)^{-1}\Gamma$. The dynamics (22) has a characteristic equation $1 + \mathcal{N}(z) = 0$, which is equivalent to (19).

**Example 9.** (*F/Nyquist/Bode plot accurately predicts PDB in COTC.*) Consider a voltage mode COTC buck converter (Fig. 64) from [32] with the following parameters: $v_s = 5$ V, $T = 3$ $\mu$s, $d = 1.2$ $\mu$s, $R = 0.5$ $\Omega$, $R_c = 20$ m$\Omega$, $L = 2$ $\mu$H, and $C = 20$ $\mu$F. Here, $D = d/T = 0.4$ and $v_r = v_s D = 2$. Also no ramp is used here and $m_a = 0$. With these parameters, instability occurs as shown in [32, Fig. 9].

**Sampled-data pole.** Based on the *exact* sampled-data analysis by calculating the eigenvalues of $\Phi$, one pole is 0 and the second pole is -1.1, implying occurrence of PDB.

**F-plot, Nyquist plot and Bode plot.** The F-plot is shown in Fig. 65 and it encircles the $m_a$ point (if it is plotted for $\theta \in [0, 2\pi]$ to complete the other half of symmetric part). The F-plot accurately predicts the instability. The Nyquist plot is shown in Fig. 66 and it encircles the -1 point. The Bode plot is shown in Fig. 67 which shows a negative gain margin of -0.135 dB. □

## X. Conclusion

By transforming an exact stability condition, a new F-plot is proposed to predict occurrences of *all three* typical instabilities in DC-DC converters in a *single* plot. The F-plot predicts *exactly* the critical parameters (such as $v_s = 24.5$ in Example 1) when the instabilities occurs because it is based on the exact switching system of Fig. 1. The F-plot is equivalent to the Nyquist plot and therefore shares similar properties. It is symmetric with respect to the real axis in the complex plane and it shows instability if it encircles the $m_a$ point in the complex plane. The Nyquist plot or Bode plot of a *particular discrete-time* dynamics (8), not reported before, also accurately predicts the instability. The F-plot offers a different perspective from the Nyquist plot or the Bode plot. Depending on the closeness of the $m_a$ point to either $F(0)$, $F(\pi)$, or $F(\theta)$ (where $\theta \neq 0$ or $\pi$), one can predict respectively which instability (SNB, PDB, or NSB) is likely to occur. The F-plot also shows the required ramp slope to avoid these instabilities and predicts the oscillation frequency associated with NSB. Nine examples are used to show the accurate prediction of F/Nyquist/Bode plot to predict PDB/SNB/NSB in the buck or boost converter with fixed or variable switching frequency under various control schemes.

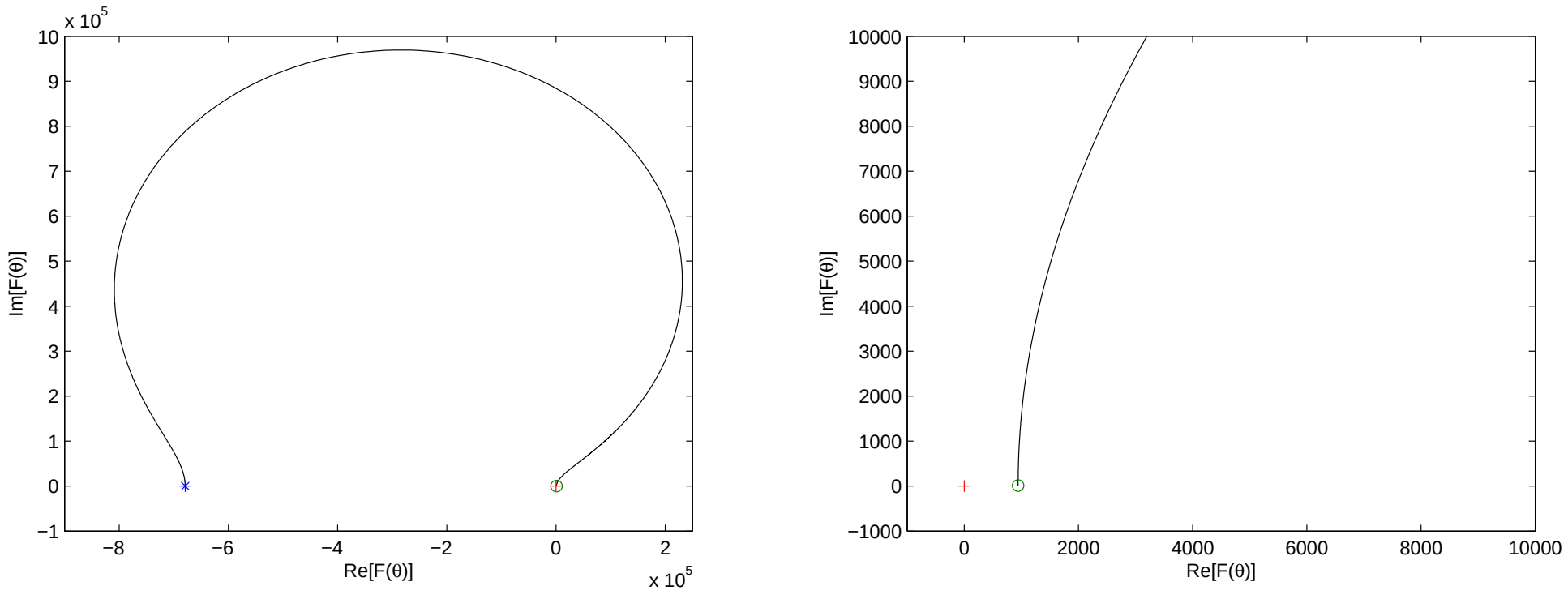

(a) (b)

Figure 65. The F-plot starts from $F(0)$ (marked as *) and ends at $F(\pi)$ (marked as $\circ$), (a) $\theta \in [0, \pi]$, (b) blown-up view of (a) shows that the $m_a$ point (marked as $+$) is encircled by the F-plot (if the lower symmetric half is plotted).

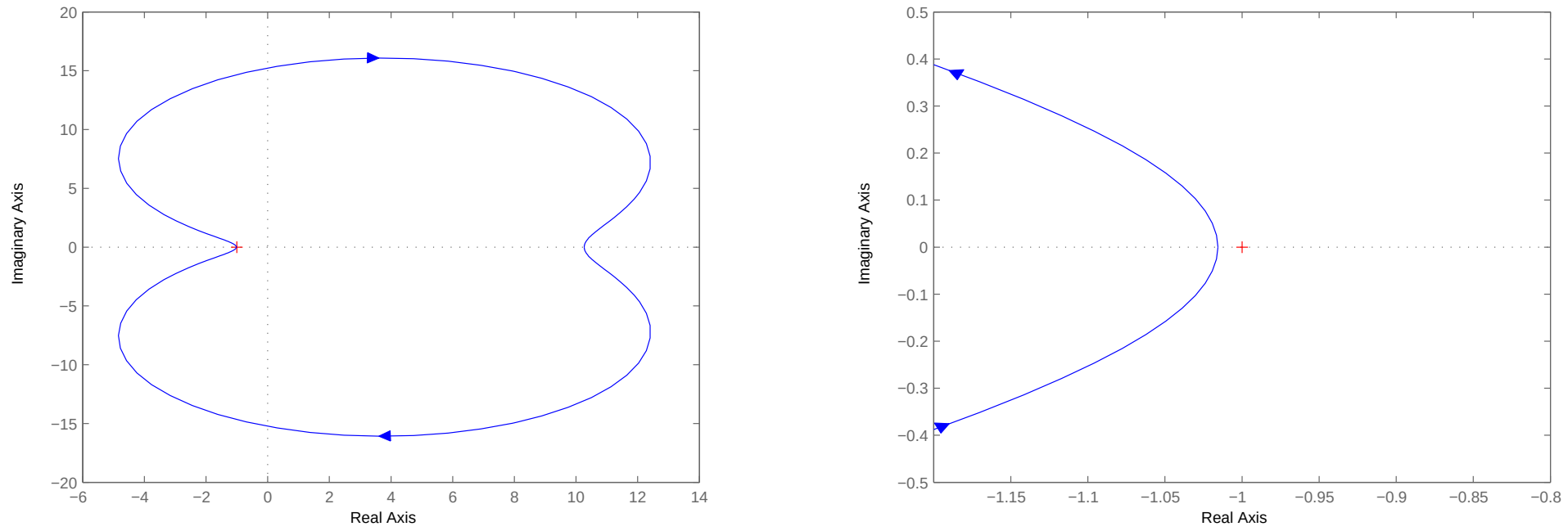

(a) (b)

Figure 66. (a) The Nyquist plot. (b) The blown-up view of Nyquist plot shows encirclement of the -1 point.

## References

[1] C.-C. Fang. *Sampled-Data Analysis and Control of DC-DC Switching Converters*. PhD thesis, Dept. of Elect. Eng., Univ. of Maryland, College Park, 1997. available: http://www.lib.umd.edu/drum/, also published by UMI Dissertation Publishing in 1997.

[2] C.-C. Fang and E. H. Abed. Saddle-node bifurcation and Neimark bifurcation in PWM DC-DC converters. In S. Banerjee and G. C. Verghese, editors, *Nonlinear Phenomena in Power Electronics: Bifurcations, Chaos, Control, and Applications*, pages 229–240. Wiley, New York, 2001.

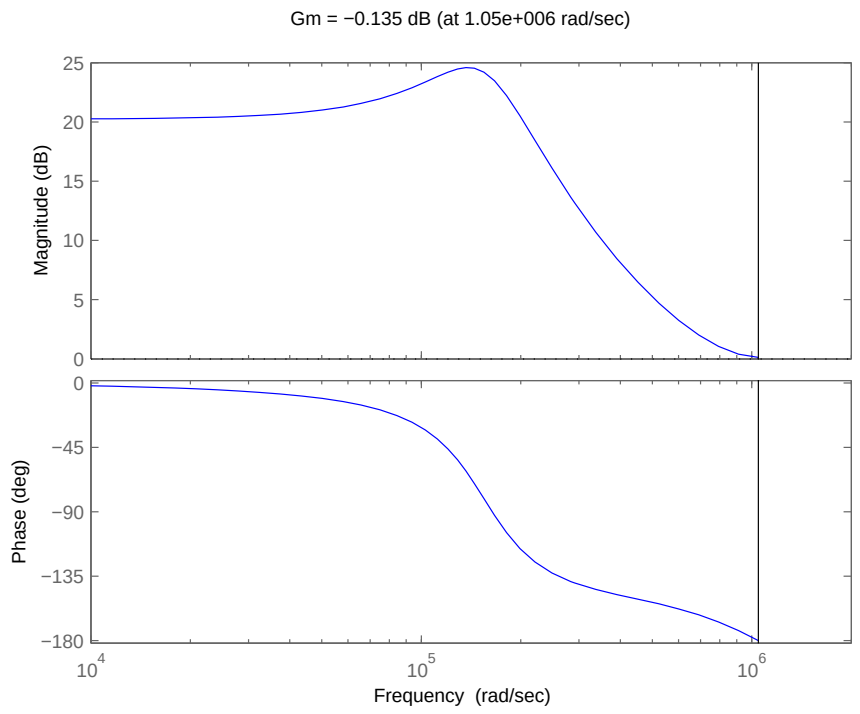

Figure 67. The Bode plot shows a negative gain margin of -0.135 dB.

[3] C.-C. Fang and E. H. Abed. Local bifurcations in DC-DC converters. In *EPE International Power Electronics And Motion Control Conference*, 2002. Paper SSIN-02, 11 pages.

[4] Y. A. Kuznetsov. *Elements of Applied Bifurcation Theory*. Springer-Verlag, New York, 1995.

[5] C.-C. Fang. Unified discrete-time modeling of buck converter in discontinuous mode. *IEEE Trans. Power Electron.*, 26(8):2335–2342, 2011.

[6] D. C. Hamill, J. H. B. Deane, and J. Jefferies. Modeling of chaotic DC-DC converters by iterated nonlinear mappings. *IEEE Trans. Power Electron.*, 7(1):25–36, 1992.

[7] C.-C. Fang. Unified model of voltage/current mode control to predict subharmonic oscillation. submitted to International Journal of Circuit Theory and Applications on Dec. 8, 2010, also available as `arXiv:1202.4232v1 [cs.SY]`, http://arxiv.org/abs/1202.4232v1.

[8] C.-C. Fang. Closed-form critical conditions of subharmonic oscillations for buck converters. submitted to an IEEE Journal on Dec. 23, 2011, resubmitted to IEEE Transactions on Circuits and Systems-I on Feb. 14, 2012, also available as `arXiv:1203.5612v1 [cs.SY]`, http://arxiv.org/abs/1203.5612.

[9] C.-C. Fang. Unified model of voltage/current mode control to predict saddle-node bifurcation. submitted to International Journal of Circuit Theory and Applications on Dec. 23, 2010, also available as `arXiv:1202.4533v1 [cs.SY]`, http://arxiv.org/abs/1202.4533.

[10] C.-C. Fang. Bifurcation boundary conditions for switching DC-DC converters under constant on-time control. submitted to International Journal of Circuit Theory and Applications on Aug. 10, 2011, also available as `arXiv:1202.4534v1 [cs.SY]`, http://arxiv.org/abs/1202.4534.

[11] C.-C. Fang. Sampled-data and harmonic balance analyses of average current-mode controlled buck converter. submitted to International Journal of Circuit Theory and Applications on Aug. 9, 2010, also available as `arXiv:1202.4537v1 [cs.SY]`, http://arxiv.org/abs/1202.4537.

[12] R. B. Ridley. A new, continuous-time model for current-mode control. *IEEE Trans. Power Electron.*, 6(2):271–280, 1991.

[13] F. D. Tan and R. D. Middlebrook. A unified model for current-programmed converters. *IEEE Trans. Power Electron.*, 10(4):397–408, 1995.

[14] A. El Aroudi, E. Rodriguez, R. Leyva, and E. Alarcon. A design-oriented combined approach for bifurcation prediction in switched-mode power converters. *IEEE Transactions on Circuits and Systems II: Express Briefs*, 57(3):218–222, Mar. 2010.

[15] B. C. Kuo. *Automatic Control Systems*. Wiley, Hoboken, NJ, Seven edition, 1995.

[16] C.-C. Fang and E. H. Abed. Limit cycle stabilization in PWM DC-DC converters. In *IEEE Conference on Decision and Control*, pages 3046–3051, Tampa, FL, USA, 1998.

[17] C.-C. Fang and E. H. Abed. Robust feedback stabilization of limit cycles in PWM DC-DC converters. *Nonlinear Dynamics*, 27(3):295–309, 2002.

[18] C.-C. Fang. Saddle-node bifurcation in the buck converter with constant current load. *Nonlinear Dynamics*, 2012. Accepted and published online, DOI: 10.1007/s11071-012-0382-6.

[19] C.-C. Fang and E. H. Abed. Sampled-data modeling and analysis of closed-loop PWM DC-DC converters. In *Proc. IEEE ISCAS*, volume 5, pages 110–115, 1999.

[20] C.-C. Fang and E. H. Abed. Sampled-data modeling and analysis of power stages of PWM DC-DC converters. *Int. J. of Electron.*, 88(3):347–369, March 2001.

[21] R. W. Erickson and D. Maksimovic. *Fundamentals of Power Electronics*. Springer, Berlin, Germany, second edition, 2001.

[22] M. K. Kazimierczuk. Transfer function of current modulator in PWM converters with current-mode control. *IEEE Trans. on Circuits Syst. I: Fundam. Theory and Appl.*, 47(9):1407–1412, 2000.

[23] C.-C. Fang. Bifurcation boundary conditions for current programmed PWM DC-DC converters at light loading. *Int. J. of Electron.*, 2011. Accepted and published online, DOI:10.1080/00207217.2012.669715.

[24] C.-C. Fang. Sampled data poles, zeros, and modeling for current mode control. *Int. J. of Circuit Theory Appl.*, 2011. Accepted and published online, DOI: 10.1002/cta.790.

[25] C.-C. Fang. Modeling and instability of average current control. In *EPE International Power Electronics And Motion Control Conference*, 2002. Paper SSIN-03, 10 pages.

[26] W. Tang, F. C. Lee, and R. B. Ridley. Small-signal modeling of average current-mode control. *IEEE Trans. Power Electron.*, 8(2):112–119, 1993.

[27] C. P. Basso. *Switch-Mode Power Supplies*. McGraw-Hill, 2008.

[28] Doug Mattingly. Designing stable compensation networks for single phase voltage mode buck regulators. Technical report, Intersil Americas Inc., 2003. available: www.intersil.com/data/tb/tb417.pdf.

[29] B. Lehman and R. M. Bass. Switching frequency dependent averaged models for PWM DC-DC converters. *IEEE Trans. Power Electron.*, 11(1):89–98, 1996.

[30] J. W. van der Woude, W. L. de Koning, and Y. Fuad. On the periodic behavior of PWM DC-DC converters. *IEEE Trans. Power Electron.*, 17(4):585–595, 2002.

[31] G. Yuan, S. Banerjee, E. Ott, and J. A. Yorke. Border-collision bifurcations in the buck converter. *IEEE Trans. on Circuits Syst. I: Fundam. Theory and Appl.*, 45(7):707–716, 1998.

[32] R. Redl and J. Sun. Ripple-based control of switching regulators - an overview. *IEEE Trans. Power Electron.*, 24(12):2669–2680, Dec. 2009.